 \documentclass[pre,amsmath,amssymb,twocolumn, showpacs, superscriptaddress,10pt]{revtex4-1}

 \usepackage{tikz,tkz-tab,tkz-graph, tkz-berge}
	\usetikzlibrary{matrix,decorations.pathreplacing,calc, positioning,automata,arrows, arrows.meta, petri, topaths, shapes, decorations.markings}
	\usepackage{pgfplots}
	
\tikzset{
    partial ellipse/.style args={#1:#2:#3}{
        insert path={+ (#1:#3) arc (#1:#2:#3)}
    }
}

\tikzset{middlearrow/.style={
        decoration={markings,
            mark= at position 0.55 with {\arrow[thick]{#1}} ,
        },
        postaction={decorate}
    }
}

\newcommand{\lattice}{
\begin{tikzpicture}[scale=1, every node/.style={scale=1, inner sep=0pt}]

\node   (0) {~};
\node[right= 200pt of 0]   (9) {~};
\node[right= 3pt of 9]{$\gamma$};

\draw[-latex] (0) --  (9) ;

  \draw[densely dotted] ($ (0)+(2.7,30pt)$) -- ($ (0)+(2.7,0pt)$);
    \draw($ (0)+(2.6,25pt)$) -- ($ (0)+(2.6,0pt)$);
    \draw($ (0)+(2.6,25pt)$) -- ($ (0)+(2.2,50pt)$);    
    
      \draw ($ (0)+(2.8,25pt)$) -- ($ (0)+(2.8,0pt)$);
            \draw ($ (0)+(2.8,25pt)$) -- ($ (0)+(3.2,50pt)$);
            
            \draw[latex-latex,dashed]($ (0)+(3.2,50pt)$)-- node [text width=1.4cm,above=1em, scale=1.15 ] {$\frac{1}{N^{1/3}\log N}$} ($ (0)+(2.2,50pt)$) ;
            
    \draw[densely dotted] ($ (0)+(4.3,-13pt)$) -- ($ (0)+(4.3,-42pt)$);

\node[below right = 3pt and 2.6 of 0]{1};
\node[below right = 3pt and 4.2 of 0]{2};

\node[above = 30pt  of 0, align = left]{\textbf{Spectrum} \\  \textbf{density}};
\node[below  = 30pt  of 9, align = right]{\textbf{Level} \\  \textbf{statistics}};

\node[above right = 14pt and 0.5 of 0]{Semi-circle};
\node[above right = 8pt and 3.7 of 0, align=left]{$\rho(a)$ column \\ energy density};

\node[below right = 17pt and 1.6 of 0]{Wigner-Dyson};
\node[below right = 17pt and 5.1 of 0, align=left]{Poisson};

\end{tikzpicture}
}

\newcommand{\MinorMatrix}{
\[
\begin{tikzpicture}[baseline,>=latex,scale=1]
\matrix[
  matrix of math nodes,
  ampersand replacement=\&,
  nodes in empty cells,
  left delimiter=(,
  right delimiter=),
  nodes={text height=12pt,text depth=3pt,text width=15pt}
] (mat)
{
\& \& \&  \\
\& \& \&  \\
\& \& \&  \\
\& \& \&  \\
};
\foreach \Valor/\Texto in 
  {
  1/{$M_1$},
  2/{$M_2$},
  3/{$M_3$},
  4/{$M_4$}
  }
{
\draw (mat-\Valor-1.south west) -| (mat-1-\Valor.north east);
\draw[->] 
  (mat-\Valor-\Valor.center) 
    to[out=60,in=180] 
  ([xshift=1cm]mat.east|-mat-\Valor-\Valor) 
    node[anchor=west] {\Texto}  ;  
}
\end{tikzpicture}
\]
}

\newcommand{\contourBBP}{

\begin{tikzpicture}[scale=1, 
						every node/.style={scale=1, inner sep=0pt},
						every axis/.append style={
                    axis x line=middle,    
                    axis y line=none,    
                    axis line style={-latex}, 
                    xlabel={~},          
                    ylabel={~}          
                    }	         ]
    \begin{axis}[ xtick={-4.5,-3,-2,0},
    xticklabels={$a_3$,$a_2$,$a_1$,$q$}, xticklabel style={yshift=-3ex, anchor=south}, ytick= {},
            xmin=-5,xmax=5,
        ymin=-5,ymax=5]
        \addplot [dashed,thick,domain=-5:5,
postaction={decorate, decoration={markings,
mark=at position 0.25 with {\arrow{latex};},
 mark=at position 0.75 with {\arrow{latex};}
      }
      }] ({1.5}, {x});
        \addplot [dotted,thick,domain=-2:2,
postaction={decorate, decoration={markings,
mark=at position 0.25 with {\arrow{latex};},
 mark=at position 0.75 with {\arrow{latex};}
      }
      }] ({-cosh(x)}, {2*sinh(x)});
      \addplot+[
  mark=+,
  only marks,
  mark size=3pt,
  mark options={line width=1pt,draw=blue, fill=blue}
] 
  coordinates
  {(-4.5,0) (-3,0) (-2,0)   };
        \addplot+[
  mark=x,
  only marks,
  mark size=3pt,
  mark options={line width=1pt,draw=red, fill=red},
] 
  coordinates
  {  (0,0) };
    \end{axis}
    
 \node[anchor=south] at (6.1,5) {contour of $J_N$};
  \node[anchor=south] at (0.3,5) {contour of $H_N$};

\end{tikzpicture}
}

\newcommand{\contourBBPreversed}{

\begin{tikzpicture}[scale=1, 
						every node/.style={scale=1, inner sep=0pt},
						every axis/.append style={
                    axis x line=middle,    
                    axis y line=none,    
                    axis line style={-latex}, 
                    xlabel={~},          
                    ylabel={~}          
                    }	         ]
    \begin{axis}[ xtick={-4.5,-3,-2,0},
    xticklabels={$a_3$,$a_2$,$a_1$,$q$}, xticklabel style={yshift=-3ex, anchor=south}, ytick= {},
            xmin=-5,xmax=5,
        ymin=-5,ymax=5]
        \addplot [dotted,thick,domain=-2:2,
postaction={decorate, decoration={markings,
mark=at position 0.25 with {\arrow{latex};},
 mark=at position 0.75 with {\arrow{latex};}
      }
      }] ({-cosh(x)}, {2*sinh(x)});
              \addplot [dotted,thick,domain=-2:2,
postaction={decorate, decoration={markings,
mark=at position 0.25 with {\arrow{latex};},
 mark=at position 0.75 with {\arrow{latex};}
      }
      }] ({cosh(x)}, {2*sinh(x)});      
        \addplot[loosely dashed] expression {-2*x};
          \addplot[loosely dashed] expression {2*x};
      \addplot+[
  mark=+,
  only marks,
  mark size=3pt,
  mark options={line width=1pt,draw=blue, fill=blue},
] 
  coordinates
  {(-4.5,0) (-3,0) (-2,0)   };
        \addplot+[
  mark=x,
  only marks,
  mark size=3pt,
  mark options={line width=1pt,draw=red, fill=red},
    mark color = black
] 
  coordinates
  {  (0,0) };
    \end{axis}
    
 \node[anchor=south,align=left] at (6,3.8) {contour of $H_N$ \\ reversed};
  \node[anchor=south] at (1.2,4) {contour of $H_N$};
    
\end{tikzpicture}
}

\usepackage{amsmath}
\usepackage{graphicx}
\usepackage{amsfonts}
\usepackage{amsthm}
\usepackage{cases}
\usepackage{bm}
\usepackage{dsfont}
\usepackage{comment}

\graphicspath{{image/}}

\newcommand{\doidoi}[2]{\href{http://dx.doi.org/#1}{#2}}

\usepackage{hyperref}
\hypersetup{
    colorlinks=true,      
    linkcolor=blue,          
    citecolor=magenta,        
    filecolor=red,      
    urlcolor=cyan          
}

\newcommand{\rmd}{\mathrm{d}}
\newcommand{\I}{\ensuremath{\mathbf{i}}}

\usepackage{color}
\definecolor{Blue}{rgb}{0.00, 0.00, 1.00}
\definecolor{Red}{rgb}{1.00, 0.00, 0.00}

\newcommand{\nn}{\nonumber}
\newcommand{\be}{\begin{equation}}
\newcommand{\ee}{\end{equation}}
\newcommand{\bea}{\begin{eqnarray}}
\newcommand{\eea}{\end{eqnarray}}

\newcommand{\inlaw}{\overset{\rm in \, law}{=}}

\newcommand{\beq}{\begin{equation}}
\newcommand{\eeq}{\end{equation}}
\newcommand{\beqn}{\begin{eqnarray}}
\newcommand{\eeqn}{\end{eqnarray}}

\DeclareMathOperator{\Ai}{Ai}

\DeclareMathOperator{\Det}{Det}

\DeclareRobustCommand{\rchi}{{\mathpalette\irchi\relax}}
\newcommand{\irchi}[2]{\raisebox{\depth}{$#1\chi$}} 

\renewcommand*{\geq}{\geqslant}
\renewcommand*{\leq}{\leqslant}

\DeclareMathOperator*{\argmax}{arg\,max}
\DeclareMathOperator*{\argmin}{arg\,min}

\begin{document}

\title{Tilted elastic lines with columnar and point disorder, non-Hermitian quantum mechanics
and spiked
random matrices: pinning and localization}

\author{Alexandre Krajenbrink}

\affiliation{SISSA and INFN, via Bonomea 265, 34136 Trieste, Italy}

\author{Pierre Le Doussal}
\affiliation{Laboratoire de Physique de l'\'Ecole Normale Sup\'erieure, CNRS,
ENS, Universit\'e PSL, Universit\'e de Paris,\\
24 rue Lhomond, 75005 Paris, France.}

\author{Neil O'Connell}
\affiliation{School of Mathematics and Statistics, University College Dublin, Ireland}

\date{\today}

\begin{abstract}
We revisit the problem of an elastic line (such as a vortex line in a superconductor) subject to both columnar disorder and point disorder in dimension $d=1+1$. Upon applying a transverse field, a 
delocalization transition is expected, beyond which the line is tilted macroscopically. 
We investigate this transition in the fixed tilt angle ensemble and
within a "one-way" model where backward jumps are neglected.
From recent results about directed polymers in the mathematics literature, and their connections to random matrix theory, we find that for a single line and a single strong defect this transition in presence of point disorder coincides with the Baik-Ben~Arous-P\'ech\'e (BBP) transition for the appearance of outliers in the spectrum of a perturbed random matrix in the Gaussian Unitary Ensemble. This transition is conveniently described in the polymer picture by a variational calculation. In the delocalized phase, the ground state energy exhibits Tracy-Widom fluctuations. In the localized phase we show, using the variational calculation, that the fluctuations of the occupation length along the columnar defect are described by $f_{\rm KPZ}$, a distribution which appears ubiquitously in the Kardar-Parisi-Zhang universality class.
We then consider a smooth density of columnar defect energies. Depending on how this density vanishes at its lower edge
we find either (i) a delocalized phase only (ii) a localized phase with a delocalization transition. We analyze this transition which is an infinite-rank extension of the BBP transition. The fluctuations of the ground state energy of a single elastic line in the localized phase (for fixed columnar defect energies) are described by a Fredholm determinant based on a new kernel, closely related to the kernel describing the largest real eigenvalues of the real Ginibre ensemble.
The case of many columns and many non-intersecting lines, relevant for the study of the Bose glass phase, is also analyzed.  The ground state energy is obtained using free probability and the Burgers equation. Connections with recent results on the generalized Rosenzweig-Porter model suggest
that the localization of many polymers occurs gradually upon increasing their lengths.
\end{abstract}

\maketitle

{\hypersetup{linkcolor=black}
\setcounter{tocdepth}{2}
    \tableofcontents
}

\newpage

.

\newpage

\section{Introduction}

\subsection{General motivation and overview}

Directed elastic lines have been used to model vortex lines in type II superconductors
\cite{blatter1994vortices,giamarchi1998statics,doussal2011novel},
aligned with an external magnetic field applied along the $z$ axis. Point impurities,
such as oxygen vacancies in high $T_c$ superconductors, provide a 
short-range correlated random potential which tends to pin the vortex lines.
Spatially correlated disorder may also arise, either planar, e.g. from 
twin boundaries, or columnar e.g. from linear defects such
as dislocation lines or damage tracks artificially created by heavy ion irradiation. 
In presence of columnar disorder along $z$ the vortex
lines tend to localize along the columns leading to the so-called Bose glass phase
(by analogy with the glass phase of interacting bosons
\cite{giamarchi1988anderson,fisher1989boson,giamarchi1996variational}), with enhanced
pinning and critical currents \cite{nelson1993boson,pldnelson1993splay,pldnelson1994splay,radzihovsky2006thermal}. 

\begin{figure}[h!]
\includegraphics[scale=0.55]{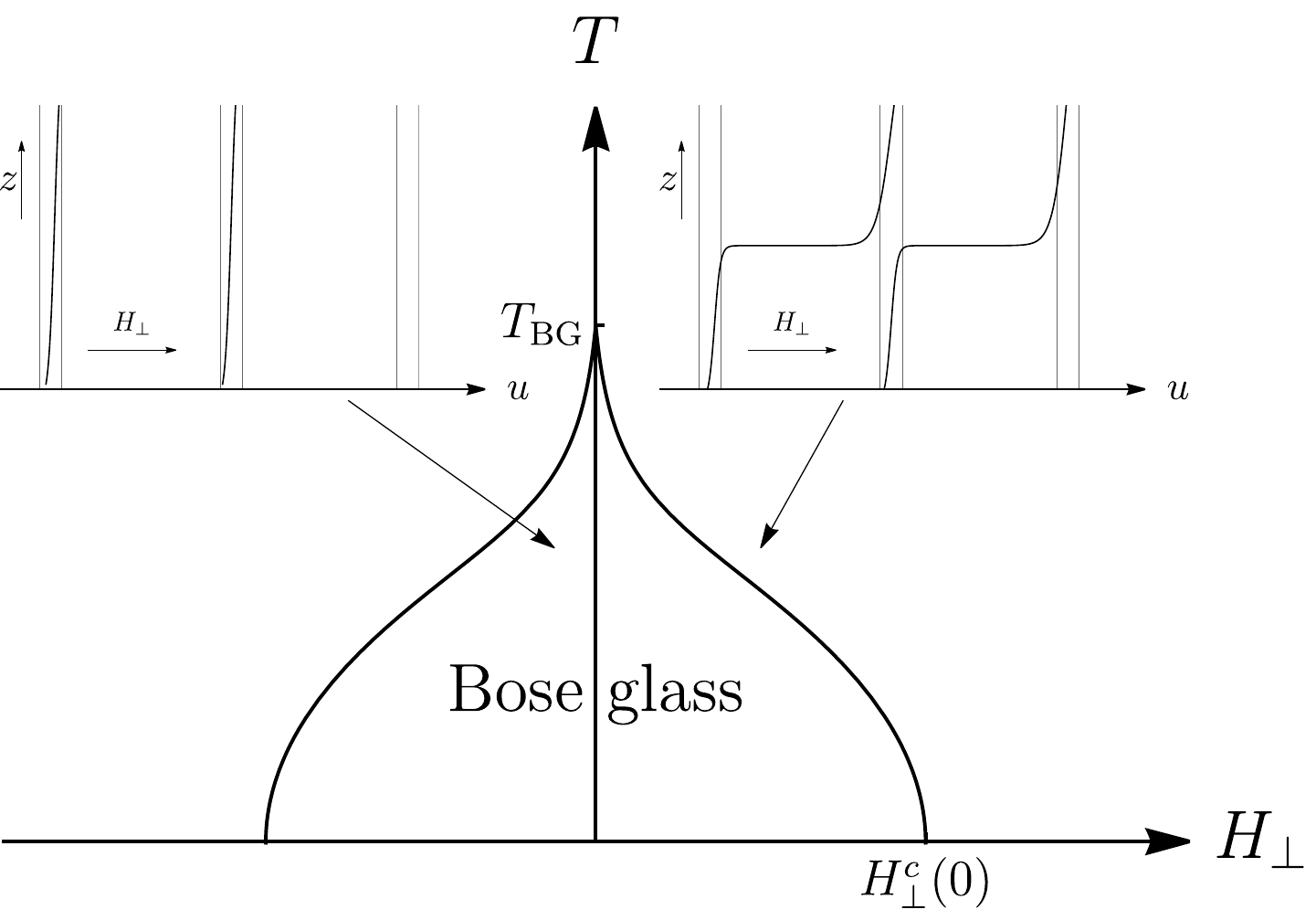}
\includegraphics[scale=0.39]{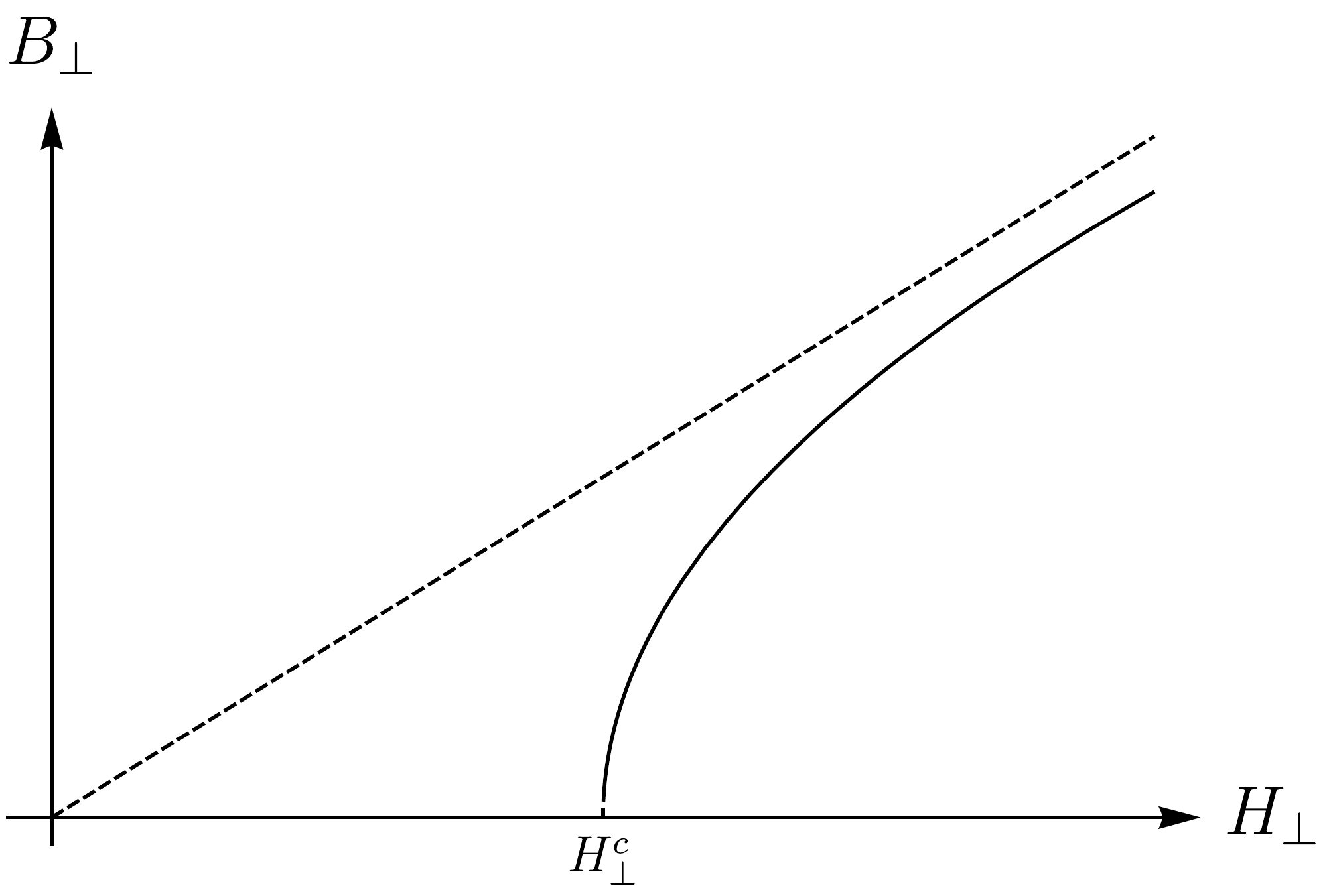}
\caption{\textbf{Top:} typical schematic phase diagram for vortex lines in the presence of columnar defects, as a function
of the temperature $T$ and of the transverse external field $H_\perp$ (noted $H$ in the text). Inset: schematic picture of the line configurations (localized along the columns in the Bose glass phase, and delocalized in the tilted phase). 
\textbf{Bottom:} behavior of the total transverse magnetic induction $B_\perp$ (proportional to the mean tilt angle of the vortex lines) as a function of  $H_\perp$ at fixed temperature. $H_\perp^c$ is the critical field above which the vortex lines begin to tilt.}
\label{fig:BG}
\end{figure}

If the external field is weakly tilted away from the $z$ direction, the response
is zero, i.e. there is a threshold transverse field needed to tilt the lines, see Fig.\ref{fig:BG}.
This effect is known
as the transverse Meissner effect and has been observed in experiments in various
geometries 
\cite{TransverseMeissnerExperiments1,TransverseMeissnerExperiments2,TransverseMeissnerExperiments3}. In the 
absence of point disorder, this transition has been described as a commensurate-incommensurate transition
\cite{nelson1993boson,hwa1993flux}. 

A continuum model for a single directed elastic line (also called directed polymer)
in dimension $d=1+1$, 
of coordinates $(u(z),z)$, is defined by the energy
\be {\cal E}[u]= \hspace{-.1cm} \label{en0} 
\int_0^L \hspace{-.2cm} \rmd z [ \frac{\gamma}{2} (\frac{\rmd u(z)}{\rmd z})^2 + U(u(z)) + V(u(z),z) - H  \frac{\rmd u(z)}{\rmd z} ]
\ee
The first term is the elastic energy cost of deforming the line away from
the $z$ axis, $\gamma$ being the line tension,
$U(u)$ is a columnar potential, $V(u,z)$ a random potential from point impurities. 
Written here in $d=1+1$, the model extends to $d=2+1$, with $u(z) \to \vec u(z)$. It is usually studied
at temperature $T$, defining the canonical partition sum 
$Z = \int \mathcal{D}u(z) e^{ - \frac{1}{T} {\cal E}[u]}$. Here $H$ is the transverse part of
the magnetic field, and the term $- H \int_0^L \rmd z \frac{\rmd u(z)}{\rmd z}= - H(u(L)-u(0))$ in the energy \eqref{en0} tends to tilt the elastic line away from the $z$ axis. In the absence of external potentials, i.e.
for $U=V=0$, the preferred slope of the line is $\frac{\rmd u}{\rmd z} = \tan \phi = H/\gamma$,
see Fig.\ref{fig:BG}.

For $V=0$ and $H=0$, the model \eqref{en0} at temperature $T$ maps onto the quantum mechanics of
a particle of position $u$ in the potential $U(u)$, described by the Hamiltonian $\hat H_0= - \frac{\gamma}{2} T^2 \partial^2_u + U(u)$ ($Z$ being its imaginary time path integral version). When $U(u)$ is a
random potential, the eigenstates of $\hat H_0$ are localized. The transverse field $H$ acts as a non-Hermitian perturbation of $\hat H_0$ and leads to delocalized states above a certain threshold field, corresponding to tilted lines \cite{hatano1996localization,hatano1997vortex,hatano1998non,shnerbnelson1998}. For the model \eqref{en0} it is easily understood by a simple argument \cite{ledou1993}. Consider a localized eigenfunction of $\hat H_0$, which decays typically as $\sim e^{-|u|/\xi}$, $\xi$ being the localization length. Since the $H$ term in \eqref{en0} is a total derivative, for $H>0$ this eigenfunction
becomes $\sim e^{-|u|/\xi - \frac{H}{T} u}$, which is normalizable (no macroscopic tilt)
for $H<H_c  = T/\xi$. For $H>H_c$ this localized state (real eigenenergy) ceases to exist and is
replaced by a delocalized state (with complex eigenenergy). Note that the higher energy, less localized states, i.e. with larger values of $\xi$, are the first one to disappear upon increasing $H$.
This problem initiated a wave of interest for the so-called non-Hermitian quantum mechanics, in particular to
study non-Hermitian localization/delocalization transitions, see e.g. Refs.~\cite{goldsheid1998distribution,feinberg1999non,brezin1998non,goldsheid2018real,hatano2020statistical}
and population dynamics \cite{shnerbnelsondahmen1998}.

The model \eqref{en0} is extended to many interacting elastic lines to study 
the transverse Meissner effect in the Bose glass phase in $d=2+1$ \cite{nelson1993boson}
or $d=1+1$ dimension \cite{giamarchi1996variational,refael2006transverse,balents1995problems}.
Schematically mimicking hard core interactions by Fermi exclusion, 
the threshold field for delocalization and macroscopic tilt is reached when the localized eigenstates 
at the "Fermi energy" start to disappear. Other situations have been
studied, such as 
many interacting lines and a single columnar defect \cite{hofstetter2004non,affleck2004non,radzihovsky2006thermal},
as well as additional (non-Hermitian) Mott phases which arise upon commensuration of the number of lines and columns, or in presence of an additional periodic potential
\cite{lehrer2000b,lehrer1998vortex,hebert2011hatano}.

The question of the additional effect of point disorder $V(u,z)$ 
is of great importance since point impurities are usually present in the experimental systems.
The competition between extended and point defects
was studied in the context of many interacting lines
and many columnar defects in \cite{hwa1993flux}. Weak point disorder was 
argued to weaken the pinning by the extended defects,
with the possibility that the Bose glass phase be unstable
to point disorder, but only beyond
an astronomical large scale. Strong point disorder
was shown to be stable to weak correlated disorder. 
The case of a single line and a single columnar defect (at $u=0$)
in dimension $d=1+1$ is rather subtle. A one-sided version 
of the model (restricted to $u \geq 0$), natural in the context of wetting, leads to 
an unbinding transition \cite{Kardarwetting}, studied 
later in the context of the half-space Kardar-Parisi-Zhang (KPZ) equation 
\cite{borodin2016directed,AlexLD,deNardisPLDTT,barraquand2020half} and of 
related models \cite{baik2001symmetrized}. However 
in the full space model it was argued that the line is always pinned (i.e. localized on
the column) at and below $d=1+1$
\cite{tang1993directed,BalentsKardar,hwa1995disorder}. The question
is now settled in the mathematics literature, it is known as the slow bond problem
\cite{basu2014last}, and it was numerically confirmed  \cite{soh2017effects}.

When the columnar defects are strong, the kink energy $E_k$, i.e. the
energy cost from going from one column to its neighbor, is large. The
polymer spends most of its length on the columns and the jumps are rare, see Appendix \ref{app:model}.
It it thus natural to study the discrete hopping model with $N$ sites in one dimension
\be \label{H1} 
\mathcal{H} = - \sum_j \frac{w_R}{2}  |j+1 \rangle \langle j | + \frac{w_L}{2} |j \rangle \langle j +1 | 
+  (\epsilon_j + \eta_j(t)) |j \rangle \langle j|
\ee 
where $w_R=w e^{h}$ and $w_L=w e^{-h}$ are the hopping rates to the 
right and the left and the $\epsilon_j$ are the on-site attractive potentials 
of the columns, which we denote for convenience in the reminder 
of this paper
\be
\epsilon_j= - a_j 
\ee 
where we often choose the column strength $a_j$ positive. The point disorder is modeled by white noise $\eta_j(t)$ i.i.d on each column.
This model without the point disorder has been much studied
\cite{hatano1996localization,hatano1997vortex,hatano1998non,shnerbnelson1998,feinbergzee1997,EfetovNHloc1997,Janik1997,Brouwer}
and the spectrum (in the complex plane) has been obtained exactly when the 
$a_j$ are i.i.d random variables from a Cauchy distribution
\cite{goldsheid1998distribution,feinberg1999non,brezin1998non}. In the absence
of columnar disorder the spectrum is concentrated along an ellipse in the complex plane,
corresponding to delocalized states, while in presence of columnar disorder 
it develops "wings" on the real axis corresponding to localized states
\cite{footnoteGinibre}.
In these works the boundary conditions at the ends of the chain of $N$ sites
are often chosen periodic. 

Since the model \eqref{H1} is quite difficult to analyze in presence of point disorder 
we will consider the simpler limit $h \to +\infty$, $w e^{h} \to 2$, where the lines can
only jump to the right (i.e. $w_L \to 0$). In that case the operator ${\cal H}$ in \eqref{H1} is the
Markov generator of the so-called O'Connell-Yor polymer (at finite temperature),
which we will study here with free boundary conditions.
Note that this "one-way" limit model, also called maximally non-Hermitian, was also studied in Refs.~ \cite{feinberg1999non} and \cite{brezin1998non} in the absence of point disorder, and retains
some of the features of the full model \eqref{H1}. In particular, for Cauchy disorder these works found that there are also localized states. 

Another motivation to study the "one-way" model is that one expects that near the
transition at $H=H_c$. e.g. just above it, the lines start tilting and the backward jumps may have a subdominant effect,
see Appendix \ref{app:model}.
Whether this model captures some of the universal features of the transition at $H=H_c$ remains to be understood. In this limit however we will present very detailed results.

\begin{figure*}[t!]
\centerline{
\includegraphics[scale=0.35]{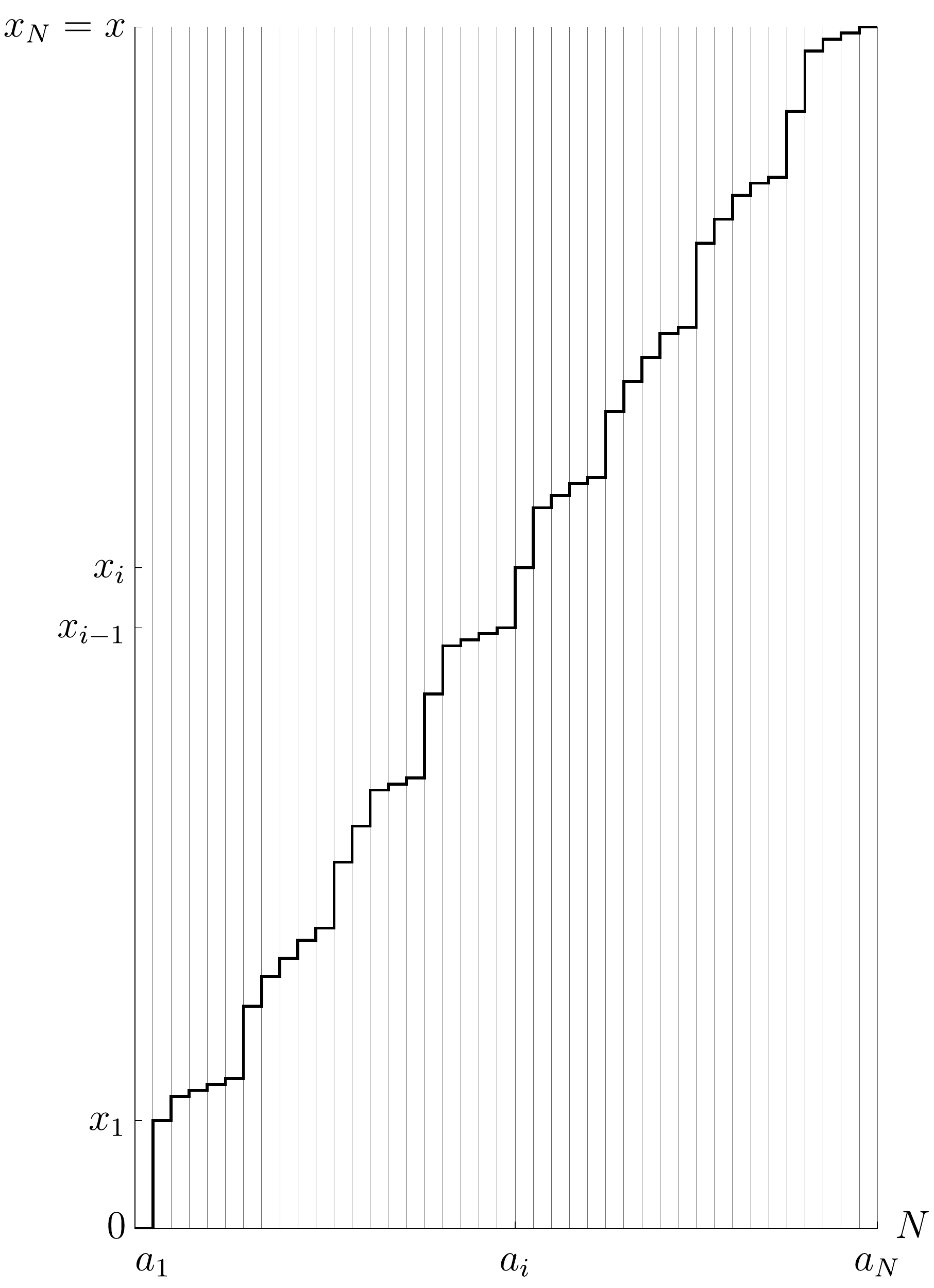}
\hspace*{2cm}
\includegraphics[scale=0.35]{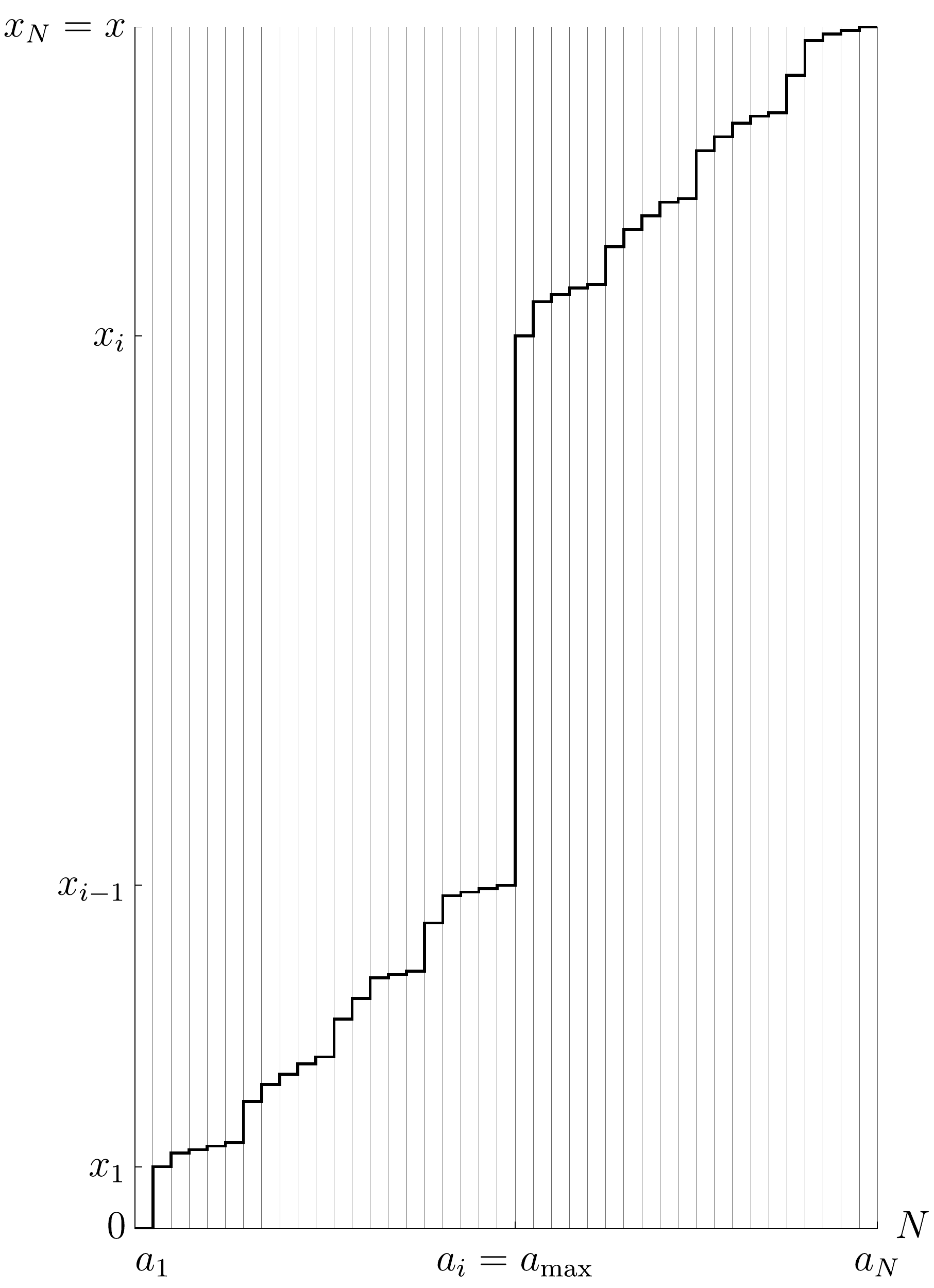}
}
\caption{\textbf{Left.} Single O'Connell-Yor polymer of total length $x$ with $N$ columnar defects,  delocalized over the space. The $x_i$ are the position of the jumps and the occupation length of column $i$ is $\ell_i=x_i-x_{i-1}$. 
\textbf{Right.} Single O'Connell-Yor polymer localized over the column $i$, where the disorder is most favorable $a_i=a_{\max}$, and which 
has a macroscopic occupation length $\ell_i = \mathcal{O}(x)$.}
\label{fig:oneline}
\end{figure*}

\subsection{Aim of the paper, model and observables}

In this paper we study a model of lines (equivalently called polymers) in $d=1+1$ in presence of
both columnar and point disorder, defined on a lattice with $N$ sites and with jumps only to the right. 
It is called the O'Connell-Yor (OY) polymer and corresponds to the one-way limit of the model \eqref{H1}.
The OY model is related to random matrix theory (RMT) and many results are known in the mathematical
context. A first aim of this paper is to review and translate these results in the language of localization/delocalization transitions for the polymers, and to make them more widely known in the physics community. 
In addition we derive some new results, in particular in the case of continuous distributions of column strengths $a_j$, or concerning the macroscopic occupation length of the columns by the lines in the localized regime, little addressed in RMT, see e.g. Fig. 
\ref{fig:oneline}. Although we briefly address finite temperature, most of our study concerns the ground state energy, and its sample to sample fluctuations in the various phases due to point disorder.

The outline is as follows.
In this subsection we first define the model and the observables for a single line. We then recall the connection to RMT in the simplest case and present a few immediate consequences for the physics of a single line. In Section 
\ref{sec:singlesingle} we study in more details the case of a single line and a single "active" column (i.e. $a_1>0$ and $a_{j \neq 1} =0$). For large $N$ there is a localization/delocalization transition related to the Baik-Ben~Arous-P\'ech\'e (BBP) transition in RMT for the appearance of outliers in the spectrum of a perturbed random matrix in the Gaussian Unitary Ensemble (GUE). In the delocalized phase, the 
fluctuations of the ground state energy due to point disorder are described by the Tracy-Widom distribution. We give a detailed description of the occupation length of the columnar defect by the line, 
see Fig. \ref{fig:oneline}, and its fluctuations, in both phases and near the transition. 
In Section \ref{sec:singlemany} we study one line and many columnar defects, in the case of a
smooth density $\rho(a)$ of column strengths. It corresponds to a perturbation of {\it infinite rank} of
a GUE random matrix. Only a few works have addressed infinite rank perturbations \cite{baik2014batch,shcherbina2011universality,Capitaine_2015}, see also
\cite[Remark 2]{borodin2008airy}, \cite[Section 5.5.4]{knizel2019generalizations},
but not in the regime of interest here. We show that if $\rho(a)$ vanish sufficiently fast near its
(finite) upper edge there is a localization transition for the polymer. We obtain the fluctuations of the 
ground state energy (for fixed column strengths) in the localized phase, and 
around criticality, We show that it is described by a new one-parameter 
universal distribution, reminiscent of the one describing the largest real eigenvalue of the real Ginibre ensemble of random matrices. In Section \ref{sec:manymany} we extend our study to many lines,
first with a few active columnar defects, then with many columnar defects. The latter case can be studied
using free probability and Burgers equation, and has connections with the Rosenzweig-Porter model,
a toy model for many-body localization much investigated recently. We find that many line localization can occur for sufficiently long polymers, via an intermediate non-ergodic delocalized phase.

The study in this paper is performed in the fixed tilt angle ensemble. In Appendix \ref{app:model}  we discuss
the fixed transverse field $H$ ensemble. We first recall the picture of the tilting transition for the continuum model \eqref{en0} of an elastic line. We discuss a possible realization of the OY model by introducing
a periodic array of columns of various strengths, and discuss the effect of point disorder.

The Appendices \ref{app:dbm}, \ref{app:BBP} and \ref{app:defpoly} recall useful results about
the Dyson Brownian motion and the BBP kernel, and define the many-line model. The Appendices
\ref{app:variational} and \ref{app:trans} give more details about a variational calculation and 
the approach of the transition from the delocalized phase.

\subsubsection{Definition of the model for a single line}

The O'Connell-Yor polymer model \cite{o2001brownian}, extended to arbitrary drifts, is defined as follows. The directed polymer path lives only on the columns $j=1,\dots,N$ and 
jumps from column $j$ to $j+1$ at a height $x_j$. There are no leftward jumps. The path 
is parametrized by the set ${\bf x}=\{x_j \}_{0 \leq j \leq N}$ 
with $x_0=0<x_1<\dots<x_N=x$, see Fig. \ref{fig:oneline}. One part of its energy is 
\be
E^p_N({\bf x}) = B_1(x_1) + B_2(x_2,x_1)  + \dots + B_{N}(x,x_{N-1})
\ee
where the $B_j(x)$ are independent unit Brownian motions with $B_j(0)=0$.
The $B_j(x,x')=B_j(x)-B_j(x')$ represent the total random 
energy from point impurity disorder 
collected along column $j$. Note that the endpoints are fixed at $(x,j)=(0,1)$ and $(x,N)$.
In addition, there is a (negative) binding energy $\epsilon_j = - a_j \leq 0$ to the columnar defects
\be
E^c_N({\bf x}) =  -  \sum_{j=1}^{N} a_j \ell_j \quad , \quad \ell_j = x_{j}-x_{j-1}
\ee 
where $\ell_j$ is the length along column $j$ occupied by the directed polymer. The $a_j$
are also called drifts since they can also be seen as (minus) the drifts for the Brownian $B_j$.

The model at temperature $T$ is defined by its canonical partition sum 
\be \label{Z} 
\mathcal{Z}_N(x,T)= \int_0^x \rmd x_1 \int_{x_1}^x \rmd x_2 \dots \int_{x_{N-2}}^x \rmd x_{N-1} e^{- E_N({\bf x})/T} 
\ee 
where $E_N({\bf x})=E^p_N({\bf x}) + E^c_N({\bf x})$ is the total disorder energy.
Its free energy is ${\cal F}_N(x,T)=- T \log \mathcal{Z}_N(x,T)$, and at $T=0$ it
equals the ground state energy defined by the minimization problem
\be
{\cal E}^0_N(x) = {\cal F}_N(x,T=0)=\min_{{\bf x}} E_N({\bf x})
\ee 
In general ${\cal F}_N(x,T)$ fluctuates from sample to sample w.r.t. the point disorder
(the Brownian motions) as well as the columnar disorder (the $\epsilon_j=-a_j$).
In this paper we will study the fluctuations w.r.t the Brownian motions, for a 
fixed values of the columnar strengths $a_j$. 
Hence we are interested in the 
mean value and probability distribution function (PDF), for a given set of $a_j$. Indeed these
observables allow to distinguish the various phases.
Note that, remarkably, one can show that the PDF of ${\cal F}_N(x,T)$ is invariant
by any permutation of the $a_j$
\cite{footnotenew1}.

To study a single line, we will be interested in the limit of both $N$ and $x$ are taken large, with 
a fixed ratio $\theta = x/N$. Denoting $\phi$ the angle of the polymer with the columns (i.e. with the
$z$ axis) we consider a fixed ratio 
\be
\tan \phi \simeq \frac{N r_0}{x} = \frac{r_0}{\theta} 
\ee 
where we work in units where the lattice spacing $r_0=1$. The case of small $\phi$ corresponds
to a field close to the $z$ axis and to a vortex line almost localized by the columns.
The case $\phi=\frac{\pi}{2} - \psi$ with $\psi$ small
corresponds to another situation (natural in layered superconductors)
where the external field is almost perpendicular
to the columns.

In presence of columnar disorder $\epsilon_j = - a_j \leq 0$ one wants to study the possible localization 
of the directed polymer along the columns. This can be quantified by the occupation length $\ell_j$ (defined above) which the polymer spends on column $j$. The statistics of this observable is one of the focus of this
paper, and has not been addressed until very recently in the mathematics literature 
\cite{noack2020central,noack2020concentration}. As we will see below the occupation fraction $\ell_j/x$ plays the role of an order 
parameter for the localization transition. Its expectation value can be obtained from the
following important relation
\be \label{occ1} 
 \partial_{\epsilon_j} {\cal F}_N(x,T)  = - \partial_{a_j} {\cal F}_N(x,T) = \langle \ell_j \rangle_T
\ee 
in each disorder configuration, i.e. for each 
realisation
of the Brownian motions $B_j$, where $\langle \dots \rangle_T$ denotes the
thermal average.
 
As emphasized above the present study is performed in the fixed tilt angle $\phi$ ensemble. 
In order to connect to models such as \eqref{en0} it is interesting to 
also consider the fixed external field $H$ ensemble, where
the exit position of the line, $u(x)=N$ can fluctuate. These
two ensembles are related by a Legendre transform. Defining
$f(\tan \phi) = \lim_{x \to +\infty} {\cal F}_{N=x \tan \phi}(x,T)/x$, the
free energy per unit length at fixed $H$ is $\min_{\phi \geq 0} [ f(\tan \phi) - H \tan \phi ]$.
This ensemble, and the connection to elastic line models and to the transverse
Meissner effect physics, is
discussed in Appendix \ref{app:model}. We argue that
the localization/delocalization transition which occurs at $\tan \phi_c=1/\theta_c$
in the fixed angle ensemble, may be associated 
to a first-order jump in the tilt response at $H=H_c$, and that
the localized phase discussed here for $\phi<\phi_c$, i.e. $\theta>\theta_c$,
can be seen actually as a coexistence region.

\subsubsection{Connection to random matrices: zero temperature}
\label{subsec:rmt} 

It has been known for some time in mathematics that the PDF of the 
optimal energy of the O'Connell-Yor model (i.e. at zero temperature) is related to the one of  
the largest eigenvalue of a random matrix from the so-called deformed GUE
\cite{Baryshnikov_2001,gravner2001limit,o2001brownian,doumerc2002asymptotic,benaych2013gue,O_Connell_2003,o2003path}. Let $V$ a $N \times N$ random Hermitian 
matrix drawn from the GUE, i.e 
with measure $Z^{-1}\exp( - \frac{1}{2} \mathrm{Tr} V^2) \rmd V$,
where $dV=\prod_i \rmd V_{ii} \prod_{i<j} \rmd {\rm Re}V_{ij} \rmd {\rm Im}V_{ij}$.
With this normalization, the spectrum of $V$ becomes, in the large $N$ limit, a 
semi-circle with support $\simeq [-2\sqrt{N},2\sqrt{N}]$. Consider the matrix ${\cal M}(x)$ defined as
\be \label{M} 
{\cal M}(x) = x \; \mathrm{diag}(a_1,\dots, a_{N})+\sqrt{x} \, V,
\ee 
and denote $\lambda_1(x) \geq \lambda_2(x) \geq \dots \geq \lambda_N(x)$ its eigenvalues 
in decreasing order.
Then one has the equality in law for the ground state energy
\be \label{la1} 
{\cal E}_N^0(x)= {\cal F}_N(x,T=0) \inlaw - \lambda_1(x)
\ee 
This characterizes the fluctuations over the point disorder, and it is valid for any fixed
configuration of the column energies $\epsilon_j=-a_j$, and for any $N$. 

As shown in Refs.~\cite{O_Connell_2002,bougerol2002paths,o2003path},
the equality in law at fixed $x$ in \eqref{la1} can be extended to an equality 
in law as a process in $x$ (i.e. as random functions of $x$ on both sides)
if one replaces $\sqrt{x} \, V \to W(x)$ where $W(x)$ is the Dyson Brownian
motion (DBM), i.e. a Brownian motion in the space of $N \times N$ Hermitian matrices, see \cite{Dyson1962,tao2012topics,mehta2004random,JPBook} for the definition of the DBM,
and the Appendix \ref{app:dbm}.

The formulae \eqref{M}, \eqref{la1} allow to 
make a bridge between the polymer representation and the random matrix representation.
First note that the occupation lengths in the ground state, denoted $\ell_j^0$, can be 
obtained as
\be \label{occT0} 
 \partial_{\epsilon_j} {\cal E}_N^0(x) = - \partial_{a_j} {\cal E}_N^0(x)= \ell_j^0
\ee
by taking the $T=0$ limit of \eqref{occ1}.
Now, denote $\psi_i$, $i=1,\dots,N$, the eigenvector associated to $\lambda_i$. A simple
perturbation theory argument shows that 
\be \label{occ2}
\partial_{a_j} \lambda_1(x) = 
x |\psi_1(j)|^2
\ee 
for each realization of the matrix $V$. Consider now the average of \eqref{occ2}
over $V$ and compare it with the average of \eqref{occT0} over the Brownian motions $B_j$,
using \eqref{la1}.
One finds for any $j$ and $\epsilon_j$ one has
\be \label{meanconnection} 
\overline{ \ell^0_j }^B  = 
x\;  \overline{  \vert\psi_1(j)\vert ^2}^V
\ee
where averages over the Brownian motions $B$ and the matrix randomness $V$, respectively are denoted by the corresponding
overbars.
Hence the average of the occupation length $\ell^0_j$
of the polymer on the column $j$ at $T=0$, can be related to the mean overlap with the column $j$
of the eigenvector $\psi_1$ of ${\cal M}(x)$ with the largest eigenvalue. In this respect, note
that recent studies address the distribution of the eigenvectors 
for such ensemble of random matrices either in the bulk \cite{benigni2017eigenvectors}, 
or near the edge but in the large deviation regime \cite{biroli2019large}. 
One would hope to determine the distribution of $\vert\psi_1(j)\vert ^2$ in the typical fluctuation regime using \eqref{occ2}, or to relate it to the distributions of $\ell^0_j$ 
further using \eqref{la1}, \eqref{occT0}, however for this one would need the knowledge of $\lambda_i(x)$ as a process with respect to the $a_i$'s, which is not currently available. 
In fact the PDF of $\ell^0_j$ and of $\vert\psi_1(j)\vert ^2$ differ in general,
despite their means being related via \eqref{meanconnection}, as seen in an explicit example in Appendix \ref{app:sub:delococc}.

\vskip .1cm

There are many interesting consequences of the result \eqref{la1}, some will be explored in this paper,
others can be already stated here.

\vskip .1cm

(i) In the absence of columnar disorder, $a_j=0$, the matrix ${\cal M}(x)=\sqrt{x} \, V$ has the same
one point distribution at fixed $x$ as the Dyson Brownian motion $W(x)$. Hence the statistics of $\lambda_1(x)$ 
is known in the limit of large $N$, which implies that 
\be \label{TW1} 
{\cal F}_N(x,T=0) \inlaw - 2 \sqrt{N x} - \frac{\chi_2}{N^{1/6}} \sqrt{x} 
\ee 
where $\chi_2$ is distributed according to the GUE Tracy-Widom distribution
\cite{tracy1994level}. Furthermore,
as discussed in the paragraph below \eqref{la1},
$\lambda_1(x)$ evolves as a function of $x$ 
as the largest eigenvalue of the DBM. At large $N$, and in a window of values of $x$
of width of order $\delta x \sim x/N^{1/3}$
it can be approximated as follows (see Appendix \ref{app:sub:convergence})
\bea \label{TW2} 
&& {\cal F}_N(x + \delta x,T=0) \nn \\
&& \inlaw - 2 \sqrt{N (x+\delta x)} - \frac{{\cal A}_2(\frac{N^{1/3} \delta x}{2 x})}{N^{1/6}} \sqrt{x} 
\eea 
where $z \mapsto {\cal A}_2(z)$ is the so-called Airy$_2$ process, a universal random
function introduced
in \cite{prahofer2002scale} in the context of a discrete growth model. It is a continuous
stationary process (i.e. statistically invariant by translation) with a slow decay of correlations $1/z^2$. All its multipoint 
correlations are known and can be expressed as Fredholm determinants
\cite{quastel2014airy}.
For $\delta x=0$, \eqref{TW2} recovers \eqref{TW1} since
${\cal A}_2(0)=\rchi_2$ i.e. the one point distribution of the
Airy$_2$ process is the GUE Tracy-Widom distribution.

\vskip .1cm

(ii) In presence of columnar disorder, the problem maps onto determining 
the largest eigenvalue of the so-called {\it deformed or spiked} GUE. Study of that problem was pioneered in physics 
by Brezin and Hikami \cite{brezin1996,brezin1997spectral,brezin1997extension,brezin1998universal,brezin1999level}
and in mathematics by Johansson \cite{Johansson_2001} and Tracy and Widom
\cite{TW1,TW3}. It can be reformulated in terms of the Dyson Brownian motion as follows. Upon redefining $ x a_j = b_j$, 
the matrix ${\cal M}(x)$ in \eqref{M} can be interpreted, for fixed values of $b_j$,
as performing a DBM in $x$, with initial condition $\lambda_j(x=0)=b_j$, see 
Appendix \ref{app:dbm}.

Various initial conditions
have been studied in the equivalent random matrix models with sources \cite{brezin2016random}.
These also admit interesting representations as non-crossing random walks, also called watermelons
\cite{johansson2005non,TW1,TW2}, and DBM with wanderers \cite{adler2010airy}.

The simplest case occurs for a single attractive columnar defect, 
$\epsilon_1=-a_1<0$ and $a_{j \geq 2}=0$.
From \eqref{M} it corresponds to a rank-one perturbation of the GUE matrix $\sqrt{x} V$. 
This was studied in a celebrated
work by Baik, Ben-Arous and P\'ech\'e for spiked covariance
matrices \cite{Baik_2005}, and spiked GUE matrices \cite{Peche_2005}.
In Ref.~\cite{Peche_2005} it was shown that the largest eigenvalue $\lambda'_1$ of the deformed GUE matrix $M'=\mathrm{diag}(\pi_1,\{0\})+\frac{V}{\sqrt{N}}$ exhibits two phases 

\begin{itemize}
\item If $\pi_1<\hspace*{0.2cm} 1$, $\lambda'_1 = 2+N^{-2/3}\rchi_2$,
\item If $\pi_1>1$, $\lambda'_1=\pi_1+\frac{1}{\pi_1}+ N^{-1/2} {\cal N}(0,\sigma^2)$ 
with $\sigma^2 =1 - \frac{1}{\pi_1^2}$.
\end{itemize}
Here ${\cal N}(m,\sigma^2)$ denotes a gaussian random variable with variance $\sigma^2$
and mean $m$.  The correspondance with our notations leads to $\pi_1=\sqrt{\frac{x}{N}} a_1$
and $\lambda_1= \sqrt{Nx}\lambda'_1$. Thus, using \eqref{la1}, this predicts the following leading behavior for the
free energy of the polymer at $T=0$

- If the column is weak, i.e. $a_1 < \sqrt{\frac{N}{x}}$, equivalently
$\theta < \theta_c= \frac{1}{a_1^2}$ (or $\tan \phi > \tan \phi_c = a_1^2$, i.e.
large angle from the $z$ axis) 
then the rank-one perturbation (the columnar defect) has little effect, i.e. 
the largest eigenvalue of ${\cal M}(x)$ still behaves as $\lambda_1(x) \simeq 2 \sqrt{N x}$
at large $N$ and the result \eqref{TW1} for the ground state free energy 
still holds. 

- If the column is strong, i.e. $a_1 > \sqrt{\frac{N}{x}}$, equivalently
$\theta > \theta_c= \frac{1}{a_1^2}$ (or $\tan \phi < \tan \phi_c = a_1^2$
i.e. small angle from the $z$ axis) then the largest eigenvalue of ${\cal M}(x)$ 
detaches from the Wigner semi-circle and becomes an outlier. This leads to
\begin{equation}
\label{loc} 
{\cal F}_N(x,T=0) \simeq - ( a_1 x +\frac{N}{a_1}) 
+\sqrt{N}\mathcal{N}(0, \theta - \frac{1}{a_1^2})
\end{equation}
As we discuss below, this BBP transition corresponds to a first-order localization
transition of the polymer on the columnar defect, which has some 
features of a freezing transition.

In the next section we will analyze in more details this transition,
and extract in particular information about the occupation length.
This will prepare us to study the case of many columnar defects 
in the following section.

\section{Single line, single columnar defect, zero temperature}
\label{sec:singlesingle} 

Let us consider now in more details the case of a single line and a single column with an attractive potential $\epsilon_1=-a_1 \leq 0$
(which with no loss of generality we can choose
in position $j=1$). All other potentials are set to $\epsilon_j = - a_j=0$ for $j \neq 1$.
We first present a variational calculation based on the polymer picture.  In a second stage we recall the kernel which describes the largest eigenvalue of the 
matrix ${\cal M}(x)$ and study the phases using Fredholm determinants.

\subsection{Approach by a variational calculation}

We now obtain a physical derivation of the localization transition of the polymer, when a
finite fraction of the length of the columnar defect becomes occupied. We first
discuss the two phases and then the critical region.

\subsubsection{Description of the two phases}

\begin{figure}[t!]
\centerline{\includegraphics[scale=0.6]{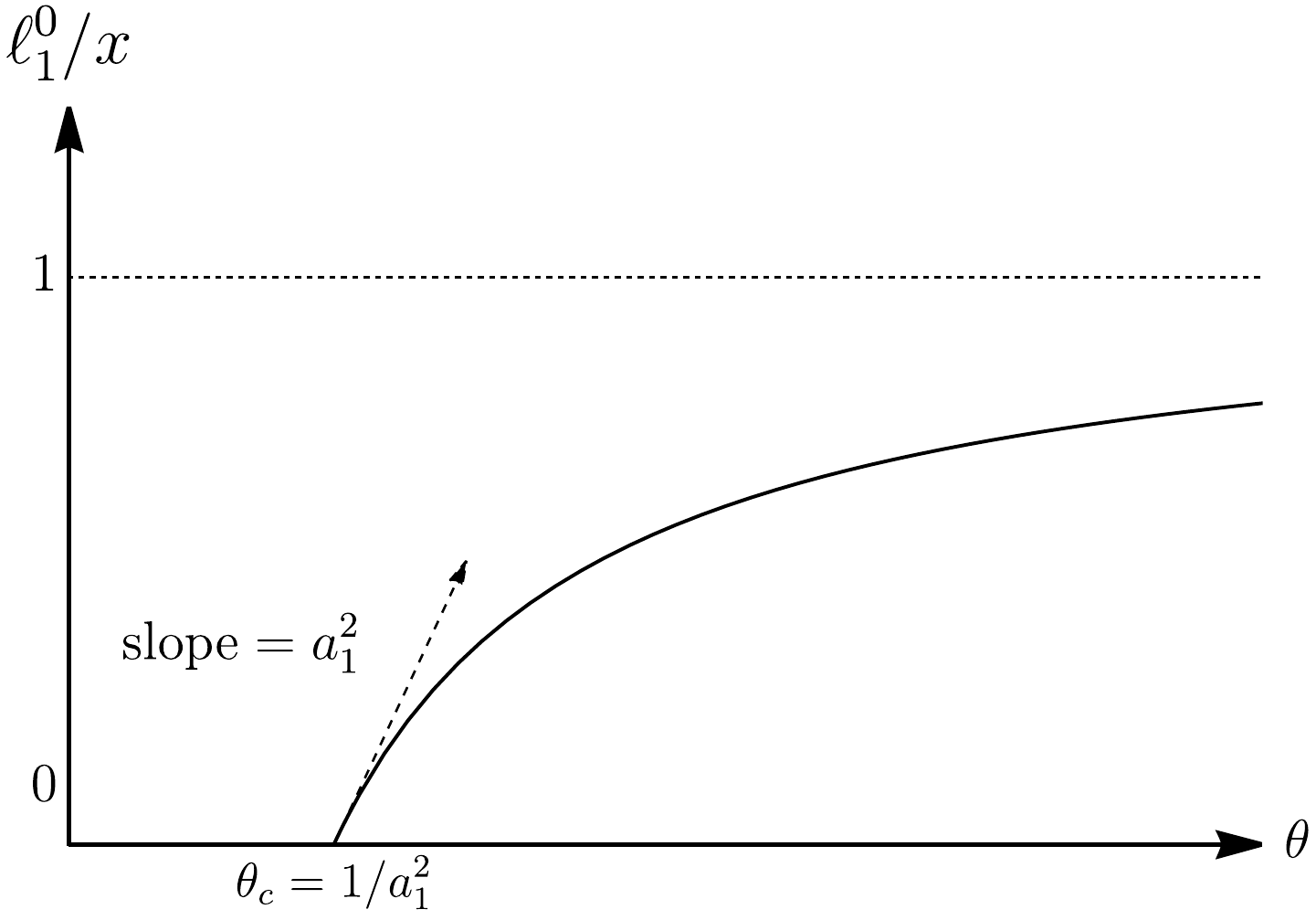}}
\caption{Occupation fraction $\ell_1^0/x$ at zero temperature for a single columnar defect
as a function of $\theta=\frac{x}{N} = 1/\tan \phi$. It undergoes a first order localization
transition at the critical value $\theta=\theta_c=\frac{1}{a_1^2}$.}
\label{Single_column_localization}
\end{figure}

For a single columnar defect, of energy $\epsilon_1=-a_1 \leq 0$, the ground state
energy ${\cal E}^0_N(x)$ is given by the following variational problem 
\begin{equation} \label{var1} 
{\cal E}^0_N(x) = \min_{\ell_1\in [0,x]} \left[B_1(\ell_1)- a_1 \ell_1+G_{N-1}(\ell_1,x)\right]
\end{equation}
where 
\begin{equation} \label{GG} 
G_{N-1}(\ell_1,x)=\min\limits_{\ell_1<x_2<\dots<x_{N-1}<x}\sum\limits_{i=2}^{N}B_i(x_{i},x_{i-1})
\end{equation}
where $x_1=\ell_1$ is the occupation length of the first column, and $x_N=x$ the total length.
As before, from \eqref{la1}, $G_{N-1}(\ell_1,x)$ for fixed $\ell_1$ and $x$ is distributed as the largest eigenvalue of $\sqrt{x-\ell_1}$ times a $(N-1) \times (N-1)$ GUE matrix. When varying $\ell_1$ at fixed $x$ it  varies as
the largest eigenvalue of a Dyson Brownian motion evolving during time $x-\ell_1$.
Hence, for large $N$, the contribution of $G_{N-1}$ is the sum of a deterministic part and a 
subdominant fluctuating part
\be \label{GG2} 
G_{N-1}(\ell_1,x) = - 2 \sqrt{N (x-\ell_1)} + \mathcal{O}\left( \frac{(x-\ell_1)^{1/2}}{N^{1/6}} \right)
\ee
where for now, to determine the phases, we do not need to specify in more details the fluctuating
part in \eqref{GG2} (it will be important only near the transition). Similarly, to leading order, we can neglect the Brownian contribution $B_1(\ell_1)$ in 
\eqref{GG},
as well as the fluctuation term. This leads to
the estimate
\begin{equation} \label{varleading} 
{\cal E}^0_N(x) \simeq \min_{\ell_1\in [0,x]}[- a_1 \ell_1 - 2\sqrt{N(x-\ell_1)}]
\end{equation}
The optimal occupation length $\ell_1^0$ of the first column is thus obtained to leading order as
\be \label{29} 
\ell_1^0 \simeq (x-\frac{N}{a_1^2})_+ = x (1 - \frac{\theta_c}{\theta})_+ 
\ee
where $(x)_+=\max(0,x)$, $\theta=\frac{x}{N}$, and $\theta_c=1/a_1^2$ 
is the critical angle, see Fig. \ref{Single_column_localization}. We see that a finite fraction of the column is occupied if and only if 
\be
a_1 = - \epsilon_1 >  \sqrt{\frac{N}{x}} 
\ee 
which corresponds to the localized phase $\theta> \theta_c$. Our result \eqref{29} coincides, using the occupation length--overlap connection
given in \eqref{meanconnection}, with the result for the 
overlap for the BBP transition obtained in \cite{benaych2011eigenvalues}.
In the localized phase the ground
state energy is, to leading order,
\be
{\cal E}^0_N(x) \simeq - ( a_1 x +\frac{N}{a_1} ) = - N a_1 (\theta + \theta_c) 
\ee
and, at the transition $\theta=\theta_c$ it reaches the value ${\cal E}^0_N(x) \simeq - 2 N \theta_c = - 2 \frac{N}{a_1}$.

It is easy to also obtain the leading fluctuations of the ground state energy in the localized phase. The leading
fluctuating part of ${\cal E}^0_N(x)$ is clearly $B(\ell_1^0)$, which is
a Gaussian random variable with variance equal to 
$\ell_1^0= N (\theta -\frac{1}{a_1^2}) = N (\theta-\theta_c)$
in full agreement with \eqref{loc}. Note that from \eqref{GG2} we see that the fluctuations originating from the
delocalized segment of the polymer of length $x-\ell_1$, i.e. the term $G_{N-1}(x,\ell_1)$ in \eqref{var1}, is 
only of order $\mathcal{O}(N^{1/3})$, hence subleading as compared to the $\mathcal{O}(N^{1/2})$ Gaussian fluctuations  
originating from the point disorder along the localized segment of length $\ell_1^0$. 
The fluctuations of $\ell_1^0$ are more subtle and we show, see \eqref{ress} below, that in the localized phase there are KPZ-like
fluctuations  of order $N^{2/3} \ll \ell_1^0 = \mathcal{O}(N)$
around the value given in \eqref{29}.

If $a_1 < \sqrt{\frac{N}{x}}$, that is $\theta<\theta_c$,
the minimum in \eqref{varleading}  is attained at $\ell_1^0=0$ and we then recover the results
in the delocalized phase of Eqs.~\eqref{TW1} and \eqref{TW2}. 

The above results are in agreement with the predictions from the BBP
transition summarized in the previous subsection.
In fact a related calculation was given in \cite[Sec. 6]{Baik_2005} 
for a fully discrete polymer model (last passage percolation 
on the square lattice) which in that case involves Wishart random matrices.
These arguments provide an intuitive way to obtain the transition 
criterion for spiked random matrices for various ensembles. In the context of polymers
this variational calculation, as we show below, allows to also explore
the fluctuations around the localisation transition.

Note that the variational calculation can be extended to several
columns located at different positions, and one can verify the
property of invariance with respect to permutation of the columns,
see Appendix \ref{app:permutation}.

\subsubsection{Fluctuations in the critical region}

\label{subsubsec:critical} 

We can now refine the variational calculation to
obtain the critical regime near the transition $\theta \approx \theta_c$.
That region will be defined by
\be \label{crit1} 
\theta = \frac{x}{N} = \theta_c (1 + \frac{\delta}{N^{1/3}}) 
\ee 
with $\delta=\mathcal{O}(1)$. In that region, as we will see, the optimal occupation
length $\ell_1^0$ will fluctuate but with a typical magnitude $\mathcal{O}(N^{2/3})$,
corresponding to a vanishing occupied fraction $\ell_1^0/x \sim N^{-1/3} \ll 1$.

\begin{figure}[t!] 
\begin{center}
\includegraphics[scale=0.6]{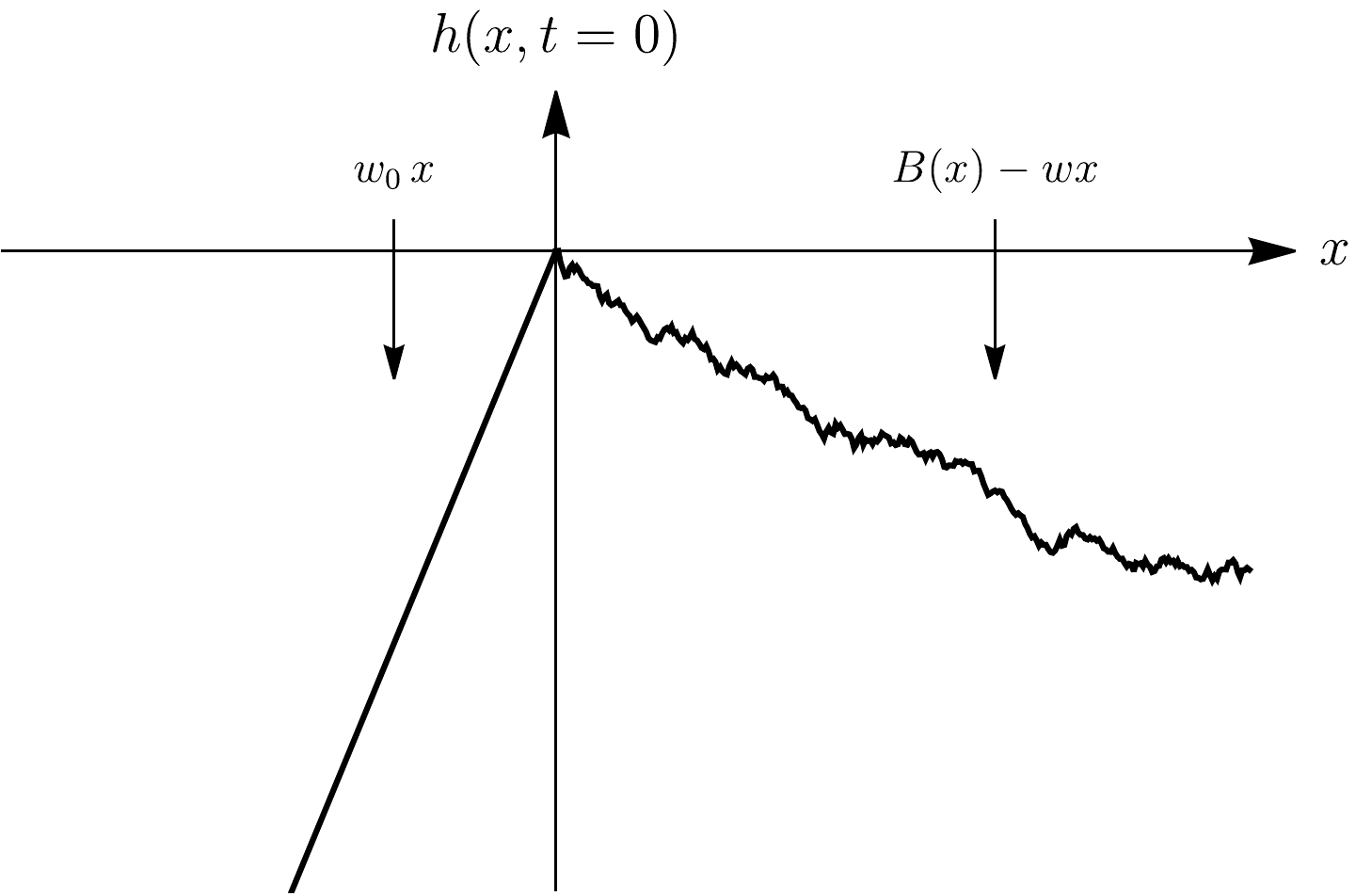}
\caption{Half-Brownian motion with drift $w$ as the initial condition $h(x,t=0)$ of the Kardar-Parisi-Zhang equation \eqref{kpzeq}. The fluctuations of the height at large time $h(0,t)$, relates to those
of the ground state energy of the polymer \eqref{upsilon} at the localization transition.}
\label{fig:Brownian}
\end{center}
\end{figure}

Let us go back to \eqref{var1} and use the estimate \eqref{TW2}
with $\delta x=-\ell_1$ 
\bea \label{var2} 
&& {\cal E}^0_N(x) \simeq \min_{\ell_1\in [0,x]} \bigg[B_1(\ell_1)- a_1 \ell_1
- 2 \sqrt{N (x - \ell_1)} \nn
\\
&& ~~~~~~~~~~~~~ - \frac{\sqrt{x- \ell_1}}{N^{1/6}}
{\cal A}_2(- \frac{N^{1/3} \ell_1}{2 x})
\bigg]
\eea
We see that if we want the argument of the Airy$_2$ process ${\cal A}_2$ to be of order unity,
we need indeed to choose $\ell_1 \sim N^{2/3}$, hence we will define 
the reduced length $\hat \ell_1$ by setting
\be \label{redl} 
\ell_1 = 2 x N^{-1/3} \hat \ell_1 = 2 N^{2/3} \theta \hat \ell_1
\ee 
We can now insert \eqref{crit1} and \eqref{redl} into \eqref{var2}
and expand at large $N$. We find
\bea \label{var33} 
&& {\cal E}^0_N(x) = -2 \sqrt{N x} \\
&& + \frac{N^{1/3}}{a_1} 
\min_{\hat \ell_1 \geq 0} [\sqrt{2} B(\hat \ell_1)  - \delta \hat \ell_1 + \hat \ell_1^2
- {\cal A}_2(- \hat \ell_1) ] + \mathcal{O}(1) \nn
\eea
where $B(s)$ is a Brownian motion obtained from $B_1(s)$ by the
rescaling \eqref{redl}. Hence we find that the ground state energy in the
critical region behaves as
\be  \label{upsilon} 
 {\cal E}^0_N(x) \simeq -2 \sqrt{N x} - \frac{N^{1/3}}{a_1}  \upsilon_\delta
\ee
where the random variable $\upsilon_\delta$ is defined by the variational problem
\be \label{var3} 
\upsilon_\delta = \max_{z \geq 0} [\sqrt{2} B(z)  + \delta z + {\cal A}_2(z) - z^2] 
\ee
where $B$ and ${\cal A}_2$ are statistically independent, 
and we have used that the process ${\cal A}_2(-z)$ is statistically identical to
${\cal A}_2(z)$. The PDF of this random variable appears in the
problem of KPZ growth with a "half-Brownian" initial condition 
\cite{imamura2011replica,corwin2013crossover} which we now briefly recall.
Consider for instance the KPZ equation \cite{KPZ} for the growth of the height field $h(x,t)$ as
a function of time $t$ 
\be \label{kpzeq}
\partial_t h = \partial_x^2 h + (\partial_x h)^2 + \sqrt{2} \xi(x,t) 
\ee 
where $\xi(x,t)$ is a space-time white noise. Denote $h_w(x,t)$ the solution with initial condition
$h_w(x,t=0)= (B(x) - w x) \Theta(x) + w_0 x \Theta(-x)$ (where $\Theta(x)$ is the Heaviside
function) which is represented in
Fig. \ref{fig:Brownian}, in the limit where $w_0 \to +\infty$. 
The random variable $\upsilon_\delta$ defined by \eqref{var3}, 
appears in the large time limit, $h_{w=- \delta/(2t^{1/3})}(0,t)= - \frac{t}{12} + \upsilon_\delta t^{1/3}$,
where it is standard from the KPZ literature to scale the drift $w$ with the observation time,
as $w=- \delta/(2t^{1/3})$. The localized phase of the polymer problem,
$\delta >0$, corresponds
 to $w<0$, leading to a Gaussian distribution in the limit $\delta \to +\infty$.
 The delocalized phase, $\delta<0$, corresponds to $w>0$,
leading to the GUE Tracy-Widom distribution for $\delta \to -\infty$,
i.e. $\upsilon_{- \infty}={\cal A}_2(0)=\chi_2$. 
The PDF of $\upsilon_\delta$ for any $\delta$ was obtained in \cite{imamura2011replica,corwin2013crossover}
and it was observed there that it coincides with the critical BBP
distribution, see Section \ref{bppcrit}. This is expected from the relation 
of the polymer problem to the Baik-Ben Arous-P\'ech\'e transition
described in Section \ref{subsec:rmt}. 

Interestingly, we now obtain a new distribution if we study the 
occupation length $\ell_1^0$ in the ground state. Indeed the 
PDF of $\ell_1^0$ is given by 
\be \label{39} 
\frac{\ell_1^0}{2 N^{2/3}/a_1^2} \inlaw \argmax_{z \geq 0} [\sqrt{2} B(z)  
+ \delta z + {\cal A}_2(z) - z^2]
\ee 
To our knowledge this distribution is not known exactly. We can study two limits:\\

(i) {\bf Limit $\delta \to - \infty$: the side of the delocalized phase.} The argmax is obtained
for small values of $z$ in \eqref{39}, so we set $z= y/\delta^2$. One can then approximate the
Airy process by a Brownian motion since \cite{hagg2008local}
${\cal A}_2(\frac{y}{\delta^2}) - {\cal A}_2(0) \simeq \frac{\sqrt{2}}{|\delta|} \tilde B(y)$ as $\delta \to - \infty$,
where $\tilde B$ is a Brownian motion independent of $B$. Hence we find
\be \label{399} 
\frac{\ell_1^0}{2 N^{2/3}/a_1^2} \inlaw \frac{1}{\delta^2} \argmax_{y \geq 0} [2 B(y)  - y ]
\ee 
We can now use the known result that the PDF of the time of the maximum
$\omega$ of a Brownian motion of variance $\sigma^2$ and negative drift $-\mu$,
with $\mu>0$, defined as
\be
\omega = \argmax_{y \geq 0}[ \sigma B(y) - \mu y ]  
\ee
is given by (e.g. \cite[Chapter IV, item 32]{HandbookBrownian}, or
taking the limit $T \to +\infty$ in \cite{Shepp1979}
or
\cite[Eq. (30)]{majumdar2008optimal}), see
also Appendix \ref{app:sub:delococc})
\be \label{pmu} 
P_{\mu,\sigma}(\omega) =\frac{\mu }{\sigma} \sqrt{\frac{2  }{ \pi \omega}}e^{-\frac{\mu ^2 \omega}{2 \sigma ^2}}
-\frac{ \mu ^2 }{\sigma ^2}\text{erfc}\left(\frac{\mu \sqrt{\omega}}{\sqrt{2} \sigma}\right)
\ee
Hence we find, setting $\sigma=2$ and $\mu=1$, that for $\delta \to -\infty$
\be \label{occ12} 
\ell_1^0 \simeq \frac{2 N^{2/3}}{a_1^2} \frac{\omega}{\delta^2} 
\ee 
where $\omega$ is a positive random variable
distributed with $p(\omega) d\omega = P_{1,2}(\omega) d\omega$.

It is shown in the Appendix \ref{app:sub:delococc} that inside the delocalized
phase far from the critical region, i.e. for $\theta < \theta_c$, the occupation
length fluctuates as $\ell_1^0 = \mathcal{O}(1)$ with the same distribution characterized
by $\omega$, scaled by an amplitude which diverges as $\sim 1/(\theta_c-\theta)^2$ at the transition,
and which matches smoothly with the result \eqref{occ12} in the critical regime, i.e.
for $\delta=\mathcal{O}(1)$.\\

(ii) {\bf Limit $\delta \to + \infty$: the side of the localized phase. }
To study that limit let us rewrite in a equivalent way the argmax in \eqref{39} by defining
\be
z = \frac{\delta}{2} + \tilde z 
\ee 
We can rewrite 
\be 
\frac{\ell_1^0}{2 N^{2/3}/a_1^2} \inlaw \argmax_{\tilde z \geq - \delta/2}[\sqrt{2} B(\tilde z)  
+ {\cal A}_2(\tilde z) - \tilde z^2]
\ee 
where we have used (i) that the ${\cal A}_2$ process is statistically invariant by translation
(ii) that $B(\frac{\delta}{2}+\tilde z)-B(\frac{\delta}{2})$ is equivalent in process
to a two-sided Brownian $B(\tilde z)$ for $\tilde z \geq -\delta/2$. In the limit $\delta \to +\infty$
we see that it becomes a (two-sided) optimization over the real axis
\be 
\frac{\ell^0_1}{2 N^{2/3}/a_1^2} \inlaw \argmax_{\tilde z \in \mathbb{R}} [\sqrt{2} B(\tilde z)  
+ {\cal A}_2(\tilde z) - \tilde z^2]
\ee 
The PDF of the r.h.s. was obtained in \cite{PLDAiry}, where it was shown that
it equals $f_{\rm KPZ}$, a function introduced in \cite{prahofer2004exact122,SasamotoStationary} to describe the two space-time point stationary correlations of the Burgers velocity field, associated to the KPZ height field. 
It also describes the midpoint distribution of a directed polymer in a stationary regime
as obtained in \cite{maes2017midpoint}. Hence we obtain that for $\delta \to +\infty$
\be \label{ress} 
\ell^0_1 \simeq \frac{2 N^{2/3}}{a_1^2} (\frac{\delta}{2} + \omega ) = x (1 - \frac{\theta_c}{\theta})  + \frac{2 N^{2/3}}{a_1^2}  \omega
\ee 
where $\omega$ is a real random number distributed with $p(\omega) d\omega = f_{\rm KPZ}(\omega) d\omega$.
We recall that $f_{\rm KPZ}(\omega)$ is an even function of $\omega$ with cubic exponential decay at large value of the 
argument $|\omega| \to +\infty$ 
\be
f_{\rm KPZ}(\omega) \simeq e^{- 0.295 |\omega|^3} 
\ee
and standard deviation $0.714$ and fourth moment $0.733$ \cite{prahofer2004exact122}.

Using the definition \eqref{crit1}, we notice
that the first term in \eqref{ress} is precisely the leading estimate of $\ell_1^0$ in the localized phase obtained in \eqref{29}. Hence our result \eqref{ress} matches smoothly the critical region with the localized phase. We thus discover that inside the localized phase there are non-trivial KPZ like fluctuations
of the occupation length around its typical value $x (1 - \frac{\theta_c}{\theta})$. 

Using \eqref{var1} and the estimate \eqref{TW2} (with $x \to x-\ell_1$),
one can indeed check that these fluctuations are described by $f_{\rm KPZ}$ in the whole localized phase, i.e.
the result \eqref{ress} in the form $\ell^0_1 \simeq x (1 - \frac{\theta_c}{\theta})  + \frac{2 N^{2/3}}{a_1^2}  \omega$, where $\omega$ is distributed with $f_{\rm KPZ}$, holds for any $\theta > \theta_c$.

In conclusion we have found that the fluctuations of the occupation length around its typical value
are of the order $N^{2/3}$ inside the localized phase, and up to and including the transition. In the
delocalized phase these fluctuations are of order $\mathcal{O}(1)$. 

\subsection{Approach using a Fredholm determinant}

We now perform the calculation of the ground state energy and its fluctuations using the
method of Fredholm determinants. The manipulations follow the analysis of the BBP
transition in RMT in \cite{Peche_2005} and
\cite{Baik_2005}, expressed here in the language of the polymer. Although they are classical 
they allow us to prepare the ground for the generalization to an infinite rank perturbation in the
next section. 

\subsubsection{Kernel at any $N$} 

A classical calculation \cite{Johansson_2001,Baik_2005,Peche_2005}, 
obtains the cumulative distribution function (CDF) of the largest eigenvalue $\lambda_1(x)$ of the matrix ${\cal M}(x)$ in \eqref{M} 
for an arbitrary set of $a_j$, $j=1,\dots, N$, as a Fredholm determinant
\be \label{PDF0} 
\mathbb{P}(\lambda_1(x) \leq \Lambda)=\Det(I- K_N)_{\mathbb{L}^2(\Lambda,+\infty)}
\ee

\begin{figure}[t!]
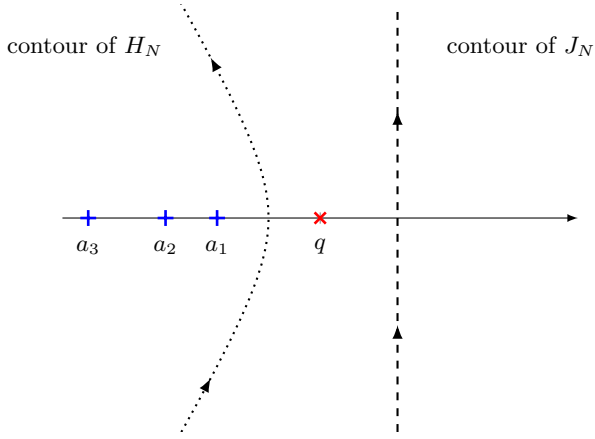

\contourBBP
\caption{Integration contours in the integrals $H_N$ and $J_N$
in \eqref{HJ}.}
\label{fig:contourBBP}
\end{figure}

where the kernel $K_N(v,v')$ is given for $v,v' \in [\Lambda,+\infty[$ by
\begin{equation}
\label{eq:GUEk}
K_N(v,v')=\int_{0}^{+\infty}\mathrm{d}r\, H_N(v+r)J_N(v'+r)
\end{equation}
along with
\bea \label{HJ} 
&& H_N(v)=\oint \frac{\mathrm{d}z}{2\I \pi }\, e^{v(z-q) - xz^2/2} \prod_{j=1}^N\frac{1}{z-a_j} 
\\
&& 
J_N(v)=\int_{\I \mathbb{R}+\varepsilon}\frac{\mathrm{d}z}{2\I \pi }e^{ -v(z-q)+x z^2/2} \prod_{j=1}^N(z-a_j) \nonumber
\eea
where $q$ is introduced for convergence purpose
(it does not change the value of the Fredholm determinant).
The contour in $H_N(v)$ is taken to enclose all $\lbrace a_j \rbrace $'s counter-clockwise.
The contour in $J_N(v)$ is chosen such that $\varepsilon > \max_j a_j$,
and passes to the right of the countour for $H_N(v)$, see Fig. \ref{fig:contourBBP}.
For more details on the kernel \eqref{eq:GUEk} and its derivation
see \cite[Proposition 2.3 and Eq. (2.18)]{Johansson_2001},
\cite[Eq. (3.19)]{brezin1997extension}
\cite[Proposition 1.3]{Peche_2005}.

Note that the functions $H_N$, $J_N$ and the kernel $K_N$ are invariant under permutations
of the $a_j$, which shows the property mentioned in the introduction. 

We first use this result to study the ground state energy in the case 
of a single active column $a_1 >0$ and $a_{j > 1}=0$.
The general case will be studied in Section \ref{sec:singlemany}. 
We recall that we focus on the limit of large $x,N$ at fixed $\theta=x/N$
and determine the asymptotic form of the kernel $K_N$ in that limit.
We start with the delocalized phase.

\subsubsection{Delocalized phase} 

\begin{figure}[t!]
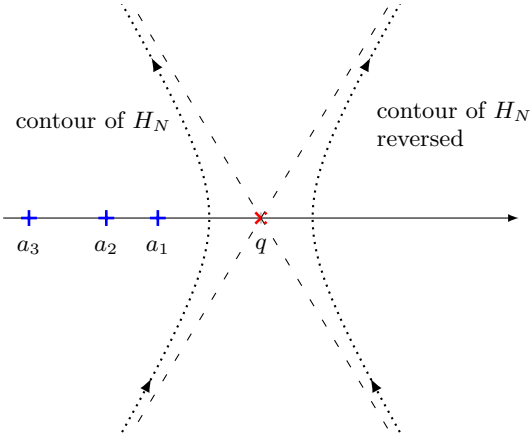

\contourBBPreversed
\caption{Contour reversal: the original contour for $H_N$ is on the left of $q$, as in
Fig. \ref{fig:contourBBP}. The reversal is performed by a rotation around $q$
of angle $\pi$ followed by a reversal of orientation of the contour.
This leads to the contour on the right. This allows to have the same contours
for $H_N$ and $J_N$ asymptotically (near the saddle point) leading to the
formula in the text.}
\label{fig:contourBBPreversed}
\end{figure}

In the delocalized phase one anticipates that at large $N$, $\lambda_1 - \mu N = \mathcal{O}(N^{1/3})$
where $\mu$ is to be determined later.
Hence we rewrite the function $H_N$ and $J_N$ in \eqref{HJ} as 
\begin{equation} 
\begin{split}
&H_N(\mu N+ \sigma N^{1/3})=
\oint \frac{\mathrm{d}z}{2\I \pi }\, e^{- N \varphi(z) + z \sigma N^{1/3}}  \\
&J_N(\mu N+ \sigma N^{1/3} )= \int_{\I \mathbb{R}+\varepsilon}\frac{\mathrm{d}z}{2\I \pi }
e^{N \varphi(z) - z \sigma N^{1/3}} \label{JJ}
\end{split}
\end{equation}
where, using $a_j=a_1 \delta_{j1}$
\be
\varphi(z)= -\mu z+ \theta \frac{z^2}{2}  + \log z + \frac{1}{N} \log(1-\frac{a_1}{z}) 
\ee
Here and below we omit the factor $q$ from \eqref{HJ} since it plays no role in
the results. Let us first examine the function $J_N(z)$, for which the integration contour obeys $\Re (z)>a_1$.
The first two derivatives read
\bea \label{der12} 
&& \varphi'(z)= - \mu + \theta z + \frac{1}{z} + \frac{1}{N} (\frac{1}{z-a_1}- \frac{1}{z}) \\
&&  \varphi''(z)= \theta - \frac{1}{z^2} + \frac{1}{N} (\frac{1}{z^2}- \frac{1}{(z-a_1)^2}) 
\eea
We now look for a degenerate saddle point $z=z^*$ such that $\varphi'(z^*)=0$ and
$\varphi''(z^*)=0$. Inside the delocalized phase one can neglects the $1/N$ terms 
in \eqref{der12} and one finds that 
\be
z^* = \frac{1}{\sqrt{\theta}} \quad , \quad \mu^* = 2 \sqrt{\theta} 
\ee
Hence we choose $\mu=\mu^*$. Since one must have $z^* > a_1$, this is possible only until the transition point 
\be
\theta < \theta_c = \frac{1}{a_1^2} 
\ee
One can now expand around the saddle point up to the third order and
write 
\be
\varphi(z) = \varphi(z^*)  + \frac{1}{3}\big(\frac{z}{z^*}-1\big)^3  + \mathcal{O}((z-z^*)^4,\frac{1}{N}) 
\ee 
Performing the change of variable
\be
z = z^* (1 + y N^{-1/3} )
\ee
and inserting into \eqref{JJ} we obtain
\be \label{JJ3} 
 J_N(\mu N+ \sigma N^{1/3} ) \simeq z^* N^{-1/3} e^{N \varphi(z^*) - z^* \sigma N^{1/3}}
{\rm Ai}(z^* \sigma) 
\ee
where
\be \label{Ai} 
{\rm Ai}(w) = \int_{\I\mathbb{R}+\varepsilon}\frac{\mathrm{d}y}{2\I \pi }
e^{\frac{y^3}{3}  - y w }
\ee
and the neglected terms are subdominant at large $N$. 
One can show \cite{Baik_2005,Peche_2005} that the 
counter-clockwise contour integral for the function $H_N(\mu N+ \sigma N^{1/3})$
is dominated by the same saddle point at $z=z^*$. This leads to
an expression for $H_N(\mu N+ \sigma N^{1/3})$ which is identical
to \eqref{JJ3} (taking into account the reversal of the contour) 
but with a prefactor
$e^{- N \varphi(z^*) + z^* \sigma N^{1/3}}$, see Fig.~\ref{fig:contourBBPreversed}.
Putting these expressions together 
in \eqref{eq:GUEk} and rescaling $r \to r N^{1/3}/z^*$ one 
obtains 
\be
 K_N(\mu N+ \sigma N^{1/3},\mu N+ \sigma' N^{1/3}) \simeq \frac{z^*}{N^{1/3}} K_{\rm Ai}(z^* \sigma, z^* \sigma')
\ee
in terms of the Airy kernel
\be
K_{\rm Ai}(v,v') = \int_0^{+\infty} \rmd r \, {\rm Ai}(v+r) {\rm Ai}(v'+r)
\ee 
We have discarded a factor $e^{ z^* N^{1/3} (\sigma-\sigma')}$ which is
immaterial in the Fredholm determinant. From \eqref{PDF0} we now
obtain that 
\begin{equation}
\begin{split}
- {\cal E}_N^0(x)=  \lambda_1(x)&= \mu^* N + \frac{N^{1/3}}{z^*} \rchi_2 \\
& = 2 \sqrt{x N} + \sqrt{x} N^{-1/6} \rchi_2 \label{resa1} 
\end{split}
\end{equation}
where $\mathbb{P}(\rchi_2 \leq s) = F_2(s) = \Det(I- K_{\Ai})_{\mathbb{L}^2(s,+\infty)}$.
This is the standard result in the delocalized phase.

\subsubsection{Critical region} 
\label{bppcrit} 

As $\theta \to \theta_c=\frac{1}{a_1^2}$ the saddle point $z^*$ of the previous calculation
moves to $a_1^+$. Hence one cannot neglect the term $\frac{1}{N} \log(1 - \frac{a_1}{z})$.
It is now more convenient to leave it outside the exponential 
\be \label{J1} 
J_N(\mu N+ \nu N^{1/3} )= \int_{\I\mathbb{R}+\varepsilon}\frac{\mathrm{d}z}{2\I \pi }
(1 - \frac{a_1}{z}) e^{N \varphi_0(z) - z \nu N^{1/3}} 
\ee
with
\be
\varphi_0(z)= -\mu z+\theta \frac{z^2}{2}  + \log z 
\ee and
$\varepsilon > a_1$. There is a similar expression for $H_N$ with
a contour which encloses $a_1$ counter-clockwise. 
The critical regime is defined by
\be \label{crit4} 
\theta = \frac{1}{a_1^2}  + \tau N^{-1/3}
\ee
which is consistent with \eqref{crit1} obtained via the variational calculation, 
with the correspondence in notations $\tau=\theta_c \delta$.

Let us again denote $z^*= \frac{1}{\sqrt{\theta}}$ the denegerate point where $\varphi'_0(z^*)=0$
and $\varphi''_0(z^*)=0$. Its position is now
\be
z^* =  \frac{1}{\sqrt{\theta}}  = a_1 - \frac{\tau a_1^3}{2}  N^{-1/3} 
\ee
One performs again the change of variable $z = z^* (1 + y N^{-1/3} )$, one obtains
\be
J_N(\mu N+ \nu N^{1/3} ) 
 = 
\frac{z^*}{a_1} N^{-1/3} e^{N \varphi_0(z^*) - z^* \sigma N^{1/3}} {\rm Ai}^{b_1}(z^* \nu) 
\ee
with 
\bea
{\rm Ai}^{b_1}(w)  = \int_{\I \mathbb{R}+\varepsilon'}\frac{\mathrm{d}y}{2 \I \pi}
(y  - b_1) e^{\frac{1}{3} y^3 - w y} \nn
\eea
where $\epsilon' > b_1 $ and $b_1=\frac{\tau a_1^2}{2}$,
which can be checked to also equal $b_1 = \lim_{N \to +\infty} 
N^{1/3} (a_1 - \frac{1}{\sqrt{\theta}})/a_1$. A very similar expression
holds for $H_N$ 
\be
H_N(\mu N+ \nu N^{1/3} ) 
= a_1
z^* N^{-1/3} e^{- N \varphi_0(z^*) + z^* \sigma N^{1/3}} {\rm Ai}_{b_1}(z^* \nu) 
\ee
taking into account the reversal of the contour, see Fig. \ref{fig:contourBBPreversed}
\be
{\rm Ai}_{b_1}(w)  = \int_{\I \mathbb{R}+\varepsilon'}\frac{\mathrm{d}y}{2\I \pi }
\frac{1}{- y  - b_1} e^{\frac{1}{3} y^3 - w y} \nn
\ee
where $\epsilon' < - b_1 $. Putting these expressions together 
in \eqref{eq:GUEk} and rescaling $r \to r N^{1/3}/z^*$ one 
obtains from \eqref{eq:GUEk} 
\be
 K_N(\mu N+ \sigma N^{1/3},\mu N+ \sigma' N^{1/3}) \simeq \frac{z^*}{N^{1/3}} K_{BBP,b_1}(z^* \sigma, z^* \sigma')
\ee
in terms of the rank $1$ BBP kernel
\be \label{BPPaa} 
K_{\rm BBP,b_1}(v,v') = \int_0^{+\infty} \rmd r {\rm Ai}_{b_1}(v+r) {\rm Ai}^{b_1}(v'+r)
\ee 
This is a particular case of the general rank $m$ BBP kernel recalled in the 
Appendix \ref{app:BBP}. In particular \eqref{BPPaa} is identical to
\eqref{bpp11}, upon simple integrations.
Hence one obtains, using \eqref{PDF0}, the ground state energy and its fluctuations
in the critical region as
\bea
&& - {\cal E}_N^0(x) = \lambda_1(x) = \mu^* N + \frac{N^{1/3}}{z^*}  \rchi_{\rm BBP,b_1}  \\
&& = 2 \sqrt{x N} + \sqrt{x} N^{- 1/6} \rchi_{\rm BBP,b_1} \label{res00} 
\eea 
where \cite{footnotePeche} 
\be \label{CDFBBP} 
\mathbb{P}(\rchi_{\rm BBP,b_1}  \leq \Lambda)=\Det(I- K_{\rm BBP,b_1})_{\mathbb{L}^2(\Lambda,+\infty)}
\ee
is the CDF of the BBP random variable. Comparing with Section \ref{subsubsec:critical} 
we see that $\rchi_{\rm BBP,b_1} = \upsilon_\delta$ with $b_1=\delta/2$. 
As discussed there, this random variable, and the kernel \eqref{BPPaa} 
also describes the half-Brownian IC for the KPZ class
\cite[Formula~(6.23)]{imamura2011replica}.

\subsubsection{Localized phase}

In the localized phase $\theta > \theta_c= 1/a_1^2$, there is no condition to determine $z^*$, hence we 
will choose $z^*=a_1$. We start with \eqref{J1} and we choose $\mu=\mu^*$
so that $\varphi'_0(z^*)=0$. Now the second derivative does not vanish and
the rescaling involves now $N^{1/2}$. 
This gives values for $\mu^*$ and $z^*$ different from those in the delocalized
phase
\be
\mu^* = (\theta + \theta_c) a_1 \quad , \quad \varphi_0''(z^*)=\theta-\theta_c
\ee 
Note that the interpretation is that the occupation of the column $a_1$ is
$\theta-\theta_c$. We insert the Taylor expansion of $\varphi_0(z)$ around
$z=z^*=a_1$ 
inside formula \eqref{J1}, and change integration variable as 
$z=a_1 + \frac{y}{\sqrt{N (\theta-\theta_c)}}$,
and obtain
\bea \label{J1n} 
&& J_N(\mu N+ \nu [(\theta-\theta_c) N]^{1/2} ) \\
&& \simeq \frac{e^{N \varphi_0(z^*) - z^* [(\theta-\theta_c) N]^{1/2} } }{N (\theta-\theta_c) a_1} 
\int_{\I\mathbb{R}+\varepsilon}\frac{\mathrm{d}y}{2\I \pi }
y \, e^{ \frac{y^2}{2} - y \nu} \nn
\eea
where the higher orders in the Taylor expansion are subdominant
at large $N$. Similarly 
\bea \label{H1n} 
&& H_N(\mu N+ \nu [(\theta-\theta_c) N]^{1/2} ) \\
&& \simeq a_1 e^{- N \varphi_0(z^*) + z^* [(\theta-\theta_c) N]^{1/2} } 
\int_{\Gamma}\frac{\mathrm{d}y}{2 \I \pi}
\frac{1}{y} \, e^{- \frac{y^2}{2} + y \nu} \nn
\eea
where $\Gamma$ passes to the right of zero and goes upward
(it encircles zero).
Putting these expressions together 
in \eqref{eq:GUEk} and rescaling $r \to r [(\theta-\theta_c) N]^{1/2}$ one 
obtains
\bea
 && K_N(\mu N+ \sigma [(\theta-\theta_c) N]^{1/2},\mu N+ \sigma' [(\theta-\theta_c) N]^{1/2}) \nn  \\
  && \simeq 
 [(\theta-\theta_c) N]^{-1/2} \int_0^{+\infty} \rmd r \label{KNN}  \\
 && \times 
 \int_{\Gamma}\frac{\mathrm{d}w}{2 \I \pi}
\frac{1}{w} e^{-\frac{w^2}{2}  + w (\sigma+r)} 
\int_{\I \mathbb{R}+\varepsilon}\frac{\mathrm{d}y}{2 \I \pi}
y e^{\frac{y^2}{2}  - y (\sigma'+r)} \nn
\eea
Explicit calculation, using the residue at $w=0$ from the first integral and
Gaussian integration from the second gives
\bea
&& K_N(\mu N+ \sigma [(\theta-\theta_c) N]^{1/2},\mu N+ \sigma' [(\theta-\theta_c) N]^{1/2}) \nn  \\
  && \simeq 
 [2 \pi (\theta-\theta_c) N]^{-1/2} e^{- \frac{(\sigma')^2}{2}} 
\eea 
This leads to 
\bea
&& \lambda_1(x) = (\theta+\theta_c) a_1 N + N^{1/2} {\cal N}(0,\theta-\theta_c) \\
&& =  x a_1 + \frac{N}{a_1} + N^{1/2} {\cal N}(0,\theta - \frac{1}{a_1^2})
\eea 
in agreement with \eqref{loc}.

Note that this is generalized in \cite{Peche_2005} to the
case of $m$ columns equal to $a_1$. The power factors in the integrals in \eqref{KNN} 
generalize
to $y \to y^m$ and $w^{-1} \to w^{-m}$. One then obtains the kernel associated
to a $m \times m$ GUE random matrix, i.e. Eqs.~\eqref{eq:GUEk}, \eqref{HJ} with $N=m$ and $a_j=0$. 


\section{Single line, many columnar defects, zero temperature} 
\label{sec:singlemany} 

In this Section we study the OY model with a single line/polymer and many columnar defects
at $T=0$, specifically with a continuum distribution of energies $\epsilon_j=-a_j$
described by a density $\rho(a)$. We assume that in the limit $N \to \infty$ the number of columns with energies $a_j \in [a,a+da]$ is $\rho(a) da$, with $\int da \rho(a)=1$.
We will assume that this density has an upper (right) edge $a_e$,
where it vanishes for $a \to a_e^{-}$ as
\begin{equation} \label{dens0} 
\rho(a) \sim A (a_e-a)^{2k+\frac{1}{2}}
\end{equation}
with $\rho(a)=0$ for $a>a_e$. We also assume that there
are no columns with $a_j>a_e$. We recall that we are interested in the limit $x,N \to +\infty$
with $\theta=x/N$ fixed and for a given configuration of the $a_i$. 

It convenient for the analysis below to assume the following convergence at large $N$: for $\Re (z) > \max_i a_i$ 
\begin{equation}
\sum_{j=1}^N \log(z-a_j) \simeq N \int_\Omega \mathrm{d}a\, \rho(a) \log(z-a) + o(N^{1/3})
\end{equation}
where $\Omega$ is the support of $\rho$.
Although this condition appears a bit restrictive (it excludes the case of i.i.d. random variables $a_i$),
we believe it is simply a technical restriction and does not impact our main results
\cite{footnoterestriction}.

We can now define the rate function for the many columns case as
\begin{equation} \label{rate} 
\varphi(z)=-\mu z+\theta \frac{z^2}{2}+ \int_\Omega \mathrm{d}a\, \rho(a) \log(z-a)
\end{equation}
so that the functions $H_N$ and $J_N$ from \eqref{HJ} become, at large $N$
(setting for simplicity the convergence factor $q=0$)
\begin{equation}
\begin{split} \label{eq2} 
&H_N(v+\mu N)=\oint \frac{\mathrm{d}z}{2 \I \pi}\, e^{vz - N \varphi(z) }  \\
&J_N(v+ \mu N)=\int_{\I \mathbb{R}+\varepsilon}\frac{\mathrm{d}z}{2 \I \pi}e^{ N \varphi(z) -vz}
\end{split}
\end{equation}

We start by studying the extensive part of the ground state energy, and then we proceed
to obtain its fluctuations.

\subsection{Ground state energy}

\label{sec:gse} 

\label{subsec:gse} 

\begin{table*}[t!]
\begin{center}
\begin{tabular}{p{2.0cm} p{4.5cm} p{4.5cm} p{3.5cm} }
  \hline
    &  Nature of the transition & Existence of localized phase  & Exponent of $\mu-\mu_e$ vs $\theta_c-\theta$ \\ [1ex]
\hline
&    &  & \\[0.5ex]
$k\leq \frac{1}{4} $ & No transition  & No & $\emptyset$ \\[1ex]
$k\in ]\frac{1}{4},\frac{3}{4}[$ & Anomalous  & Yes & $\frac{4 k+1}{4 k-1} \in \, ]2,+\infty[$  \\[1ex]
$k>\frac{3}{4}$& Airy  & Yes & 2 \\[1ex]
  \hline 
\end{tabular}
  \caption{Existence and nature of the phases for a single line and a continuous column density which vanishes near its upper edge as $\rho(a) \sim (a_e-a)^{2 k + \frac{1}{2}}$. The delocalized phase with Tracy-Widom fluctuations of the ground state energy always exist. "Airy" means that critical fluctuations are described by the new-one parameter distribution described in \eqref{TWTW2}-\eqref{Ktau}.}
  \label{table:phase_transition}
\end{center}
\end{table*}

In this Section we determine the ground state energy, ${\cal F}_N(x,T=0)= {\cal E}_N^0(x)$ to leading order
in $N$ as a function of the angle $\theta$ (equivalently the position of the edge of the spectrum of ${\cal M}(x)$ in \eqref{M}). It reads
\be
{\cal E}_N^0(x) \simeq - \mu N    \quad , \quad \mu=\mu(\theta) 
\ee
To determine $\mu$
 we study the integrals in 
\eqref{eq2} setting $v=0$, and look for a saddle point at large $N$. 

The critical points of the rate function \eqref{rate}
are defined as the solutions of $\varphi'(z^*)=0$, i.e. with ${\rm Re}(z^*) > a_i$
\begin{equation} \label{eq1} 
-\mu +\theta z^* + \int_\Omega \mathrm{d}a\, \frac{\rho(a)}{z^*-a}=0
\end{equation}
For finite $N$ this is a polynomial equation of degree $N+1$ and studying the
graph of the function one finds that there is a unique real solution
$z^*$ in the interval $]a_e , +\infty[$. Since we have two unknown,
$z$ and $\mu$ we need an extra condition. This extra condition is
the degeneracy condition is given by the condition that $\varphi''(z^*)=0$, i.e.
\begin{equation} \label{eqtheta} 
\theta -\int_\Omega \mathrm{d}a\, \frac{\rho(a)}{(z^*-a)^2}=0
\end{equation}
There are thus two cases, see Table \ref{table:phase_transition}. Either 

(i) the following integral diverges
\be \label{int} 
\int_\Omega \mathrm{d}a\, \frac{\rho(a)}{(a_e-a)^2} =+\infty
\ee
which happens if $k \leq 1/4$, and there is always a solution to \eqref{eqtheta}, in which case
there only a delocalized phase.

(ii) or, the integral in \eqref{int}
is finite, which happens for $k > 1/4$, then there exists a critical angle $\theta_c < +\infty$ such that
\begin{equation}
\forall \, \theta <\tilde \theta_c, \exists (z^*,\mu) \, \, \, \text{s.t.} \quad   \varphi'(z^*)=\varphi''(z^*)=0
\end{equation}
and $\theta_c$ is determined by
\begin{equation} \label{thetac} 
\theta_c=\int_\Omega \mathrm{d}a\, \frac{\rho(a)}{(a_e-a)^2} \quad , \quad k>1/4
\end{equation}
For $\theta < \theta_c$ this is a delocalized phase, while for 
$\theta > \theta_c$ it is a localized phase, as we will discuss below.

In the delocalized phase, the ground state energy is given by ${\cal E}^0_N(x) \simeq - \mu N$, with $\mu$ the solution
of \eqref{eq1} where $z^*$ is the unique root of \eqref{eqtheta} in the
interval $]a_e, +\infty[$. 

In the localized phase, $\theta > \theta_c$ the saddle point freezes at $z^*=a_e$,
with now $\varphi''(z^*)=\theta-\theta_c$ which plays the role of an order
parameter for this freezing transition. The ground state energy is now
given by ${\cal F}_N(T=0) \simeq - \mu N$ with $\mu = \mu(\theta)$ determined from
\eqref{eq1} as
\begin{equation} \label{mutheta} 
\begin{split}
\mu(\theta) &=a_e \theta + \int_\Omega \mathrm{d}a\, \frac{\rho(a)}{a_e-a}\\
&=(\theta-\theta_c)a_e+\mu(\theta_c)
\end{split}
\end{equation}
and grows ballistically as a function of $\theta=x/N$ in the localized phase,
consistent with the fact that a long enough polymer, i.e. with $x> x_c = N \theta_c$, is localized on
columns of energy $-a_e$ (their localized length is $x - x_c$).
In fact in both phases one has
\be \label{dmu} 
\frac{\rmd  \mu(\theta)}{\rmd  \theta} = z^* 
\ee 

It is important to note that upon approaching the transition from inside the
delocalized phase there are two cases, leading to different
critical behaviors, see Table \ref{table:phase_transition}.

(a) If the following integral is finite
\be \label{int3} 
\int_\Omega \mathrm{d}a\, \frac{\rho(a)}{(a_e-a)^3} < + \infty
\ee 
i.e. $k>3/4$, then $z^*-a_e$ vanishes linearly at the transition
\be \label{zstrans} 
z^*-a_e \simeq \frac{\theta_c- \theta}{2 \int_\Omega \mathrm{d}a\, \frac{\rho(a)}{(a_e-a)^3}}
\ee
From \eqref{dmu} one also has $\mu-\mu(\theta_c) \simeq - (\theta_c-\theta)a_e$ near the
transition. In addition, defining $\mu_e(\theta)=(\theta-\theta_c)a_e+\mu(\theta_c)$ the 
continuation to the delocalized phase of the formula valid in the localized
phase one finds for $\theta<\theta_c$
\be
\mu(\theta) - \mu_e(\theta) \simeq - \frac{(\theta_c- \theta)^2}{4 \int_\Omega \mathrm{d}a\, \frac{\rho(a)}{(a_e-a)^3}}
\ee 
i.e. $z^*-a_e$ vanishes linearly iff $\int_\Omega \mathrm{d}a\, \frac{\rho(a)}{(a_e-a)^3} < + \infty$
which is the case if $k>3/4$. 

(b) If the integral \eqref{int3} diverges, i.e. for $1/4 < k < 3/4$ the transition is in a different, anomalous,
universality class. Indeed one finds in that case (see Appendix \ref{app:trans} for details)
\be \label{new} 
z^*-a_e \sim (\theta_c- \theta)^{\frac{2}{4 k - 1}} ~ , ~ 
\mu(\theta) - \mu_e(\theta) \simeq - (\theta_c- \theta)^{\frac{4 k + 1}{4 k - 1}} 
\ee
The behavior of the fluctuations of the ground state energy in this new universality class is
at present open. Below, we will only study the critical region in the case $k>3/4$. 

\begin{figure*}[t!]
  \centering
\includegraphics[scale=0.52]{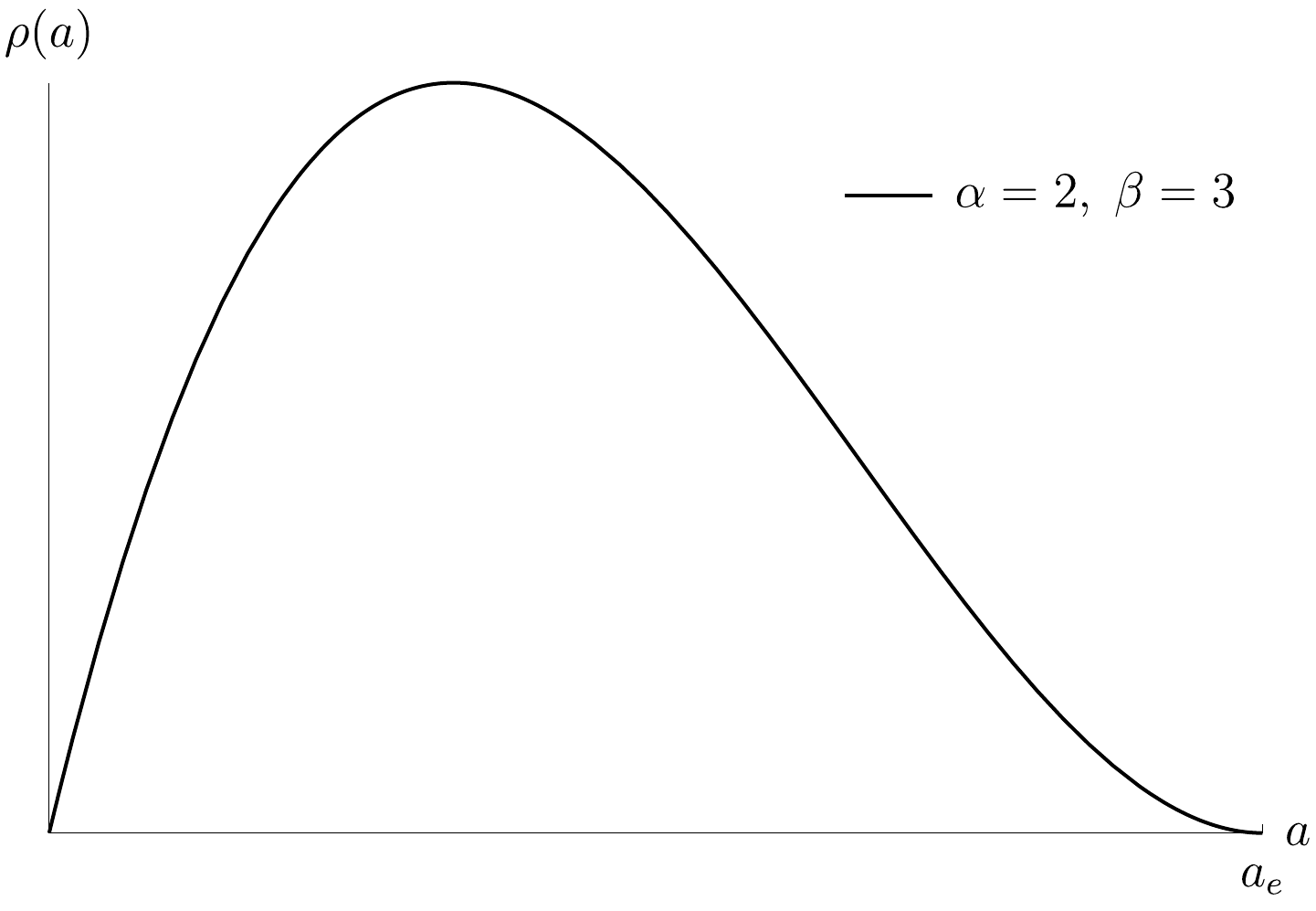}
  \hspace{-5pt}
\includegraphics[scale=0.52]{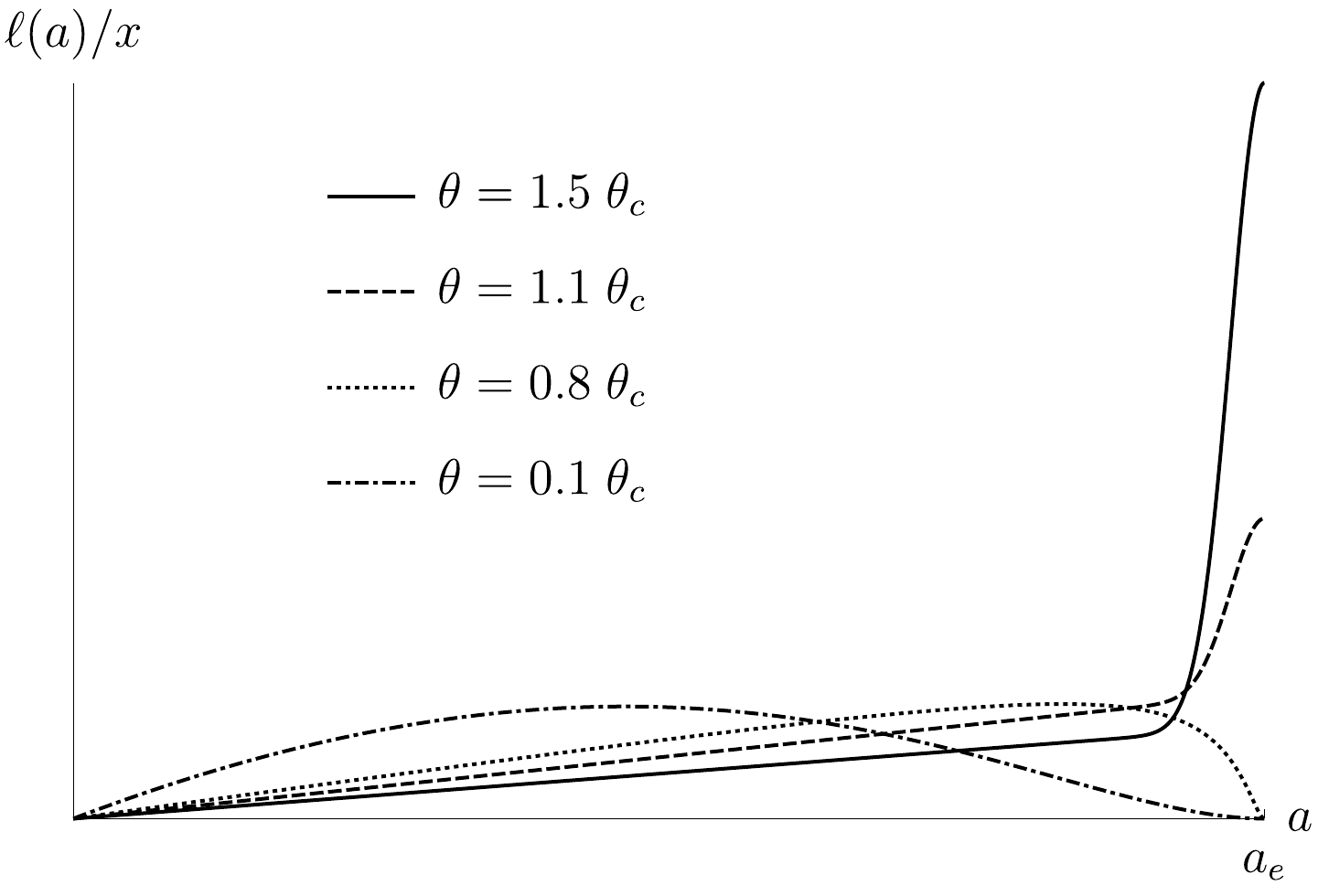}
\includegraphics[scale=0.52]{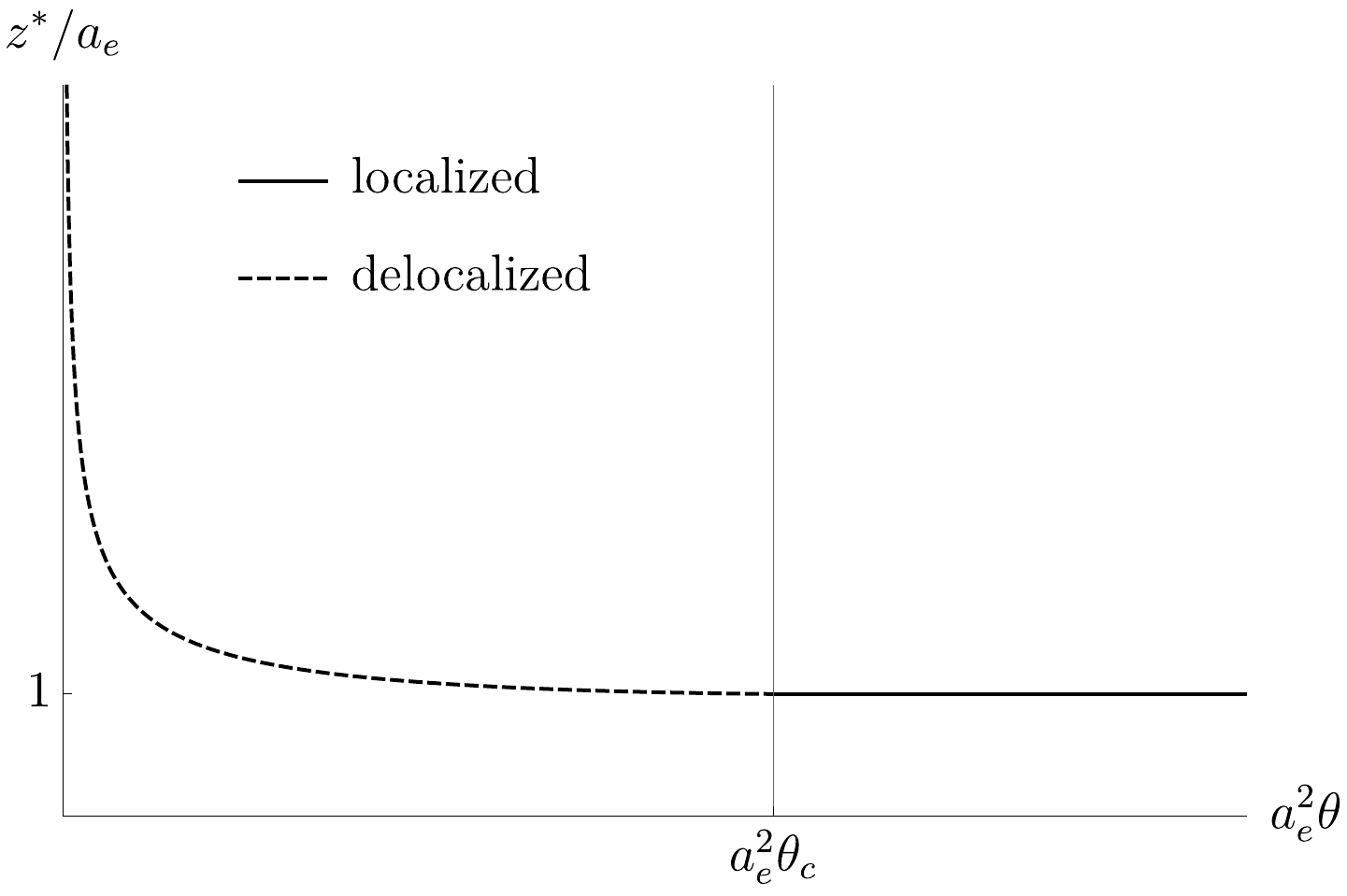}
  \hspace{5pt}
\includegraphics[scale=0.52]{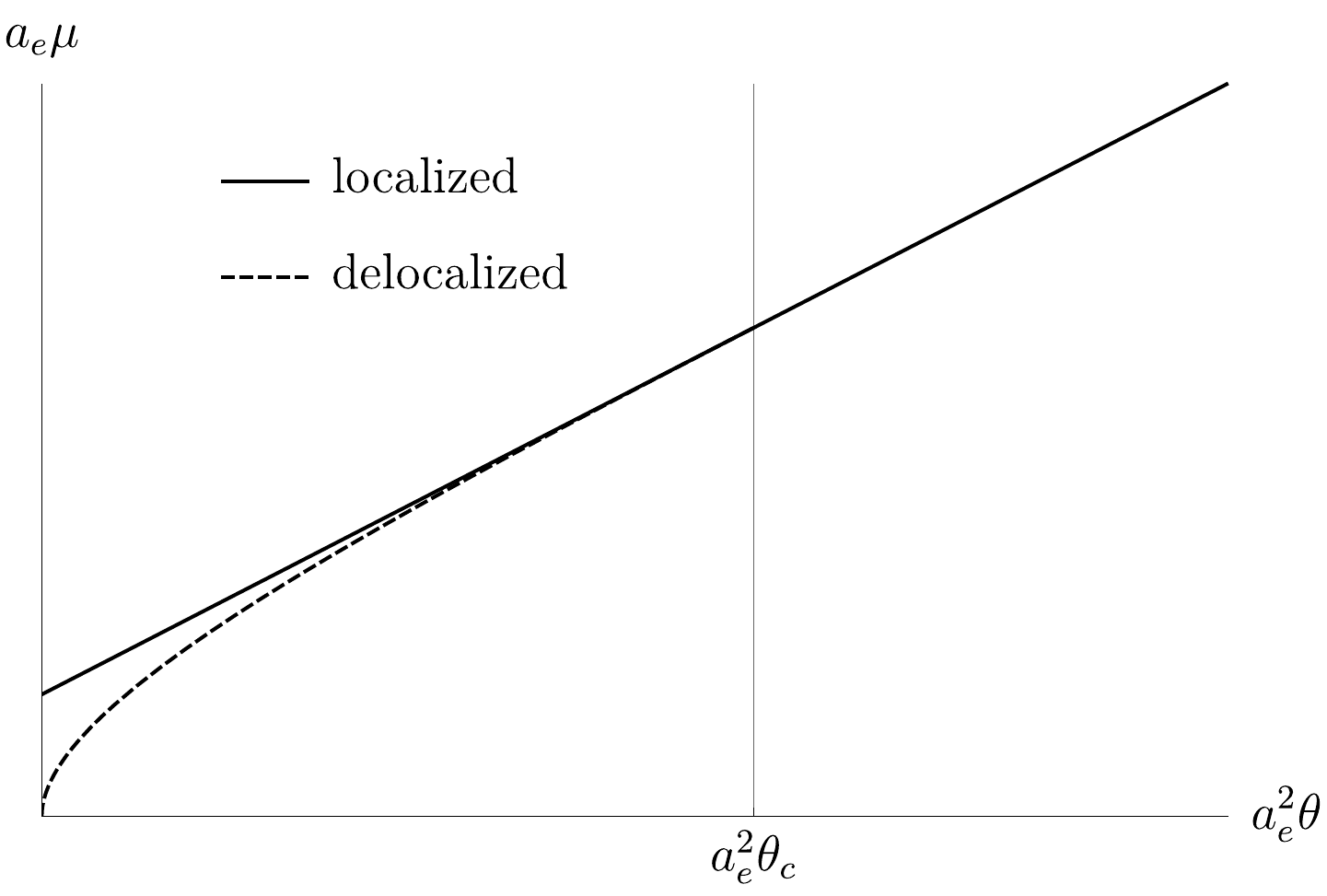}
 \caption{Results for the model \eqref{BetaModel} for $\alpha=2$ and $\beta=3$, corresponding
 to $k=3/4$.
Top left: The density $\rho(a)$ of column strengths $a$. Top right: the occupation length measure $\ell(a) da/x$ normalized to unity (defined in the text) plotted as a function of $a$ for various angles
$\theta$. The localized phase corresponds to $\theta>\theta_c$ and the expected delta function peak in Eq. 
\eqref{laloc} is represented for clarity as a narrow half-Gaussian with the same weight.
Bottom left: plot of the saddle point $z^*= d\mu(\theta)/d\theta$ as a function of $\theta$ (in the two phases).
Its value freezes at $z^*=a_e$ at the localization transition. Bottom right: (minus) the ground state energy per column $\mu(\theta)$ as a function of $\theta$. The dashed line is the result in the delocalized
phase $\theta<\theta_c$. The solid line is the result in the localized
phase for $\theta>\theta_c$ and its continuation for $\theta<\theta_c$ denoted $\mu_e(\theta)$ in the text. }
  \label{fig:gamma2}
\end{figure*}

These results are summarized in the Table \ref{table:phase_transition},
and for some choice of $\rho(a)$, plotted in Fig. \ref{fig:gamma2}.

\subsection{Occupation length} 
\label{subsec:occup}

Let us now discuss the occupation length of the columnar defects in the ground state in both phases. 
In the case of a continuum distribution of column strength the occupation 
length becomes a measure. Let us define
$\ell(a)$ such that the total occupation length of columns in any interval $[a_-, a_+]$
is given by 
\be
\int_{a_-}^{a_+} \ell(a) da = \sum_j \ell^0_j \, \Theta(a_-<a_j<a_+)
\ee
To obtain $\ell(a)$ we can use the equation \eqref{occT0} in the ground state, which gives
that $\ell^0_j \simeq N \partial_{a_j} \mu$, together with the expression of
$\mu$ given above from \eqref{eq1} and \eqref{eqtheta}. This leads
to $N \partial_{a_j} \mu = \frac{1}{(z^*-a_j)^2}$.
We thus obtain that 

(i) in the delocalized phase $\ell(a) da$ is
a continuous measure given by
\be \label{delocdeloc} 
\ell(a) \rmd a = N \frac{\rho(a) \rmd a}{(z^*-a)^2}  \quad , \quad \int_{\Omega} \ell(a) \rmd a = N \theta = x 
\ee 
where $z^*$ is the solution of Eq. \eqref{eqtheta}. Since $z^* > a_e$, this measure
is bounded everywhere. 

(ii) in the localized phase, the occupation length measure, in the large $N$ limit, 
exhibits a smooth part for $a<a_e$, and an atomic part at $a=a_e$
\bea \label{laloc} 
&& \ell(a) \rmd a = N \frac{\rho(a) \rmd a}{(a_e-a)^2} 
+ (x-x_c) \delta(a-a_e) \rmd a 
\eea
Using \eqref{thetac} we see that 
\be
 \int^{a_e^-} \ell(a) \rmd a = N \theta_c = x_c < x
\ee 
The occupation length measure is plotted for illustration in Fig. \ref{fig:gamma2}
for some choice of $\rho(a)$. 

In summary, the localization transition here is a 
"condensation" on the (few) columns of lowest energy very close to the edge $a=a_e$. 
The existence of this transition requires that the density $\rho(a)$ vanishes sufficiently fast
near the edge, $k>1/4$ (see Fig. Table \ref{table:phase_transition}), in other
words it requires an "elitist" population of columns. If there are two many columns
near the edge, the overwhelming competition between them results in a delocalized phase 
only. A workable example of this situation is given in the Appendix \ref{app:2groups}.\\

{\bf Application.} A nice tractable example is provided a Beta distribution of column strengths
with parameters $(\alpha,\beta)$
\be \label{BetaModel} 
\rho(a) = \frac{\Gamma(\alpha + \beta)}{\Gamma(\alpha) \Gamma(\beta)} \frac{a^{\alpha-1} (a_e-a)^{\beta-1}}{a_e^{\alpha + \beta -1}}\mathds{1}_{[0,a_e]}
\ee 
The correspondence with the decay exponent \eqref{dens0} of the density at the upper edge
is $\beta= 2 k + \frac{3}{2}$. We consider the case $\beta>2$, which corresponds to $k>1/4$
such that there is a localization transition. Using the above formula one finds that all integrals can be evaluated in terms of hypergeometric functions. One finds the critical
point and the ground state energy per column
\bea \label{res10} 
 \theta_c &=& \frac{1}{a_e^2} (1 + \frac{\alpha}{\beta -1}) (1 + \frac{\alpha}{\beta -2}) \\
 \mu(\theta_c) &=& \frac{1}{a_e} (1 + \frac{\alpha}{\beta -1}) (2 + \frac{\alpha}{\beta -2}) 
\eea
The results are plotted in Fig. \ref{fig:gamma2} for $\alpha=2$ and $\beta=3$, which 
corresponds to $k=3/4$, the limiting case between the Tracy-Widom and anomalous
localized phase. It is interesting to note that for  $\beta \to +\infty$ 
one recovers from \eqref{res10} the results of a single column for $\theta_c$ 
and $\mu(\theta_c)$.

\subsection{Fluctuations of the ground state energy}

Let us go back to the equations \eqref{eq2}. To study the fluctuations of the ground state energy we 
set $v = \nu N^\beta$, with as yet unknown $\beta$, and study $J_N(\mu N + \nu N^\beta)$
and $H_N(\mu N + \nu' N^\beta)$. As in the previous section we consider the 
saddle point at $z=z^*$, such that $\varphi'(z^*)=0$, where $\varphi(z)$ is given in \eqref{rate}.
We can generally perform an expansion of the rate function $\varphi(z)$ around $z^*$ as
\begin{equation} \label{Taylor} 
\varphi(z^*+w)=\varphi(z^*)+\frac{1}{2}\varphi''(z^*)w^2+\frac{1}{6}\varphi^{(3)}(z^*)w^3+ \dots
\end{equation}
where the successive derivatives, obtained from \eqref{rate}, read
\be
\varphi''(z^*)=\theta -\int_\Omega \mathrm{d}a\, \frac{\rho(a)}{(z^*-a)^2} ~,~
\varphi^{(3)}(z^*)=2\int_\Omega \mathrm{d}a\, \frac{\rho(a)}{(z^*-a)^3}
\ee
The same value of $z^*$ is chosen in both $J_N$ and $H_N$, hence the first term in \eqref{Taylor} cancels out in the product and we can ignore it and write
\bea \label{eqJH} 
&& H_N(\mu N+\nu N^\beta)= \!\!
\oint \frac{\mathrm{d}w}{2 \I \pi}e^{ \nu N^\beta w - \frac{N}{2}\varphi''(z^*)w^2-\frac{N}{6}\varphi^{(3)}(z^*)w^3 +\dots} \nonumber \\
&& J_N(\mu N+\nu N^\beta) \\
&& ~~~~~~~~~~~~~=\int_{\I \mathbb{R}+\varepsilon}\frac{\mathrm{d}w}{2 \I \pi}e^{ \frac{N}{2}\varphi''(z^*)w^2+\frac{N}{6}\varphi^{(3)}(z^*)w^3 -\nu N^\beta w + \dots} \nonumber
\eea

\subsubsection{Localized phase}

Consider now the localized phase. From the previous section we choose $z^*=a_e$. 
Hence the second derivative in the expansion in \eqref{eqJH} is $\varphi''(z^*)=\theta- \theta_c >0$.
We can thus rescale the integration variable $w$ by $N^{-1/2}$ and choose $\beta=\frac{1}{2}$. In the large $N$ limit the higher order terms in the expansion can be neglected.
The saddle point integral
reads 
\bea
J_N(\mu N+\nu N^{1/2}) &\simeq& \frac{1}{N^{1/2}}\int_{\I \mathbb{R}}\frac{\mathrm{d}w}{2\I \pi }e^{ \frac{\theta-\theta_c}{2}w^2 -\nu w}\\
&&= \frac{1}{\sqrt{2\pi (\theta-\theta_c)N}}e^{-\frac{\nu^2}{2(\theta-\theta_c)}} \label{resJN} 
\eea
The same result \eqref{resJN} holds for $H_N$ since, under rescaling, the contour opens
to be parallel to the imaginary axis.

The kernel \eqref{eq:GUEk} then becomes, after rescaling $r \to N^{1/2} r$
\bea
&& K_N(\mu N+\nu' N^{1/2},\mu N+\nu N^{1/2}) = N^{1/2}\int_{0}^{+\infty}\mathrm{d}r\, \nn \\
&& \times H_N(\mu N+(\nu'+r) N^{1/2})J_N(\mu N+(\nu+r) N^{1/2}) \nn \\
&&= \frac{1}{2\pi (\theta- \theta_c)\sqrt{N}}\int_{0}^{+\infty}\mathrm{d}r\,  e^{-\frac{(\nu'+r)^2}{2(\theta-\theta_c)}-\frac{(\nu+r)^2}{2(\theta - \theta_c)}}\\
&&=\frac{1}{4\pi^2 \sqrt{N(\theta-\theta_c)}}   e^{-\frac{\left(\nu -\nu '\right)^2}{4 (\theta-\theta_c)}} \text{Erfc}\Big(\frac{\nu '+\nu }{2
   \sqrt{\theta-\theta_c}}\Big)
\eea
Hence, defining the kernel
\begin{equation} \label{kT} 
\mathcal{T}(x,y)=\frac{1}{\pi} \int_0^\infty \mathrm{d}u \, e^{-(x+u)^2}e^{-(y+u)^2}
\end{equation}
upon rescaling, we find that the ground state energy in the localized phase
fluctuates as
\begin{equation} \label{om}
- {\cal E}_N^0(x) = \lambda_1(x) = \mu N + N^{1/2} \sqrt{2(\theta - \theta_c)} \, \omega
\end{equation}
where $\mu=\mu(\theta)$ was obtained in \eqref{mutheta} 
and $\omega$ is a random variable whose CDF is given by
the Fredholm determinant
\begin{equation}
\mathbb{P}(\omega\leq s)=\mathrm{Det}(I- {\cal T})_{ \mathbb{L}^2(s,+\infty)}
\end{equation}
Although this distribution is, to our knowledge, novel, the kernel $ {\cal T}(x,y)$ already appeared
in the study of the largest real eigenvalue of the real Ginibre 
ensemble \cite{Rider_2014,Poplavskyi_2017,BaikBothner,footnoteZabo}.
More precisely the kernel which appears there is $ {\cal T}(x,y)+ g(x) G(y)$ where
$g(x)=\frac{1}{\sqrt{2 \pi}} e^{-x^2/2}$ and $G(y)= \int_{-\infty}^y g(x) \rmd x$.
The asymptotics are obtained as
\begin{equation}
\begin{split}
\mathbb{P}(\omega\leq s) &\simeq_{s\to +\infty} 1-\frac{e^{-2 s^2}}{16 \pi s^2} \\
& \simeq_{s\to -\infty} e^{- \frac{\zeta(3/2)}{\sqrt{2 \pi}}|s|} 
\end{split}
\end{equation}
combining, in the second case \cite[Eqs.~(1.9) and (1.11)]{Poplavskyi_2017}.

We note that to obtain this result we have (i) assumed that $\theta_c$ is finite, i.e. that 
$\int_\Omega \rmd a \frac{\rho(a)}{(a_e-a)^2} < + \infty$ (equivalent to the existence of a localized phase)
which holds for $k>1/4$ (ii) neglected, after rescaling, the higher derivatives
in the saddle point. To be more precise one can check that the condition
for the above analysis to work is that for fixed $w$
\bea
&& \lim_{N \to +\infty} N \int_\Omega \rmd a \rho(a) \bigg( \log(1 + \frac{w N^{-1/2}}{a_e-a}) \\
&& - \frac{w N^{-1/2}}{a_e-a}
+ \frac{1}{2} [\frac{w N^{-1/2}}{a_e-a}]^2 \bigg) = 0 \nn
\eea
which is weaker than the condition of the existence of the third derivative
$\varphi'''(z^*)$.

\subsubsection{Delocalized phase}

Consider now the delocalized phase, $\theta < \theta_c$. Then one can
choose $z^*>a_e$ so that $\varphi''(z^*)=0$. The saddle point will then be 
of cubic type. One now rescale $w \to  w N^{-1/3}$ and choose $\beta=\frac{1}{3}$ so that
the function $J_N$ in \eqref{eqJH} takes the form
\bea
J_N(\mu N+\nu N^{1/3})&=&\frac{1}{N^{1/3}}\int_{\I \mathbb{R}+ \epsilon}\frac{\mathrm{d}z}{2 \I \pi}e^{ \frac{\varphi^{(3)}(z^*)}{6}z^3 -\nu z} \nn \\
&&=\left(\frac{2}{N\varphi^{(3)}(z^*)}\right)^{1/3} \mathrm{Ai}\left(\frac{2^{1/3}}{\varphi^{(3)}(z^*)^{1/3}}\nu \right) \nn \\
\label{resJN2} 
\eea
Similarly one finds that $H_N(\mu N+\nu' N^{1/3})$ is also given by 
\eqref{resJN2}.

The kernel \eqref{eq:GUEk} then becomes, after rescaling $r \to N^{1/3} r$
\begin{equation}
\begin{split}
& K_N(\mu N+\nu' N^{1/3},\mu N+\nu N^{1/3}) \\
&= \Big(\frac{2}{N\varphi^{(3)}(z^*)}\Big)^{1/3}K_{\mathrm{Ai}}\left(\frac{2^{1/3}}{\varphi^{(3)}(z^*)^{1/3}}\nu,\frac{2^{1/3}}{\varphi^{(3)}(z^*)^{1/3}}\nu'\right) \\
\end{split}
\end{equation}
Hence we find Tracy-Widom fluctuations for the ground state energy at leading order
\be \label{TWTW} 
- {\cal E}^0_N(x) =\lambda_1(x) = \mu N + \Big(N\frac{\varphi^{(3)}(z^*)}{2}\Big)^{1/3} \chi_2
\ee 
We recall that to obtain the coefficients $\mu$ and $\varphi^{(3)}(z^*)$ we 
must first find $z^*$ as a function of $\theta$ from the second equation 
and then insert its value in the first and third equations of the parametric
system
\begin{equation}
\begin{split}
&\mu =\theta z^* + \int_\Omega \mathrm{d}a\, \frac{\rho(a)}{z^*-a}\\
&\theta =\int_\Omega \mathrm{d}a\, \frac{\rho(a)}{(z^*-a)^2}\\
&\varphi^{(3)}(z^*)=2\int_\Omega \mathrm{d}a\, \frac{\rho(a)}{(z^*-a)^3}
\end{split}
\end{equation}

\subsubsection{Critical region}
\label{sec:critreg} 

We now study the critical region between the localized and delocalized phases, near $\theta \approx \theta_c$. We assume
that $k> 3/4$ so that $\theta_c < +\infty$ and 
\be \label{trois} 
\varphi^{(3)}(a_e)=2  \int_\Omega \mathrm{d}a\, \frac{\rho(a)}{(a_e-a)^3} < +\infty 
\ee 
To be able to describe the critical region starting from the localized
phase, we choose $z^*=a_e$. We also
impose $\varphi'(a_e)=0$ by choosing 
\be
\mu = \mu_e(\theta) = \theta a_e + \int_\Omega \mathrm{d}a\, \frac{\rho(a)}{(a_e-a)}
\ee
so that the first derivative is zero. In \eqref{eqJH} we choose $\beta=1/3$ and obtain
\bea \label{JJ100} 
&& J(\mu N+\nu N^{1/3}) \\
&& = 
\int_{\I \mathbb{R} + \epsilon}\frac{\mathrm{d}z}{2 \I \pi}e^{ \frac{N}{2} (\theta-\theta_c) (z-a_e)^2+\frac{N}{6}\varphi^{(3)}(a_e)(z-a_e)^3 -\nu N^{1/3} z + \dots} \nn
\eea
where we used that $\varphi''(a_e)=\theta-\theta_c$. Here $\Gamma$ is an appropriate
contour parallel to the imaginary axis where the integral converges (see below).
It is clear from \eqref{JJ} that in order to balance the quadratic and cubic terms, the critical region 
is defined at large $N$ as
\be \label{critreg} 
\theta - \theta_c = \tau N^{-1/3} 
\ee 
where $\tau$ is fixed. We write $z=a_e + w N^{-1/3}$ and obtain 
\begin{equation}
J_N(\mu N+\nu N^{1/3}) \simeq N^{-1/3} \int_{\I \mathbb{R} + \epsilon}\frac{\mathrm{d}w}{2 \I \pi}e^{ \frac{1}{2} \tau w^2+\frac{1}{6}\varphi^{(3)}(a_e)w^3 -\nu w}
\end{equation}
where, after the rescaling, the higher order derivatives can be (naively) neglected. Note that
this integral is convergent only for $\tau \geq 0$, which we assume for now.

Let us recall
the useful formula, for any 
$a \in \mathbb{R}^*$, which we use here for $b \geq 0$
\be \label{Phi}
\int_{\I \mathbb{R} + \epsilon} \frac{\rmd z}{2 \I \pi} e^{a \frac{z^3}{3} + b z^2 + c z} = \frac{e^{\frac{2 b^3}{3 a^2} - \frac{b c}{a}} }{|a|^{1/3} }
{\rm Ai}(\frac{b^2}{|a|^{4/3} } - \frac{c ~ {\rm sgn}(a)}{|a|^{1/3}}) 
\ee 
using the definition of the Airy function \eqref{Ai} and shifting the integration contour.

Hence we find for $\tau \geq 0$
\bea
&& J_N(\mu N+\nu N^{1/3}) \simeq (\frac{2}{N \varphi^{(3)}(a_e)})^{1/3} 
e^{\frac{\tau^3}{3 \varphi^{(3)}(a_e)^2} + \frac{\tau \nu}{\varphi^{(3)}(a_e)}} \nn \\
&& 
~~~~~~~~~ \times {\rm Ai} \big(\frac{\tau^2}{2^{2/3} \varphi^{(3)}(a_e)^{4/3} } + \frac{2^{1/3} \nu}{\varphi^{(3)}(a_e)^{1/3}} \big) 
\eea
One can show that $H_N$ is given by the same formula, using contour reversal and orientation
reversal (see discussion around \eqref{Ai}). It amounts to use the same formula \eqref{Phi}
with $(a,b,c) \to (-a,b,-c)$, which leaves it invariant. 

This leads to the kernel in the form
\bea
&&K(\mu N+\nu' N^{1/3},\mu N+\nu N^{1/3}) \\
&& \simeq \frac{1}{N^{1/3}}\left(\frac{2}{\varphi^{(3)}(a_e)}\right)^{2/3}
e^{2 \frac{\tau^3}{3 \varphi^{(3)}(a_e)^2} }
\int_{0}^{+\infty}\mathrm{d}r\, 
e^{\frac{\tau (\nu+ \nu'+ 2 r)}{\varphi^{(3)}(a_e)}} \nn \\
&& \times 
 \mathrm{Ai}\left(\frac{\tau^2}{2^{2/3} \varphi^{(3)}(a_e)^{4/3} }  + \frac{2^{1/3}}{\varphi^{(3)}(a_e)^{1/3}}(\nu'+r) \right) \nn \\
 && \times \mathrm{Ai}\left(\frac{\tau^2}{2^{2/3} \varphi^{(3)}(a_e)^{4/3} } + \frac{2^{1/3}}{\varphi^{(3)}(a_e)^{1/3}}(\nu+r) \right) \nn
\eea

Hence the final result is as follows. For $\tau \geq 0$, i.e. on the localized side
of the critical region, the ground state
energy fluctuates to leading order as 
\be \label{TWTW2} 
 - {\cal E}^0_N(x)  =\lambda_1(x)=  \mu N + 
 \Big(\frac{\varphi^{(3)}(a_e)}{2}N\Big)^{1/3} \chi
\ee 
where the CDF of the random variable $\chi$ is given by the following Fredholm
determinant
\be
\mathbb{P}( \chi\leq  s) = {\rm Det}( I - K_{\tilde \tau})_{\mathbb{L}^2(s,+\infty)}
\ee 
where the dimensionless parameter $\tilde \tau$ measures the distance to
criticality
\be \label{tildetau} 
\tilde \tau=
\frac{N^{1/3} (\theta - \theta_c)}{2^{1/3} \varphi^{(3)}(a_e)^{2/3}} 
\ee
and the kernel $K_{\tilde \tau}$ is given for $\tilde \tau \geq 0$ as
\bea \label{Ktau} 
&& K_{\tilde \tau}(v,v') \\
&& = e^{\frac{4}{3} \tilde \tau^3} \int_0^{+\infty} \rmd r e^{\tilde \tau (v+v'+ 2 r) } 
{\rm Ai}(v+ r + \tilde \tau^2) {\rm Ai}(v'+ r + \tilde \tau^2) \nn
\eea

One now notes that for $\tilde \tau =0$, this kernel recovers simply the Airy kernel
$K_{\tilde \tau=0}=K_{\rm Ai}$. This indicates that for $\tilde \tau<0$, on the delocalized
side, one should instead choose, as usual to describe the Tracy Widom phase, 
$z^*$ such that $\varphi''(z^*)=0$. Hence for $\tilde \tau<0$, 
the result \eqref{TWTW} holds, i.e. the fluctuations are GUE Tracy-Widom, $\chi=\chi_2$. 
It is quite remarkable that there is no precursor of the transition in the leading fluctuations
of the ground state energy
on the delocalized side (fluctuations remain Tracy-Widom all the way to $\tilde \tau=0$),
while their CDF varies continuously on the localized side.

The above kernel $K_{\tilde \tau}(v,v')$ interpolates between the Airy kernel
for $\tilde \tau=0$ and the kernel $\mathcal{T}$ in \eqref{kT}, which describes the fluctuations in the
localized phase. It happens as follows. In the limit $\tilde \tau \to +\infty$ one can use in
\eqref{Ktau} the asymptotics of the Airy function for large positive argument 
\be
\Ai(z)\sim \frac{1}{ \sqrt{4 \pi} z^{1/4}} e^{ - \frac{2}{3} z^{3/2}} 
\ee 
and obtain, expanding up to quadratic order in the exponential
\be
K_{\tilde \tau}(v,v') \underset{\tilde \tau \to + \infty}{\to} \frac{1}{\sqrt{4 \tilde \tau}} 
\mathcal{T}\left(\frac{v}{\sqrt{4 \tilde \tau}}, \frac{v'}{\sqrt{4 \tilde \tau}}\right)
\ee 
We thus find that in this limit the random variable $\chi$ in \eqref{TWTW2} 
\be
\chi \to \sqrt{4 \tilde \tau} \, \omega
\ee 
where $\omega$ is the random variable in \eqref{om}, hence given the
definition of $\tilde \tau$ in \eqref{tildetau} we find that 
\eqref{TWTW2}  and \eqref{om} match deep in the localized side
of the transition.

\subsection{Additional relations to RMT: ground state energy as a function of the 
polymer endpoint position}

One can ask how the ground state energy ${\cal E}_N^0(x)$ for a polymer of length $x$
and fixed entry point position on the first column $j=1$, depends on the choice of the exit point position $j=N$.
This is asking about ${\cal E}_N^0(x)$ at fixed $x$, as a process in $N$ (in the same
disorder environment). It is indeed important to study how the ground
state responds to a small perturbation (here moving the endpoint by one unit). 
Indeed, glassy systems often have broadly distributed, intermittent response,
i.e. rare but large response called avalanches.

It turns out that this question is also related to random matrices. More precisely,
for a given configuration of the point disorder, i.e. the Brownian motions $B_j(x)$, and the column
strengths, the $a_j$, the joint PDF (JPDF) of $- {\cal E}_{N-1}^0(x)$ and $- {\cal E}_N^0(x)$
is the same as the JPDF of the largest eigenvalue $\lambda_1^{(N)}$ of a $N \times N$ GUE
random matrix and of $\lambda_1^{(N-1)}$, the largest eigenvalue of  
its $N \times N$ minor matrix (obtained by erasing one line and one column). 

It is known that ${\cal E}_N^0(x)$ as a process in $N$, is identical to the so-called minor GUE process
\cite{Baryshnikov_2001}, which is determinantal (i.e. all its correlations are given by determinants involving
a kernel) and that these properties extend in presence of drifts, in relation to the deformed GUE minor process \cite{adler2013random}.

\begin{figure}[t!]
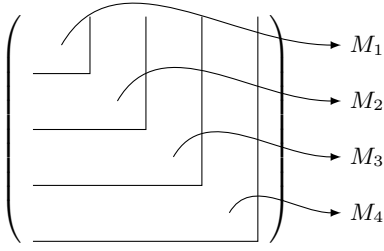

\MinorMatrix
\caption{Representation of the principal minors $M_n$ of the matrix ${\cal M}$}
\label{fig:minors} 
\end{figure}

Let us recall the properties of the deformed GUE minor process described in Ref. \cite{adler2013random}.
Define the $n$-th principal minor (top left) as shown in Fig. \eqref{fig:minors} 
\be
M_n = [{\cal M}(x=1)]_n = [{\rm diag}(a_1,\dots,a_N) + V]_n
\ee
where $V$ is the same GUE matrix as in \eqref{M}.

For simplicity we consider here a polymer of fixed length $x=1$
but arbitrary $x$ is easily obtained by rescaling. 
We denote the eigenvalues of $M_n$ as $\lambda^{(n)}=(\lambda_1^{(n)},\dots,\lambda_n^{(n)})$.
Note that here we ordered them in increasing order.

A first result is the transition probability. Suppose the eigenvalues with $N-1$ columns are known, called 
$\mu=(\mu_1,\dots,\mu_{N-1})$. Then the JPDF of the $(\lambda_1,\dots,\lambda_N)$
satisfy  \cite[Theorem 1]{adler2013random}
\bea
P_\mu(\lambda) = C_\mu \Delta_N(\lambda) e^{\sum_{i=1}^N (a_N \lambda_i - \frac{\lambda_i^2}{2})} 
\mathbf{1}_{\mu \preceq  \lambda }
\eea 
where the normalization constant $C_\mu=\frac{e^{-a_N^2/2}}{\sqrt{2 \pi} \Delta_{N-1}(\mu)} 
e^{\sum_{i=1}^{N-1} \frac{1}{2} \mu_i^2 - a_N \mu_i}$. Here
with $\mu \preceq \lambda$ stands for the interlacing property $\lambda_1 \leq \mu_1 \leq \lambda_2\leq  \dots \leq \lambda_n$,
since an important property of the eigenvalues of a matrix and of its largest minor is that they 
are interlaced. Another property is that the JPDF of the complete interlacing set of eigenvalues 
is given by  \cite[Corollary 1]{adler2013random}.
\bea
&& P(\lambda^{(1)},\dots,\lambda^{(n)},\dots,\lambda^{(N)}) \\
&& =\Delta_N(\lambda^{(N)}) (2 \pi)^{-N/2} 
e^{- \frac{1}{2} \sum_{n=1}^N a_n^2}
\prod_{n=1}^N e^{- \frac{1}{2}  (\lambda_n^{(N)})^2 + a_N \lambda_n^{(N)})} \nn \\
&& \times e^{ \sum_{n=1}^N (a_n - a_{n+1}) \sum_{i=1}^n \lambda_i^{(n)} }
\mathbf{1}_{ \lambda^{(n)} \preceq  \lambda^{(n+1)} }
\eea
In order to perform actual calculations it is useful to note that the
deformed GUE minor process is a so-called extended determinantal process.
The definitions and the explicit expression for its extended kernel $K(n_1, x ; n_2,y) \equiv K_{n_1,n_2}(x,y)$ are given in 
 \cite[Theorem 3]{adler2013random}. It is an extended
 version of the kernel \eqref{eq:GUEk}, for different values of $N$.
 
Standard formula for determinantal process then allow to write the
joint CDF of minus the ground state energies for two polymer endpoint positions at $N_1$ and $N_2$, i.e. of $\lambda_n^{(N_1)}= - {\cal E}^0_{N_1}$ and $\lambda_n^{(N_2)}= - {\cal E}^0_{N_2}$
with $N_2<N_1$ as a matrix Fredholm determinant
\bea
&& \mathbb{P}( - {\cal E}^0_{N_1}<s_1 , - {\cal E}^0_{N_2} < s_2) \\
&&= {\rm Det} \begin{pmatrix} I - P_{s_1} K_{N_1,N_1} P_{s_1} & - P_{s_1} K_{N_1,N_2} P_{s_2} \\
P_{s_2} K_{N_2,N_1} P_{s_1} & I - P_{s_2} K_{N_2,N_2} P_{s_2}  \end{pmatrix} \nonumber
\eea
where $P_s(x)= \Theta(x-s)$ is the projector of $[s,+\infty[$. We will not attempt
here to analyze this formula, but in principle it can be done along similar
lines as in this paper.

\section{Many lines, zero temperature}
\label{sec:manymany}

\begin{figure*}[t!]
\centerline{
\includegraphics[scale=0.35]{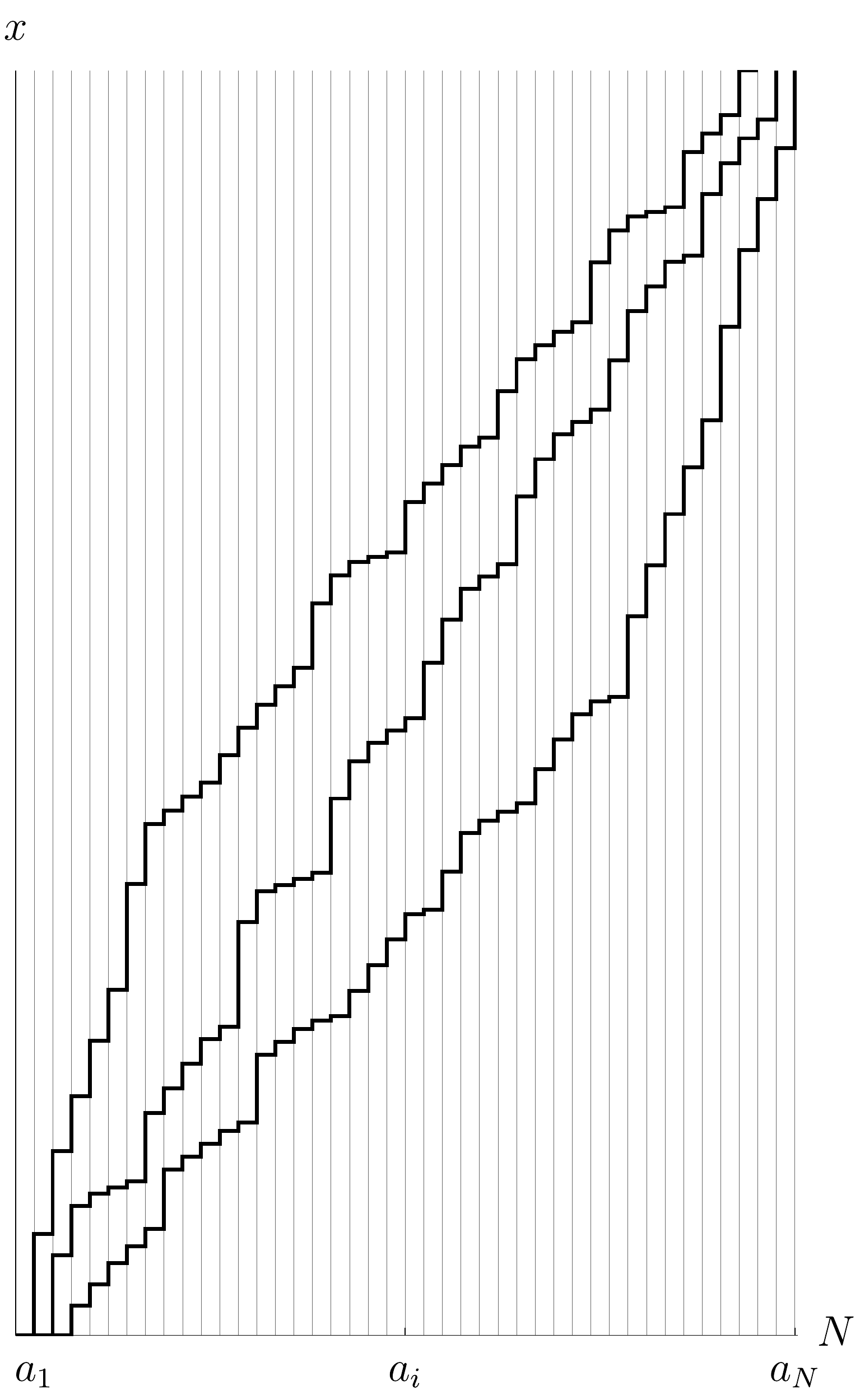}
\hspace*{2cm}
\includegraphics[scale=0.35]{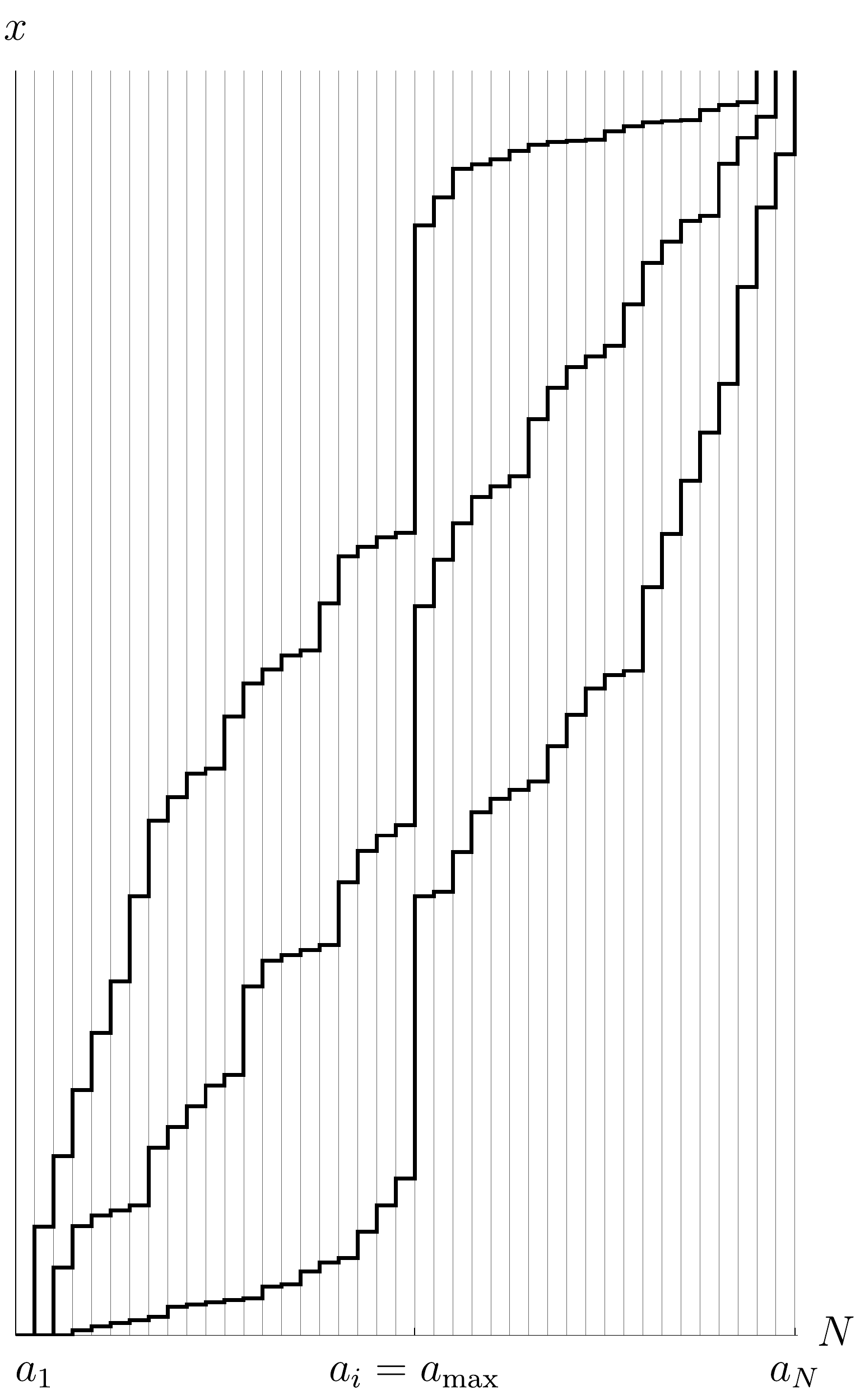}
}
\caption{\textbf{Left.} Three lines ($M=3$) in the O'Connell-Yor polymer model (i.e. constrained not
to cross) 
of length $x$ with $N$ columnar defects, in the phase where they are delocalized over the space. \textbf{Right.} Three lines localized over the column $i$ where the disorder is most favorable. The total
occupation length of column $i$, $L_i$, is the sum of the occupation lengths of each line. }
\label{fig:3lines}
\end{figure*}

In this section we explore the problem of several interacting elastic lines in the same columnar and
point impurity disorder. The "solvable" case studied here corresponds to imposing an infinite
hard core repulsion between the lines, enforced by a non-crossing condition.
In Ref. \cite{nelson1993boson} a hard core repulsion between the lines
was also considered, leading to the prediction of a "Bose glass"
phase. In the absence of point disorder, it was 
implemented in a heuristic way by filling the lowest energy 
columns (a.k. the best localized states) one by one, very much like non-interacting fermions, 
until reaching a "Fermi energy". In the case of a commensuration 
between the number of lines and of (active) columns it was predicted that one can reach a "Mott insulator" phase.

Here we describe exact results in presence of point disorder, for the
"one-way" model with many lines. Since it neglects all jumps backward 
this model is a priori more relevant to describe the delocalization transition away from the Bose glass which occurs upon tilting the lines (see Figure 5 in \cite{hwa1993flux} and Fig. \ref{fig:BG}).
The results described below also assume 
that the endpoints of the lines are
closely packed on neighbouring columns. In practice, it means that the
entry and exit positions of the flux lines are constrained within a narrow region. 
That could be enforced in experiments by inducing channels 
where the superconductivity, i.e. the critical field $H_{c1}$ and/or the vortex core energy, is locally reduced. Note that inside the sample the lines will expand and form a limit shape 
This is similar to the limit shape of a set of non-intersecting random walks, also called
watermelons, but here in presence of a columnar and point disorder. For usual
watermelons the shape can be inferred from the connection with the DBM and the 
GUE, as a semi-circle with a time-varying width, while in presence of
disorder much less is known (see discussion of related problems and continuum limits in \cite{CorwinMultilayer,johnston2019scaling, basu2020interlacing}). Note that some properties for more general endpoint configurations have also been obtained \cite{footnote2}.

\subsection{Relation to RMT} 

The relation between the O'Connell-Yor polymer at $T=0$ and the deformed GUE extends to several non-crossing polymers.
Consider now $M \geq 1$ O'Connell-Yor polymers which see the same $N \geq M$ columns, with energies $\epsilon_j=-a_j$,
and the same Brownian impurity disorder defined on each columns, $B_j(x)$, $j=1, \dots, N$. The polymers are furthermore constrained not
to intersect. If one denotes $0< x_j^{(k)}<x$ the successive jump positions of polymer $k=1,\dots,M$, they are thus contrained by an interlacing condition (see the definition of the model in Appendix
\ref{app:defmany}). 

The boundary conditions are the following.  
The polymer endpoints
at $x=0$ are in column positions $j=1,\dots,M$. The endpoints at $x$ are in 
column positions $j=N-M+1,\dots,N$. This is illustrated in Fig. \ref{fig:3lines}.
Note that as a consequence there is a 
global "tilt" of the lines by an angle $\phi_M$ such that 
\be
\tan \phi_M = \frac{N-M}{x} = \frac{1}{\theta_M} 
\ee
The total
energy to be minimized is the sum of the energies $\sum_{k=1}^M E^{(k)}_N$ of each polymer.
Let us call ${\cal E}^0_{N,M}(x)$ the minimum energy (ground state energy)
under the above constraints (non-intersection and boundary conditions). 
The theorem states that \cite{o2003path,adler2013random} 
\be \label{totalenergy} 
{\cal E}^0_{N,M}(x) \inlaw - \, \sum_{\alpha=1}^M \lambda_\alpha(x) 
\ee 
where $\lambda_1(x) > \lambda_2(x) > \dots > \lambda_M(x)$ are the $M$ largest eigenvalues
of the deformed $N \times N$ GUE matrix ${\cal M}(x)$ defined in \eqref{M}.

It turns out that the joint PDF of these eigenvalues $\lambda_j(x)$ for
the matrix ${\cal M}(x)$ at fixed value of $x$ is known explicitly, so we indicate it here.
As shown in \cite{brezin1998universal,Johansson_2001}, the symmetrized joint PDF of the $\lambda_j(x)$
(i.e. here with no ordering), $j=1,\dots,N$ is given by
\be \label{JPDFDeformed} 
P_x(\{\lambda\})=\frac{1}{(2\pi x^N)^{N /2}}\frac{\Delta_N(\lambda)}{\Delta_N(a)}\mathrm{det}(e^{-(\lambda_i 
- x a_j)^2 /2x})_{1 \leq i,j \leq N} 
\ee
where the Vandermonde determinant is defined with the following convention $\Delta_N(\lambda)=\prod_{1\leq i<j\leq N}(\lambda_i-\lambda_j)$. This JPDF has a determinantal form, with a kernel given in \cite[Proposition (2.3) and formula (2.18)]{Johansson_2001}.
This result is the starting point for obtaining the formula \eqref{PDF0}-\eqref{eq:GUEk}
for the PDF of the largest eigenvalue $\max_j \lambda_j(x)$ (which above and
below is denoted $\lambda_1(x)$).

Again, as for the case $M=1$, one can show that \eqref{totalenergy} holds as a process in $x$, replacing $\sqrt{x} V \to W(x)$ the hermitian Brownian motion \cite{o2003path}. Again the $\{ \lambda_i(x) \}_{i=1,\dots,N}$ form a determinantal point process and its extended kernel is known \cite{KurtPrivate}
following earlier works \cite{TW3,Johansson_2001},
and can be found in explicit form 
in e.g.
\cite[Proposition 2.1]{KatoriExtendedDBMKernel} and
in \cite[Eq. (1.13)]{claeys2019critical} (see also \cite{footnotenew2}).

From the above property \eqref{totalenergy} we see that if $M$ is fixed and $N$ becomes large 
(i.e. few lines, many columns) one is probing the edge of the spectrum of the matrix ${\cal M}(x)$. By contrast if 
$M$ and $N$ grow with a fixed ratio, one is probing the bulk of the spectrum. We will
study both cases below. 

On can define the total occupation length of a given column $j$ by all lines
as $L_j= \sum_{k=1}^M \ell^k_j$ where $\ell^k_j=x_{j+1}^{(k)}-x_j^{(k)}$ is the occupation length of column $j$ by the $k$-th line. These occupation lengths satisfy the sum rule $\sum_{j=1}^N L_j = M x$. 
The optimal total occupation length, denoted $L_j^0$, can again be obtained as a derivative of the ground state energy
\be \label{dercol} 
- \partial_{a_j} {\cal E}^0_{N,M}(x) = L^0_j = \sum_{k=1}^M \ell^{k,0}_j 
\ee 
where $\ell^{k,0}_j$ denote the corresponding values in the ground state.

Note that there is again a non-trivial property of permutation invariance of the column strengths $a_j$
(see remark after Theorem 8.3 of \cite{o2003path}).

Finally, note that the model can also be solved at finite-temperature \cite{O_Connell_2012,BorodinMacdo}
for $M$ non-crossing O'Connell-Yor polymers, which define a
hierarchy of partition sums.

\subsection{Few lines and few columnar defects: independent BBP transitions}

Suppose first that there are only a fixed and finite number $n$ of active columns,
with $a_1>a_2>\dots >a_n$, and all other $a_{j > n}=0$. One can ask how the 
system of $M$ lines will localize on these $n$ columns. 

Using the above relation \eqref{totalenergy} we see that this problem 
 corresponds to a rank-$n$ perturbation of a $N \times N$ GUE matrix.
It is known that as $N \to +\infty$, there are in that case 
$n$ successive and distinct BBP transitions as $\theta=x/N$ is
increased from $0$ (see \cite[Theorem 2.1 and Section 3.1]{benaych2011eigenvalues}
with the correspondence $\sigma \to \sqrt{N x}$ and $\theta_i \to x a_i$). They occur successively
at $\theta= \theta_c^{(i)} = 1/a_i^2$, $i=1,\dots,n$. For $\theta<\theta_c^{(1)}$ 
the density of eigenvalues $\lambda_i(x)$ is given by the semi-circle law with an upper
edge at $2 \sqrt{x N}$. At the first transition the largest eigenvalue 
detaches from the semi-circle, at the second transition the second largest eigenvalue detaches and so on.
Thus one has, to leading order in $N$ (up to subleading fluctuations)
\be \label{cases1} 
\lambda_i(N) \simeq
\begin{cases}
 2 \sqrt{x N} \quad &, \quad \theta < 1/a_i^2 \\
a_i x + \frac{N}{a_i} \quad &, \quad \theta > 1/a_i^2 
\end{cases}
\ee 
These transitions are "decoupled" from each others
as long as the column strengths verify $\vert a_i-a_{i+1} \vert \gg N^{-1/3}$. 

Let us consider now a fixed number of lines $M$, which remains
finite as $N \to +\infty$. From \eqref{totalenergy} and \eqref{cases1} the ground state energy (to leading
order in $N$) reads

\be
- {\cal E}^0_{N,M}(x) \simeq N \Big( 2 M \sqrt{\theta} + \sum_{i=1}^{\min(M,n)} a_i \,(\sqrt{\theta}- \frac{1}{a_i})^2_+\Big)
\ee
where we recall that $(x)_+ = \max(x,0)$. 
The second derivative of the free energy as a function of $\theta$ will thus have the form of a staircase, 
with a jump at each transition point $\theta=\theta_c(a_i)$. The total occupation
length of each active column $i=1,\dots,n$ in the ground state is obtained from
differentiation according to \eqref{dercol} to leading order in $N$ as, for 
$i \leq \min(M,n)$

\be \label{Li} 
L^0_i = \sum_{k=1}^M \ell^k_i \simeq (x - \frac{N}{a_i^2})_+ = N (\theta- \frac{1}{a_i^2})_+
\ee
and $L^0_i=0$ for $i >  \min(M,n)$. This 
coincides with the result for the single line, single column problem
\eqref{29}.
Hence the total occupation length of a given column is completely independent of
the values of the other column strength $a_i$'s (as long as all strengths
are distinct). One can also define the effective number of active columns for a given $\theta$, i.e. the
number of columns where localization occurs (that is those which have a macroscopic occupation, 
$L_i^0/N>0$ at large $N$),
as $n(\theta)= \sum_{i=1}^{\min(n,M)}
\Theta(\theta - \frac{1}{a_i^2})$ which is always smaller or equal to $M$. 
Note that the sum of the occupation lengths of the macroscopically occupied
columns is smaller than the total length
\be 
\sum_{i=1}^{\min(n,M)} L_i^0 \simeq 
\sum_{i=1}^{\min(n,M)} N (\theta- \frac{1}{a_i^2})_+ < M x = \sum_{i=1}^{N} L_i^0 
\ee 
This is because each of the other columns 
have occupation lengths $L^0_j = \mathcal{O}(1)$, which sum up to the remainder. 
In the case of a single line $M=1$, the PDF of the $\mathcal{O}(1)$ occupation
length was obtained in Appendix \ref{app:sub:delococc} in the simpler case where all $a_j=0$.

If the positions of the $n$ active columns are permuted among 
$N$, the ground state energy is unchanged but the actual optimal configuration
of the polymers, i.e. the set of $\ell_i^{k,0}$, is different.
The constraint however is that the total occupation lengths for each column
if given by the above formula \eqref{Li} (for $a_i$ any of the active columns).

Finally, if instead the active columns are all close in energies, within $a_1-a_n \sim N^{-1/3}$, there is a single localization transition,
described by the $M$ largest eigenvalues of the rank $n$ deformed GUE, that is, upon rescaling, by the $M$ largest points of the determinantal point process
described by the kernel BBP$_n$ recalled in Appendix \ref{app:BBP}.

\subsection{The limit $M,N \to \infty$ with a fixed "density" $r=\frac{M}{N}$}

In the limit $M,N \to \infty$ with a fixed "line density" $r=\frac{M}{N}$ 
one can calculate the leading term in the ground state
energy. We consider the case of a continuous density of column strengths
$\rho(a)$. First one obtains the density of eigenvalues of ${\cal M}(x)$ to
leading order, from free probability. Let us define the scaled matrix, from \eqref{M}
\bea \label{tildeM} 
&& \tilde M(x) = \frac{{\cal M}(x)}{x} = \frac{{\cal M}(x)}{\theta N} = {\rm diag}(a_1,\dots a_j)  + \sqrt{\tau} \, \tilde V \, , \nn \\
&& ~~~~~~~~~~~~~~~~~~\tau=\frac{1}{\theta} \quad , \quad  \theta=\frac{x}{N} 
\eea 
where $\tilde V=V/\sqrt{N}$ is a GUE matrix with the semi-circle density $\nu_{sc}(y) = \frac{1}{2 \pi}  \sqrt{4-y^2}$, and we recall that we study $N,x \to +\infty$ with $\theta$ fixed.
We introduce the variable $\tau=1/\theta$ which is more convenient. 
The eigenvalues of ${\cal M}(x)$ are thus $\lambda_i(x)= N \theta \tilde \lambda_i(x)$, where $\tilde \lambda_i(x)$ are the
eigenvalues of $\tilde M(x)$. Their density in the large $N$ limit 
$\nu_\tau(\tilde \lambda) = \lim_{N \to + \infty} \frac{1}{N} \sum_{i=1}^N 
\delta(\tilde \lambda - \tilde \lambda_i(x))$, is determined by the free additive convolution
 (see \cite{Biane,Claeys_2018} and references therein)
\be \label{free}
\nu_\tau = \rho \boxplus \nu_{sc,\tau} 
\ee 
where $\nu_{sc,\tau}(y) = \frac{1}{2 \pi \tau}  \sqrt{4 \tau -y^2}$. Let us define $G_\tau(z)$ as
the Stieljes transform of the density $\nu_\tau(\tilde \lambda)$ 
\be
G_\tau(z)= \lim_{N \to + \infty} \frac{1}{N} \sum_{i=1}^N \frac{1}{z-\tilde \lambda_i(x)} 
= \int_{-\infty}^{+\infty} \rmd \tilde \lambda \frac{\nu_\tau(\tilde \lambda)}{z- \tilde \lambda} 
\ee 
from which the density can be extracted as 
\be \label{dos}
\nu_\tau(\tilde \lambda) = - \frac{1}{\pi} {\rm Im} \,  G(\tilde \lambda + \I 0^+) 
=
\frac{1}{\pi} {\rm Im} \,  G(\tilde \lambda - \I 0^+) 
\ee 
Then, \eqref{tildeM}, \eqref{free} imply that $G_\tau(z)$ satisfies the Burgers equation
\be \label{burg0} 
\partial_\tau G_\tau(z) = - G_\tau(z) \partial_z G_\tau(z)
\ee 
see Appendix \ref{app:dbm}, 
of solution 

\cite{Pastur1972,BookGuionnetRMT,Biane,Claeys_2018,JPBook,Menon}, 
\cite[Theorem 5]{DeformedGUECapitaine},
which obeys for $z \in \mathbb{C}^+$ 

\be \label{resolv} 
G_\tau(z) = \int_\Omega \frac{\rmd a \rho(a)}{z - \tau G_\tau(z) - a} 
\ee 
For $\tau \to 0$, equivalently for $\theta \to +\infty$, one can neglect the term $\tau G_\tau(z)$
inside the denominator of \eqref{resolv} and the corresponding initial condition of the Burgers equation is $G_{\tau=0}(z)=\int_\Omega \rmd a  \frac{\rho(a)}{z- a}$, i.e. 
$\nu_{\tau=0}= \rho$, and the spectrum of $\tilde M(x)$ is given by the $\{a_i\}$, the strengths of the columnar defects, i.e.  as clear from \eqref{tildeM}.
In the opposite limit $\tau \to +\infty$, equivalently $\theta \to 0$ (a limit equivalent to
choosing $\rho(a) \to \delta(a)$) the equation \eqref{resolv} becomes 
$G_\tau(z)(z - \tau G_\tau(z))=1$ and the solution takes the form 
\be
G_\tau(z) \simeq \frac{1}{\sqrt{\tau}} G_{sc}(\frac{z}{\sqrt{\tau}}) 
\ee 
in terms of the resolvant associated to the unit semi-circle density
\cite{footnote3} 
\be
G_{sc}(z)=\frac{1}{2} \left(z - z \sqrt{1- \frac{4}{z^2}}\right) 
\ee  
Note that \eqref{resolv} can be written as $G_\tau(z)=G_0(w)$ with,
equivalently, $z=w + \tau G_0(w)$ or, $w=z-\tau G_\tau(z)$.

As discussed in Appendix \ref{app:dbm}, the scaled eigenvalues $\tilde \lambda_i$ perform a Dyson Brownian motion as a function of the parameter $\tau=1/\theta$. This is valid for any $N$, and
leads to the above equations at large $N$. The initial condition is $\tilde \lambda_i = a_i$ at $\tau=0$, and at large $\tau$ the density converges to the semi-circle shape.

One can ask how the density $\nu_\tau$ evolves between these two limits. 
We will assume again that the columnar defect energies have a density $\rho(a)$
with a soft right edge $a_e$ where it vanishes as in \eqref{dens0},
i.e. $\rho(a)= A (a_e-a)^{2k+\frac{1}{2}}$, with $\rho(a)=0$ for $a>a_e$.
In \cite{Claeys_2018} it is proved that $\nu_\tau(\tilde \lambda)$ vanishes at its edge
with the same exponent as long as
\be \label{int4} 
\tau< \tau_c \quad , \quad \frac{1}{\tau_c}=\theta_c = \int_\Omega \rmd a \frac{\rho(a)}{(a_e-a)^2} 
\ee 
This corresponds to the localized phase for $\theta>\theta_c$, that we have studied
by other methods in Section \ref{subsec:gse} with exactly the same value for $\theta_c$ in \eqref{thetac} 
(there we focused on the {\it largest} eigenvalue $\tilde \lambda_1$). More precisely
in \cite[Theorem 1.3]{Claeys_2018} it is shown that, denoting $\tilde \lambda_e=
\tilde \lambda_e(\tau)$ the upper
edge of $\nu_\tau$, one has
\bea \label{resnew} 
&& \nu_\tau(\tilde \lambda) \simeq A \left(\frac{\tau_c}{\tau_c-\tau}\right)^{2 k+3/2}  (\tilde \lambda_e(\tau) - \tilde \lambda)^{2 k +1/2} \\
&& \tilde \lambda_e(\tau) = a_e + \tau \int_\Omega \rmd a \frac{\rho(a)}{a_e-a} \quad , \quad
\tau=\frac{1}{\theta} 
\eea  
The position of the edge is exactly what was found in Section \ref{subsec:gse} 
from the study of the largest eigenvalue $\lambda_1$, i.e. for the single line 
problem $M=1$, with the
correspondence $\tilde \lambda_e(\tau) = \mu(\theta)/\theta \simeq \lambda_1/(N \theta)$, 
where $\mu(\theta)$ is
given in \eqref{mutheta} in the localized phase and recalling that $\tau=1/\theta$. The results \eqref{resnew} are
established in \cite{Claeys_2018} for integer $k \geq 1$, but in view of our results in Section \ref{subsec:gse} 
it is reasonable to conjecture them to be valid for any real $k >1/4$, so that the integral in
\eqref{int4} converges and $\tau_c>0$ (i.e. so that there is a localized phase
for a single line).

At criticality $\tau=\tau_c$, Ref.~\cite[Theorem 1.4. (d)]{Claeys_2018}
states that the semi-circle shape holds near the edge
\be \label{crit12} 
\nu_\tau(\tilde \lambda) \simeq \frac{1}{\pi \tau_c^{3/2} \sqrt{g}} |\tilde \lambda_e(\tau_c)-\tilde \lambda|^{1/2} 
\ee 
where 
\be
g = \int_\Omega \rmd a \frac{\rho(a)}{(a_e-a)^3} 
\ee 
Note that this parameter $g$ is identical to the one defined in \eqref{trois}, 
i.e. $g=\frac{1}{2} \varphi^{(3)}(a_e)$ in the study of the localization transition for
a single line $M=1$. There is was assumed to be finite, equivalent to
the convergence of the
integral which holds for $k>3/4$, and implied a critical regime for the fluctuations of $\lambda_1$ (equal to minus the ground state energy) 
interpolating between the Tracy Widom distribution, and a new distribution. Again we can conjecture that
the above results at criticality, shown in \cite{Claeys_2018} for integer $k \geq 1$, 
extend to any real $k>3/4$.

In conclusion, we see that the localization transition obtained for a single
line $M=1$ shows up in the many line problem $M>1$ as a transition in the
behavior of the density of eigenvalues $\tilde \lambda_i$ near its upper edge:
it vanishes with the same exponent $2k+1/2$ as the column strengths
in the (single-line localized) phase $\theta>\theta_c$, and with the semi-circle exponent $1/2$ in the 
(single-line delocalized) phase $\theta<\theta_c$. 

We can now apply these results to make predictions about the
ground state energy for a fixed density $r=M/N$ of lines versus
columns. One introduces the "Fermi" level $\tilde \lambda_f$ as the solution of the equation
\be \label{fermi} 
\int_{\tilde \lambda_f}^{\tilde \lambda_e} \rmd \tilde \lambda \, \nu_\tau(\tilde \lambda) = r = \frac{M}{N} 
\ee 
where we recall that $\tilde \lambda_e$ is the upper edge of the eigenvalue density $\nu_\tau(\tilde \lambda)$ of 
the matrix $\tilde M(x)={\cal M}(x)/x$.
The ground state energy of the system of $M$ lines is then given for large $N$, with 
$r=M/N$ fixed, as
\be \label{gsM} 
- {\cal E}^0_{N,M}(x) \simeq \theta N^2 
\int_{\tilde \lambda_f}^{\tilde \lambda_e} \rmd \tilde \lambda \, \tilde \lambda \, \nu_\tau(\tilde \lambda) 
\ee 
where we recall that $\nu_\tau(\tilde \lambda)$ is determined from solving
\eqref{resolv} and \eqref{dos} and $\tilde \lambda_f$ from \eqref{fermi}. Using the above results one finds that:

(i) \textbf{in the (single-line localized) phase} $\tau<\tau_c$, i.e. $\theta>\theta_c$ one has, to leading order for
small $r=\mathcal{O}(1)$, from \eqref{resnew}
\be
\tilde \lambda_e - \tilde \lambda_f \simeq a_k \left(1-\frac{\theta_c}{\theta}\right)  \times r^\frac{1}{2 k+\frac{3}{2}} 
\ee 
with $a_k=(\frac{2 k + \frac{3}{2}}{A})^\frac{1}{2 k+\frac{3}{2}}$. This leads to the small
$r=\mathcal{O}(1)$ expansion of the ground state energy
\be \label{resE1} 
- {\cal E}^0_{N,M}(x) \simeq \theta N^2  \left( \lambda_e r - 
b_k (1-\frac{\theta_c}{\theta})  \times r^\frac{2 k + \frac{5}{2}}{2 k+\frac{3}{2}} \right)
\ee 
with $b_k = \frac{1}{2 k + \frac{5}{2}} (2 k + \frac{3}{2})^\frac{2 k+\frac{5}{2}}{2 k+\frac{3}{2}}
A^{-\frac{1}{2 k + \frac{3}{2}}}$. The first term linear in $r$ corresponds to independent
lines, and the second, singular term arises from the non intersection constraint (i.e. the
interactions between the lines).

(ii) \textbf{at criticality} $\tau=\tau_c=\frac{1}{\theta_c}$ one finds, from \eqref{crit12}
\be
\tilde \lambda_e - \tilde \lambda_f \simeq (\frac{3}{2} \pi \sqrt{g})^{\frac{2}{3}} \frac{1}{\theta_c} \times r^{2/3} 
\ee 
leading to to the small
$r=\mathcal{O}(1)$ expansion of the ground state energy at the critical point
\be \label{resE2} 
- {\cal E}^0_{N,M}(x) \simeq \theta N^2  \left( \lambda_e r - 
\frac{2}{5} (\frac{3}{2})^{\frac{5}{3}} (\pi \sqrt{g})^{\frac{2}{3}} \frac{1}{\theta_c} \times r^{5/3} \right)
\ee
We have not attempted to study the crossover near criticality between
\eqref{resE1} and \eqref{resE2}. 

(iii) \textbf{inside the (single-line delocalized) phase} $\tau>\tau_c$, i.e. $\theta<\theta_c$
the density of states vanishes as a semi-circle and again one has $\lambda_e-\lambda_f \sim r^{2/3}$ 
and ${\cal E}^0_{N,M} + \theta N^2 \lambda_e r \propto N^2 r^{ 5/3}$. In the 
limit $\tau \gg 1$, i.e. $\theta \ll 1$, the density converges to the
semicircle, $\nu_\tau(y)\simeq \nu_{sc,\tau}(y) = \frac{1}{2 \pi \tau}  \sqrt{4 \tau -y^2}$,
which leads to $\tilde \lambda_e \simeq 2 \sqrt{\tau}$ and 
\bea
&& \tilde \lambda_e - \tilde \lambda_f \simeq (\frac{3 \pi}{2})^{\frac{2}{3}} \frac{1}{\sqrt{\theta}} r^{2/3} \\
&& - {\cal E}^0_{N,M}(x) \simeq N^2  \left( 2 \sqrt{\theta} r - 
\frac{2}{5 \pi} (\frac{3 \pi}{2})^{\frac{5}{3}} \theta^{3/2} \times r^{5/3} \right) \nonumber 
\eea 

Note that from the equations \eqref{resolv}, \eqref{dos}, \eqref{fermi} and \eqref{gsM} one could in principle access, using \eqref{occT0}, to the total occupation lengths $L_i$
in the ground state. We have not attempted that calculation.

Another case of great interest is when there are two families of columnar defects
such that the support of $\rho(a)$ consists of two intervals, separated by a gap.
It is known in the context of the deformed GUE random matrix ${\cal M}(x)$, that there is usually 
a critical value $\tau_c=1/\theta_c$ at which the gap of $\nu_\tau(\tilde \lambda)$ closes, and such that the two half-supports merge for $\tau>\tau_c$, i.e. $\theta<\theta_c$. 
The behavior around that point is quite non-trivial, for a recent review see \cite{claeys2019critical}.
In Ref. \cite{Claeys_2018} and \cite{Capitaine_2015} there are results concerning the case where the support of $\rho(a)$ has an interior singular point $a^*$ where the density vanishes. If $\rho(a)$ vanishes as $|a-a^*|^\kappa$ with 
$\kappa>1$, this singular point survives for $\tau<\tau_c$, while if $\kappa<1$ it immediately
disappears. These critical phenomena can be explored in the present problem, by varying the filling $r=M/N$ and the tilt angle of the lines (we recall that at large $N$, 
$\theta \simeq x/N$, and $\theta_M=x/(N-M) \simeq \theta/(1-r)$). 
One needs to vary $r$ near the critical filling 
where one of the half-support is fully occupied and the other empty. 
It would be of great interest to see whether it gives some description of the
the Mott insulator phase predicted in \cite{nelson1993boson},
in presence of point disorder and upon tilting the lines near the critical transverse field.
Note that the stability of a band insulator for the two-way model 
in presence of columnar disorder was studied in \cite{hebert2011hatano}.

\subsection{The Rosenzweig-Porter model and fractal delocalization of interacting lines.}

\subsubsection{The generalized Rosenzweig-Porter model} 

The problem studied here is closely related to the so-called generalized Rosenzweig-Porter (GRP) model
recently studied in physics ~\cite{kravtsov2015random,facoetti2016non,de2019survival,kravtsov2020localization} and
mathematics \cite{benigni2017eigenvectors,von2019non,von2018phase,von2019delocalization}. 
The GRP model is a cousin of the Anderson models on the Bethe lattice and on the random regular graph, themselves studied \cite{KravtsovBethe2017} as simpler settings for investigating the many-body localization transition \cite{MBL}.
In particular the GRP allows to investigate the existence of a non-ergodic delocalized phase, or bad metal, predicted in this context \cite{KravtsovBethe2017,MBL}.
These phases are also of great interest for glassy quantum dynamics in models such as
the quantum random energy model \cite{faoro2019non}, with applications to 
quantum computing \cite{smelyanskiy2020nonergodic,kechedzhi2018efficient}.

The GRP model studied in ~\cite{kravtsov2015random,facoetti2016non,benigni2017eigenvectors,von2019non,von2018phase,von2019delocalization} is defined by the deformed GUE matrix
\be
H = {\rm diag}(a_1,\dots,a_N) + \frac{\sqrt{t}}{N^{\gamma/2}} V 
\ee 
where the GUE matrix $V$ has the same distribution as 
in \eqref{M}. The $a_i$ are i.i.d random variables drawn from a distribution $\rho(a)$ with
a compact support. The connection with \eqref{M} and \eqref{tildeM} is thus
\be
H = \tilde M(x) = \frac{{\cal M}(x)}{x} \quad , \quad x = \frac{N^\gamma}{t} \quad , \quad \tau= t N^{1-\gamma}
\ee 
The GRP model is studied for $N \to +\infty$ at fixed $t$ and $\rho(a)$, equivalently for $\tau = \mathcal{O}(N^{1-\gamma})$.
Note that some works consider instead $V$ as drawn from the GOE, but there are no important 
differences in the main features discussed below.

The case $\gamma=1$ is thus the same as studied until now in this paper, 
with the correspondence $t=\tau=1/\theta$, see also \cite{RefCiteesParBiroli1,RefCiteesParBiroli2,RefCiteesParBiroli3,RefCiteesParBiroli4}.
As we discussed above, at large $N$ the mean eigenvalue density of $H$,
$\nu_\tau$, interpolates from $\rho(a)$ at small $\tau$ to a
semi-circle at large $\tau$, as described by \eqref{dos} and the self-consistent equation
\eqref{resolv} or the Burgers equation \eqref{burg0}.
If $\rho(a)$ vanishes fast enough near its upper edge, it retains its
shape for $\tau<\tau_c$ and exhibits a transition to a semi-circle shape 
at $\tau=\tau_c$, i.e. $\theta=\theta_c$.

It is thus not surprising that for $\gamma<1$ at large $N$, which corresponds to $\tau \gg 1$, the spectrum of $H$ is a
semi-circle, while for $\gamma>1$, i.e. $\tau \ll 1$, is it exactly $\rho(a)$ \cite{facoetti2016non,von2019non,von2018phase,von2019delocalization}. However the transition
in the local spectral correlations of $H$ between the Wigner-Dyson statistics
and the Poisson statistics takes place at a different value, $\gamma=2$. If $\gamma>2$, i.e. 
On the other hand, if $\gamma<2$, i.e. $\tau \gg 1/N$, the local level statistics falls into the Wigner-Dyson class \cite{LandonSosoeYau2016,LandonYau2017,von2019delocalization}.

The most interesting case is $1 < \gamma < 2$, i.e. $1/N \ll \tau \ll 1$.
Although the mean density of $H$ is $\rho(a)$, the local level
statistics is Wigner-Dyson. It was conjectured in
\cite{kravtsov2015random} that the eigenvectors
are delocalized, but only in $N^{2-\gamma}$ sites close in energy, leading to a
"fractal dimension" for the eigenstates. The mass of each eigenfunction was predicted to spread to a large number of sites, which nevertheless form a vanishing fraction of the entire volume, the "sites" $\{1, \dots , N \}$.
This phase was called a non-ergodic delocalized phase. It was then proved in \cite{von2019non} 
(for GOE matrices, $T$ there being $\tau$ here) 
that each normalized eigenfunction $\psi_{\alpha}$ delocalizes across a set of approximately $N \tau \sim N^{2-\gamma} \gg 1$ sites for which $a_i$ is closest to $\lambda$.
More precisely these sets are such that $|\tilde \lambda_\alpha-a_i|$ is of order $N^{1-\gamma}$, hence they
contain $\sim \tau N \sim N^{2-\gamma}$ sites, on which $|\psi_{\alpha}(x)|^2 \leq N^{\gamma - 2} = (N \tau)^{-1}$. Hence they are maximally delocalized on these sites. In \cite{facoetti2016non} DBM and perturbative arguments were used to explain why the abrupt transition 
in the local statistics does not contradict the gradual transition 
in the degree of eigenfunction localization, 
by arguing that the statistics retain a Poissonian character 
on mesoscopic scales greater than $\tau$. Other results concerning the eigenvalue statistics can be found in \cite{HuangLandonBetaDBM,DuitsJohansson2017} and near the edge in \cite{LandonYauEdgeDBM,claeys2019critical}.
Note that as $\gamma \to 2$, and for $\gamma>2$, the eigenstates become localized on one site,
while as $\gamma \to 1$ they become fully delocalized over the $N$ sites. 

Some of these properties can be understood using the DBM
\cite{facoetti2016non,von2019non,von2018phase,von2019delocalization}. As recalled in Appendix \ref{app:dbm},
from \eqref{eq23} the eigenvalues $\tilde \lambda_i$ of $H$ expressed as functions of $\tau$ satisfy the standard $\beta=2$ DBM
\bea \label{dbm10} 
 \rmd \tilde \lambda_i &&= \frac{1}{N} \sum_{j \neq i} \frac{\rmd \tau}{ \tilde \lambda_i - \tilde \lambda_j} + \frac{1}{\sqrt{N}} \, \rmd \tilde b_i(\tau) \\
&& = \frac{1}{N^\gamma} \sum_{j \neq i} \frac{\rmd t}{ \tilde \lambda_i - \tilde \lambda_j} + \frac{1}{N^{\gamma/2}} \, \rmd b_i(t) 
\eea
with initial condition $\tilde \lambda_i(0)=a_i$ and $b_i$ and $\tilde b_i$ are i.i.d Brownian motions. The DBM expressed in the variable $t$
thus has parameters $a=N^{-\gamma}$ and $b=N^{-\gamma/2}$.

The resolvant expressed as a function of $t$, $\hat G_t(z)=\frac{1}{N} \sum_{i=1}^N \frac{1}{z-\tilde \lambda_i}$ obeys
\be \label{burg3} 
\partial_t \hat G_t(z) = - \frac{1}{N^{\gamma-1}} \hat G_t(z) \partial_{z} \hat G_t(z)
- \frac{1}{N^{\frac{1+\gamma}{2}}} \partial_{z} \eta_{z,t} 
\ee 
with ${\rm Cov}(\eta_{z,t} \eta_{z,t} ) = - \delta(t-t') \frac{\hat G_t(z)- \hat G_t(z')}{\hat z- \hat z'}$. 
Note the misprint in the amplitude of the noise in Eq. (18) in \cite{facoetti2016non}. The resolvant 
$G_\tau(z)=\hat G_t(z)$ as
a function of $\tau$ also obeys \eqref{burg3}, but with $\gamma=1$. 

From the DBM equation \eqref{dbm10} it is clear that at very short times $\tau$ the
eigenvalues $\tilde \lambda_i$ remain close to their starting points $a_i$. As long
as they have not moved by more than the typical interparticle distance $\sim 1/N$ 
(since $\rho(a)$ has a compact support of order unity) they simply perform independent
diffusion $\tilde \lambda_i - a_i \simeq (\tau/N)^{1/2}$. Equating both scales one
see that it corresponds to $\tau = \mathcal{O}(1/N)$, i.e. $\gamma=2$, and it is thus natural to expect local
Poisson statistics below that time and Wigner Dyson above, when the neighboring particles
start interacting to avoid collisions. 
Then it takes a much longer time $\tau=\mathcal{O}(1)$ to reach a steady state 
at the global scale, and for the density to change from $\rho(a)$ to the semi-circle, which corresponds to the transition $\gamma=1$. More detailed arguments using eigenvectors are required
to understand the nature of these regimes \cite{facoetti2016non,benigni2017eigenvectors,von2019non,von2018phase,von2019delocalization}.

Note that the above analysis deals with eigenstates in the bulk. Very near the upper edge 
of the spectrum of $H$ it may be a bit different. Indeed, one cannot find $N^{2-\gamma}$ columns
too close to the edge, more precisely when $a>a_e-\delta a$ where $N^{2-\gamma} \simeq N \int_{a_e-\delta a}^{a_e} \rho(a) da
\simeq N (\delta a)^{2 k+ 3/2}$. Hence the eigenstates near the edge should localize on fewer columns.

{\bf Remark}. For $\gamma=1$ we studied in section 
\ref{sec:critreg} the transition near the edge (for $M=1$, i.e. for the largest eigenvalue), 
at $\theta=\theta_c$.
From \eqref{critreg} the critical region is defined by the scaling variable $u = N^{1/3} (\theta-\theta_c)=\mathcal{O}(1)$ (denoted by the letter $u$ to avoid confusion). We note that this critical region can also
be explored in the GRP model, at fixed $t=1/\theta_c$, if one chooses, as $N$ becomes large,
\be
\gamma = 1 + \frac{u}{\theta_c} \frac{1}{N^{1/3} \log N}
\ee
which is equivalent to 
to $\theta=\frac{N^{\gamma-1}}{t_c} \simeq \theta_c + u N^{-1/3}$. 

\begin{figure}
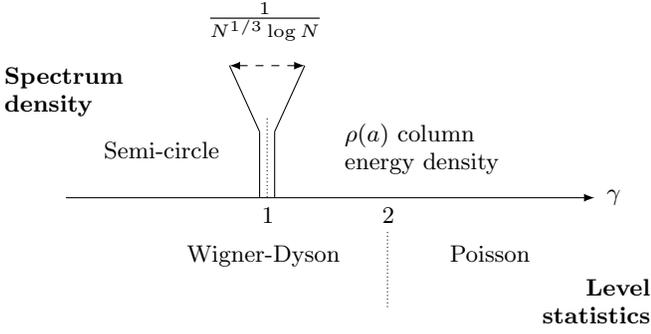

\lattice
\caption{Sketch of the different phases of the generalized Rosensweig-Porter (GRP) model in terms of the spectral density and the level statistics in the bulk. There are two cross-overs (i) at $\gamma=1$ from a delocalized phase to a delocalized non-ergodic phase where the order parameter is the spectrum density and (ii) at $\gamma=2$ from the delocalized non-ergodic phase to a standard localized phase where the order parameter is the local level statistics. The critical region associated to the 
phase transition studied in section \ref{sec:critreg},
which occurs at the edge of the spectrum near $\gamma=1+\mathcal{O}(1/N^{1/3} \log N)$ is also
indicated schematically.
}
\label{fig:phasePR}
\end{figure}

\subsubsection{From the eigenvectors to the polymers} 

Let us go back to the polymer picture, using the relation \eqref{totalenergy} 
between the ground state energies of $M$ polymers/lines and the
$M$ largest eigenvalues of $H$. The GRP model corresponds to
polymers with $x = N^\gamma/t = \mathcal{O}(N^\gamma)$, equivalently an "angle"
variable $\tan \phi = \frac{1}{\theta}=\frac{N}{x}=\tau=t N^{1-\gamma}$
at fixed $t$ and $N$ large. Thus $\gamma>1$ corresponds to very long polymers,
i.e. very small angle $\phi$, which tend to be more localized along the columns,
while $\gamma<1$ corresponds to large angles which lead to delocalization.

One should distinguish between the localization of the eigenstates
$\psi_\alpha$, $\alpha=1,\dots,N$, of $H$, which states how the normalized
measure $|\psi_\alpha(j)|^2$ tends to concentrate on a few sites (i.e. columns) $j$, and 
the localization of the $M$ polymers on the columns, measured by the total occupation
length $L^0_j=\sum_{k=1}^M \ell_j^{k,0}$ of each column $j$ in the ground state.
There are however connections between the two. Indeed a perturbation theory
argument as in \eqref{occ2} shows that each eigenvalues $\tilde \lambda_\alpha$ 
of $H$ (here ordered as a decreasing sequence) obey
\be \label{occ5}
\partial_{a_j} \tilde \lambda_\alpha = |\psi_\alpha(j)|^2
\ee 
for each realization of the matrix $V$ (and any given set of $\{a_i\}$). 
We can now combine this with \eqref{dercol} and \eqref{totalenergy}
and obtain the relation between the averages over respectively the
Brownian point disorder (indicated by $B$) and the GUE matrix $V$
\be \label{meanconnection2} 
\overline{L^0_j}^B =  \sum_{k=1}^M \overline{ \ell_j^{k,0}}^B = x \sum_{\alpha=1}^M \overline{  |\psi_\alpha(j)|^2 }^V
\ee 
which generalizes \eqref{meanconnection}. It is valid for any $j$
and for any given set of $\{a_i\}$. We recall that $\ell_j^k$ is the
occupation length along column $j$ of the polymer starting at column $i=k$ and ending at column $i'=N-M+k$. Equation
\eqref{meanconnection2} implies that the change in the mean occupation
length in the ground state, when adding one polymer, is given by
$\overline{L^0_j}^B|_M- \overline{L^0_j}^B|_{M-1}=x \overline{|\psi_M(j)|^2}^V$. It does not imply however
a relation between $\overline{\ell_j^{k,0}}^B$ and $\overline{|\psi_k(j)|^2}^V$ because
the joint distributions of the $\overline{\ell_j^{k,0}}^B$ also depend on the number of polymers $M$. 

For a single polymer $M=1$ 
we already know from Section \ref{sec:gse} 
that for $\gamma=1$ there is a localization transition of the polymer 
at $\theta=\theta_c$ (if $\rho(a)$ vanishes fast enough at $a_e$). The polymer visit all columns with the
occupation length measure $\ell(a) da$ given in \eqref{laloc}, which for $\theta>\theta_c$ develops a delta peak on the few best columns at the edge $a=a_e$. From \eqref{meanconnection2}, i.e.
\eqref{meanconnection}, it indicates that $\psi_1$, the eigenstate of $H$ with the largest
eigenvalue is localized on one or a few $\mathcal{O}(N^0)$ sites/columns for $\theta>\theta_c$, and
delocalized for $\theta<\theta_c$. As one increases $M$ the eigenstates of highest eigenvalues are successively "filled". Presumably for finite $M=\mathcal{O}(1)$ the picture is similar to $M=1$ with localized eigenstates 
near the edge.

When $M$ increases such that $M/N=r = \mathcal{O}(1)$, the above results
about the eigenvectors in the bulk of the GRP model at large $N$ apply. Let us give some
qualitative arguments. For $\gamma<1$, i.e. $\theta \ll 1$
the eigenvectors are fully delocalized so that typically $|\psi_\alpha(j)|^2 = \mathcal{O}(1/N)$. 
From \eqref{meanconnection2} one thus surmises that the $L_j$ are all of order $x M/N$
and the set of $M$ polymers are delocalized. For $\gamma>2$, i.e. $\theta \sim N^{\gamma-1} \gg 1$,
the eigenvectors $\psi_\alpha$ are typically localized on a single site/column,
say $j_\alpha$. From \eqref{meanconnection2} 
one expects that the $L_{j_\alpha} = \mathcal{O}(x)$ for $\alpha=1,\dots,M$, and that the set of $M$ polymers 
are well localized on the best columns. For $1 < \gamma < 2$, i.e. the non ergodic phase,
$1 \ll \theta \ll N$, adding an extra polymer leads to a reorganisation of the total occupation lengths
$L_j$ given by $\overline{|\psi_M(j)|^2}^V$, i.e. this excitation is
localized on $N^{2-\gamma}$ sites around $|\lambda_M - a_j| \sim N^{1-\gamma}$. Hence
it is reasonable to expect that the set of lines are again localized, but on a macroscopic set
of columns. The above heuristic arguments 
assume that the typical and average $|\psi_\alpha(j)|^2$ behave in
a similar way, and it would be interesting to make them more rigorous. 

In conclusion, when the number of polymers is $M = \mathcal{O}(N)$, we can learn
from the GRP model in the bulk. We see that because of the (non-crossing) interactions
the set of polymers tends to be more easily delocalized. It seems that the polymer localization transition
which, as we found, occurs for a single (or a few) polymers for $x \sim N$, i.e.
$\theta = \frac{x}{N} = \theta_c= \mathcal{O}(1)$,
now requires much longer polymers.
The nature of the transition changes significantly. For $\tan \phi = \mathcal{O}(1)$
the set of $M = \mathcal{O}(N)$ polymers are delocalized. For $\tan \phi \ll 1$ they start to 
localize, but the polymer localization occurs gradually. For 
for $N \ll x \sim N^\gamma  \ll N^2$, one can say that the set of $M = \mathcal{O}(N)$ polymers are
localized but in the weaker sense, related to the non-ergodic phase of the GRP,
and excitations are delocalized on subsets of $N^{1-\gamma}$ sites. 
One then expects to recover linear response when varying $H$ or $\tan \phi$ on scales 
$N^{1-\gamma}$. For $\gamma \geq 2$, i.e. $\tan \phi \leq 1/N$ they are fully localized.

\section{Some known results at finite temperature} 

We finish this paper by mentioning a few available results at finite temperature
for the single polymer OY model $M=1$. The free energy
is ${\cal F}_N(x,T)=- T \log \mathcal{Z}_N(x,T)$ where $\mathcal{Z}_N(x,T)$ is defined in \eqref{Z}. 
First let us note that the rescaling $x \to x T^2$ and $a_j \to a_j T$ in the partition function 
\eqref{Z} leads to the relation
\be \label{scaling} 
{\cal F}_N(x,T,\{a\}) = T {\cal F}_N(x/T^2,1,\{a T\})
- 2 (N-1) T \log T 
\ee

Exact results for the free energy at $T>0$ and its fluctuations were obtained 
in \cite{O_Connell_2012,imamura2016determinantal,Imamura_2017}.
Let us discuss now the results of \cite{borodin2014free,BorodinMacdo}
in presence of drifts $a_j$ which corresponds to columnar defects. 
They reproduce the ones discussed in Sections \ref{subsec:rmt} 
and \ref{sec:singlesingle} in the $T \to 0$ limit.

Consider a single active columnar defect $a_1>0$ and $a_{j \geq 1} =0$
and ask about the localization transition. We again consider the limit $x,N \to +\infty$ 
with $\theta=x/N$ fixed. Let us denote 
$t^*$ the unique positive solution of 
\be
\psi'(t^*) = \frac{x}{T^2 N}  = \frac{\theta}{T^2} 
\ee 
where $\psi(t) = \frac{d}{dt} \log \Gamma(t)$ is the di-gamma function, and denote
$c = (- \frac{1}{2} \psi''(t^*))^{1/3}$ a temperature dependent 
constant. Then, it was shown 
in \cite{borodin2014free} that in the region where 
\be \label{p1} 
a_1 T <  t^* 
\ee
the following asymptotics holds at large $N$ for the free energy
\bea
{\cal F}_N(x,T,a_1) &\simeq& - T N ( \frac{\theta}{T^2} t^* - \psi(t^*) + 2 \log T ) \nn
\\
&& - c \, T N^{1/3} \chi_2 
\eea 
The region delimited by \eqref{p1} thus corresponds to the delocalized
phase, where the free energy exhibits Tracy-Widom fluctuations. 
When one approaches the boundary of this phase,
more precisely in the critical region near the transition defined by
\cite{footnoteBoro}
\be \label{b1T} 
a_1 T = t^* + \frac{b_1}{c}  N^{-1/3} 
\ee
with a fixed value of $b_1$, 
the following asymptotics for the free energy holds \cite{borodin2014free}
\bea
 {\cal F}_N(x,T,a_1) &\simeq& - T N ( \frac{\theta}{T^2} t^* - \psi(t^*) + 2 \log T ) \nn
\\
&& - c T N^{1/3}  \rchi_{\rm BBP, b_1}  
\eea 
where $\rchi_{\rm BBP,b_1}$ is the BBP random variable, distributed according
to the BBP$(b_1)$ distribution (of CDF $F_1(x;b_1)$ 
in \cite[Definition 1.3]{Baik_2005}) which we recalled in \eqref{CDFBBP}
when we studied $T=0$.
Hence we see that the critical behavior of the delocalization/localization
transition at finite temperature is the same as for $T=0$. 

One recovers the results for $T \to 0$ as follows. Using the small $t$ behavior
$\psi(t) \simeq - \frac{1}{t}$, we obtain that as $T \to 0$ for a fixed $\theta$ one has 
\be
t^* \simeq T/\sqrt{\theta} 
\ee 
In the delocalized phase $\theta< \theta_c=1/a_1^2$,
 this leads to
\bea
&& {\cal F}_N(x,T,a_1) \simeq - 2 N \sqrt{\theta}  - N^{1/3} \sqrt{\theta} \rchi_2 
\eea 
recovering \eqref{resa1}, equivalently \eqref{TW1}.
The critical region is defined by taking the limit $T \to 0$ of \eqref{b1T},
using $c \simeq 1/t^* \simeq 1/(a_1 T)$ which leads to 
$a_1  = \frac{1}{\sqrt{\theta}} + a_1 b_1 N^{-1/3}$,
equivalent to the $T=0$ definitions in 
\eqref{crit1} and \eqref{crit4} (with $\delta=2 b_1=\tau a_1^2$). 
In that region
one obtains
\bea \label{bppp} 
&& {\cal F}_N(x,T,a_1) \simeq  - 2 N \sqrt{\theta}  - N^{1/3} \sqrt{\theta} 
\rchi_{\rm BBP, b_1} 
\eea 
which recovers the $T=0$ result \eqref{res00}.

Although there are no results for the localized phase, the methods in \cite{borodin2014free}
can be used to investigate this phase as well. In the case where there is a finite number $k$ of columns with strictly negative energies $\epsilon_j = - a_j$ within the
critical region, the above result generalises with the replacement \cite{borodin2014free} with the
rank $k$ BBP distribution, $\rchi_{\rm BBP,{\bf b}}$, see Appendix \ref{app:BBP}. 

At finite temperature for a single line there is another interesting regime: 
in a different scaling limit, $x \sim \sqrt{N}$ one obtaines the continuum directed polymer, 
i.e. the model in \eqref{en0},
equivalent to the KPZ equation, is (in the absence of drifts) 
as 
\be
{\cal Z}_N(\sqrt{\tau N} + \tilde x,T=1) \sim Z^{\rm KPZ}(\tau,\tilde x).
\ee

The case of many non-crossing lines at finite temperature can also be investigated. One interesting
question is whether there is phase transition as a function of temperature. 

\section{Conclusion} 

In this paper we have analyzed results originating from random matrix theory in the context
of the localization/delocalization of polymers/lines in $d=1+1$ dimension in 
presence of both columnar and point disorder.
The central solvable model that we revisited is the O'Connell Yor (OY) polymer for either one line
or $M>1$ lines, in presence of $N$ columns of arbitrary strengths $a_j$. 
At zero temperature this model has a direct connection to the deformed
GUE random matrices and to the Dyson Brownian motion. An immediate consequence,
that we have explored in details, is that for one line and one column there is
a localization/delocalization phase transition in the universality class of the BBP 
spiked random matrix transition. In the localized phase the occupation length
of the column by the line is macroscopic. We have developed a variational calculation 
to calculate the occupation length in the localized phase and across the transition, 
which led to new results for its fluctuations. We found that these fluctuations are of order
$N^{2/3}$ and described in the localized phase by the distribution $f_{\rm KPZ}$, which appears ubiquitously in the KPZ class. 

We have recalled and then extended classical methods using Fredholm determinants, which allowed us to solve the case of one line and many columns with a continuous distribution
of column strengths $\rho(a)$. We have found that if $\rho(a)$ vanishes sufficiently fast
near its upper edge, there is a localization/delocalization transition. We have shown that this transition
belongs to a new universality class (hence the same applies for spiked random matrices
with a full rank perturbation). We have shown that the fluctuations of the ground state energy 
in the localized phase are non trivial, at variance with the usual BBP case where they are Gaussian. We
have obtained the universal distribution for these
fluctuations and those in the critical region. It is expressed in terms of a Fredholm determinant 
involving a new one-parameter kernel. This kernel is reminiscent of the one 
appearing the elliptic Ginibre ensemble. 

In the case of many non-intersecting lines with specific (packed) boundary conditions, we
could use some known results about the OY model. The case of a few active columns
and a few lines was discussed and leads to superpositions of BBP type transitions. In the case of 
a thermodynamic number of lines $r=M/N >0$, we calculated the ground state energy. It exhibits a
change of behavior in the very dilute limit $r \to 0$, but no true phase transition at fixed 
angle $\theta=x/N$ and $r>0$. As we argued, based on known results on the 
generalized Rosenzweig-Porter (GRP) model, full polymer localization on the columns occurs for long polymers, $x \sim N^\gamma$
with $\gamma \geq 2$ 
and only partial localization occurs for $1 \leq \gamma <2$. The latter
case is related to the celebrated "non-ergodic" delocalized phase of 
the GRP model. Translated to the polymer side of the model, it suggests that
the polymers delocalize over subsets of $\sim N^{2-\gamma}$ columns,
retaining some glassy features. 

Many other interesting questions remain such as the crossover from finite-rank
to infinite-rank pertubation (in the RMT context), 
the fluctuations of the ground state energy as a function of the column
strengths $a_j$
(we have worked here for a fixed set of $a_j$). We have also
unveiled transitions with an anomalous critical behavior quite different
from the Tracy-Widom/Airy family, which we
were not able to analyze.  Although our discussion
focused on zero temperature we analyzed the existing results at $T>0$ in the case
of one line and one (or a few) columns. The conclusion there was that
temperature does not change the universality class of the $T=0$ localization/delocalization
transition. It would be interesting to generalize the present study with many columns, and possibly many lines, in the case of finite temperature, where the thermal effects are expected to be more subtle.

In a broader context, the results in this paper concern a one-way model of polymer/line (i.e. which can
jump only to the right) in $d=1+1$. One expects on heuristic grounds that this model
could serve as some kind of approximation of the more general two-way model (such as \eqref{H1}
in the introduction, or the continuum model \eqref{en0}) near its delocalization (i.e. tilting) transition (e.g. in presence of an applied field) since left jumps may be subdominant there. Such one way
models have been investigated in the context of non-Hermitian localization but not in presence of
point disorder. 
 In Appendix \ref{app:model} we have 
presented a first step in that direction, by studying the OY model in the 
fixed transverse field ensemble $H$ (rather than the fixed angle ensemble)
where the tilt angle can fluctuate. We have presented some
connections to the models of elastic lines discussed in the introduction
which exhibit a tilt angle transition at $H=H_c$ and the transverse Meissner effect.
Although more details remain to be understood, it appears, roughly,
that for a single line the localization/delocalization transition which occurs at $\tan \phi_c=1/\theta_c$
in the fixed angle ensemble, is naturally associated 
to a first order jump in the tilt response at $H=H_c$, and that
the localized phase discussed here for $\phi<\phi_c$, i.e. $\theta>\theta_c$,
can be seen as a coexistence region. In the many line case,
the fact that the polymer localization occurs gradually on scales $x \sim N^\gamma$ 
as $\gamma$ varies within $1 \leq \gamma \leq 2$ suggests that the Bose-glass transition at $H_c$ is (i) continuous
since for any fixed $\tan \phi=N/x$ the system is delocalized (ii) has non trivial
features for vanishingly small angle scales $\phi \sim N^{1-\gamma}$. It remains an outstanding question of whether some 
universal property of this transition in presence of point disorder (for one or many lines)
may be captured by the present model, for which
many analytical results have been obtained.

\acknowledgments

We thank G. Barraquand, G. Biroli, T. Gauti\'e, D. R. Nelson and N. Hatano for useful discussions.
 AK acknowledges support from ERC under Consolidator grant number 771536 (NEMO),
PLD from ANR grant ANR-17-CE30-0027-01 RaMaTraF, and NOC from ERC under
Advanced grant number 669306 (IntRanSt).
AK and PLD thank the University of Bristol and University College Dublin
mathematics departments for hospitality, where part of this
work was completed.

\appendix

\begin{widetext}

\section{Fixed $H$ ensemble and model of vortex lines}

\label{app:model} 

In this Section we discuss a possible realization of the polymer model of the text in the context of vortex lines. We focus on a single line. We first recall in an elementary way the tilting transition (transverse Meissner effect) for an elastic line, upon an applied field $H$. Then we apply the results obtained for the one-way model to investigate the tilting transition and the localization/delocalization transition in presence of columnar and point disorder.
\\

\subsection{Single columnar defect}

Let us consider the model \eqref{en0} for a single continuum vortex line 
(without point disorder $V=0$) and a
single columnar defect modeled by a local potential well $U(u)$ centered at $u=0$ of depth $-U_0$ as in Fig. \ref{fig:appA1} (Left). The line enters at $u(0)=0$ (which is fixed) and exits at $u(x)=u_f>0$. The tilt, i.e.
the mean angle with the $z$ axis, is $\phi$ with $t=\tan \phi=u_f/x$.
Since the line consists in a pinned segment of length $\ell$ and a depinned segment of length $x-\ell$ the ground state energy is approximately
\be \label{Ex01} 
{\cal E}^0 \simeq - U_0 x + \min_{0 \leq \ell \leq x} \big[ U_0 (x-\ell) + \frac{\gamma u_f^2}{2(x-\ell)} - H u_f \big] 
\ee 
Consider first the fixed angle $\phi$ ensemble discussed in this paper (setting $H=0$).
Minimizing \eqref{Ex01} over $\ell$, one finds that the depinned fraction of the line is $1- \frac{\ell^0}{x} = \sqrt{\gamma/(2 U_0)} \tan \phi$. This solution is valid for $\phi < \phi_c$ with $\tan \phi_c=\sqrt{2 U_0/\gamma}$. For $\phi \geq \phi_c$ the
line is fully depinned with $\ell^0=0$, and ${\cal E}^0=\gamma u_f^2/(2 x)$. The first solution is an analog of the 
localized phase as defined in the text, and the second of the delocalized phase (although for
quite different models, and there is no point disorder here). The energy at fixed angle is
\be \label{f00} 
\frac{{\cal E}^0}{x} = f_0(\tan \phi) \simeq \begin{cases} - U_0 + \sqrt{ 2 \gamma U_0} \tan \phi \quad , \quad \tan \phi < \sqrt{2 U_0/\gamma} \\
 \frac{\gamma}{2} (\tan \phi)^2 \hspace*{1.9cm} ,  \quad \tan \phi > \sqrt{2 U_0/\gamma}  
 \end{cases}
\ee 
In the fixed $H$ ensemble the energy is given by minimizing 
$\min_{\phi \geq 0} [ f_0(\tan \phi) - H \tan \phi]$
over $\tan \phi$, i.e. by the Legendre transform of $f_0$. It gives here 
\be \tan \phi = 
   \begin{cases}0 \quad \hspace*{2.9cm}, \quad H < H_c = \sqrt{2 U_0/\gamma} \\
 \tan \phi_c +  \frac{H-H_c}{\gamma} = \frac{H}{\gamma} \quad , \quad H > H_c =  \sqrt{2 U_0/\gamma} \end{cases} 
\ee

\begin{figure}[t!]
\begin{center}
  \includegraphics[scale=0.5]{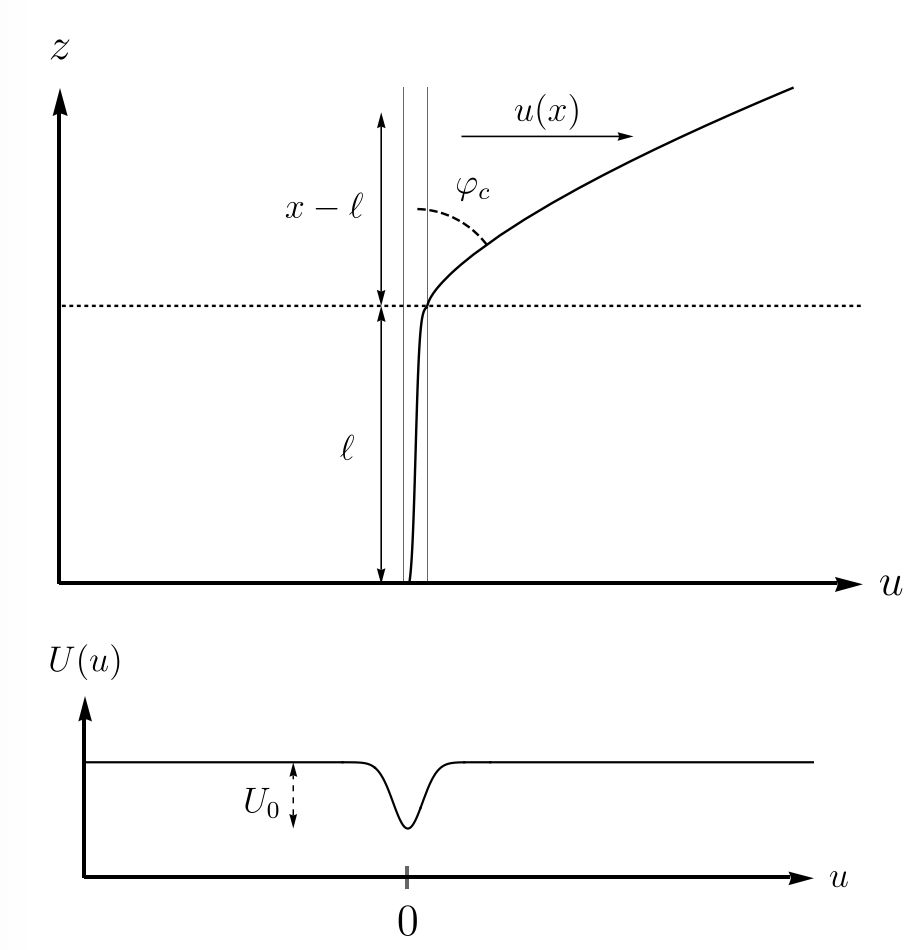} 
  \includegraphics[scale=0.6]{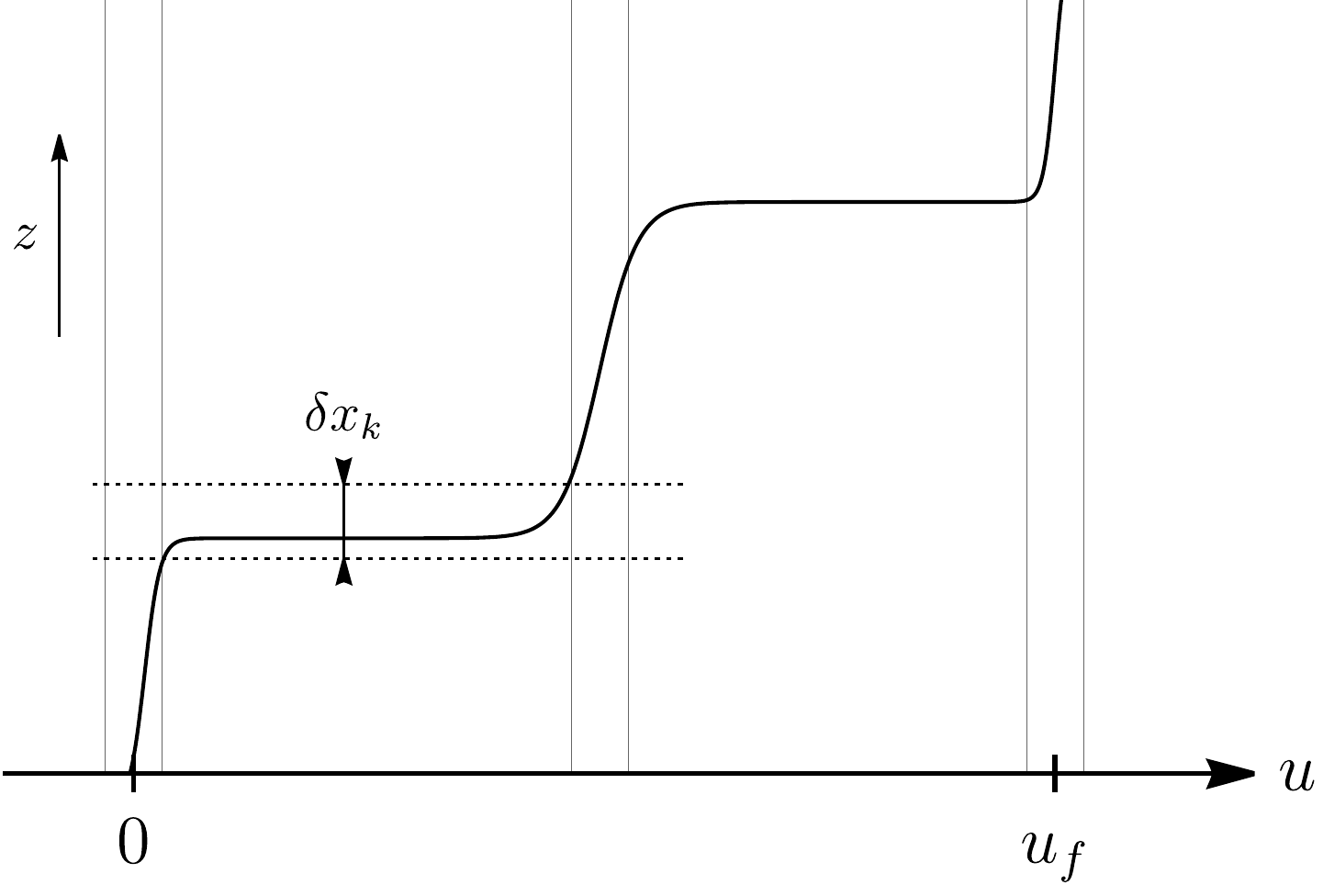} 
 \caption{\textbf{Left:} Elastic line of length $x$, ending in $u(x)=u_f$, in presence of a columnar defect
 studied in \eqref{Ex01}. At $T=0$, for a fixed tilt angle $\phi$ with the $z$ axis, the line is partially depinned when
 $\tan \phi=u_f/x < \tan \phi_c$. This is the localized phase (analog to the one discussed in the text for the OY model) where the pinned and depinned segments coexist. The tilt angle of the depinned segment is
 also $\phi_c$. \textbf{Right:} Cartoon of a series of kinks of width $\delta x_k$ and energy $E_k$, for an elastic line with a positive tilt angle in the periodic or disordered columnar model which mimics the OY model, as
 studied in \eqref{Esx}.}\label{fig:appA1}
\end{center}
\end{figure}

Hence there is a tilting transition at $H_c$ (and transverse Meissner effect for $H<H_c$),
a generic feature in presence of columnar disorder. Since $f_0'(t)$ is generically increasing, it happens when $f'_0(0)>0$, so that there is no solution to the minimization equation $H=f'_0(t)$ for $H<H_c=f'_0(0)$.
However there is a second feature in the model \eqref{Ex01}, i.e. the 
tilt angle $\phi$ jumps from $0$ to a finite value
$\phi_c$ at $H_c$. This first order jump of the tilt angle was also obtained
in the finite temperature $T>0$ version of the model \eqref{en0}, using the mapping to non-Hermitian quantum
mechanics ($\hbar$ playing the role of $T$) \cite{hatano1997vortex,ChenBalents1996}.
In that language the localized state (around the column) disappears at $H \geq H_c$
and is replaced by a pair of eigenstates (with complex conjugated energies) with
finite imaginary current (analog to the tilt). This was shown for some solvable cases
like the delta potential.
In the above $T=0$ estimate \eqref{f00}, $f_0(t)$ is linear on an interval, leading to a jump in
the Legendre transform. This calculation however can be made in a more exact manner (see
subsection \ref{subsec:exact}) and one finds that if $U(u)$ is smooth near its minimum $f_0(t)$ is not exactly linear and
the jump is rounded into a very steep
smooth curve with singularity (see e.g. \eqref{inverselog} below)
\begin{equation} \label{log2} 
\tan \phi \sim \frac{\Theta(H-H_c)}{|\log(H-H_c)|},
\end{equation}
 i.e. $\phi_c$ vanishes. The (partially localized) optimal configuration at small fixed angle will still look like in Fig. \ref{fig:appA1} (Left), the line being simply slightly shifted w.r.t. to $u=0$. These two cases (jump or no jump)
are illustrated in Fig. \ref{fig:appA2}.

Note that in the OY model studied in the text for a single active column, there is also a localized phase for $\theta>\theta_c$, i.e. $\phi<\phi_c$ (with some different, unrelated value for $\phi_c$) where a finite fraction of the column is occupied. While in \eqref{Ex01} it is simply the (trivial) elastic energy of a free depinned segment which is in competition with the columnar energy, in the OY model the competition is with the point disorder energy. 
Although the competition has a different origin we note that in both cases the tilt angle of the segment which is depinned from the column is precisely $\phi_c$. 
\begin{itemize}
\item For \eqref{Ex01} it follows since 
at the optimum: $\frac{u_f}{x-\ell^0}=\sqrt{2 U_0/\gamma}=\tan \phi_c$.
\item For the OY model
it can be seen in \eqref{29}:
$\frac{N}{x-\ell^0}=1/\theta_c=\tan \phi_c$.
\end{itemize}

Some of the above considerations extend to the many column cases. Jumps in the tilt angle
were also observed for model \eqref{H1} (with $\eta_j(t)=0$). As discussed in Section 
\ref{subsec:occup}, in the fixed tilt angle ensemble the signature of the localized
phase in this case is that the occupation length density $\ell(a)$ acquires a delta function peak
on the best columns. Below we will see an analog phenomenon for the model 
\eqref{en0} at $T=0$ in the absence of point disorder.

The OY model does not contain "free space" elastic energy. To connect
to the elastic line model \eqref{en0} one must introduce there a columnar
potential $U(u)$ with a lattice structure. Elastic deformations occur
as kinks, and one can then define an extension of the OY model in the 
fixed $H$ ensemble (with fluctuating $N$) by adding the 
 additional energy term $(E_k- H) r_0 N$. Let us describe it
in more details.

\begin{figure}[t!]
\begin{center}
  \includegraphics[scale=0.5]{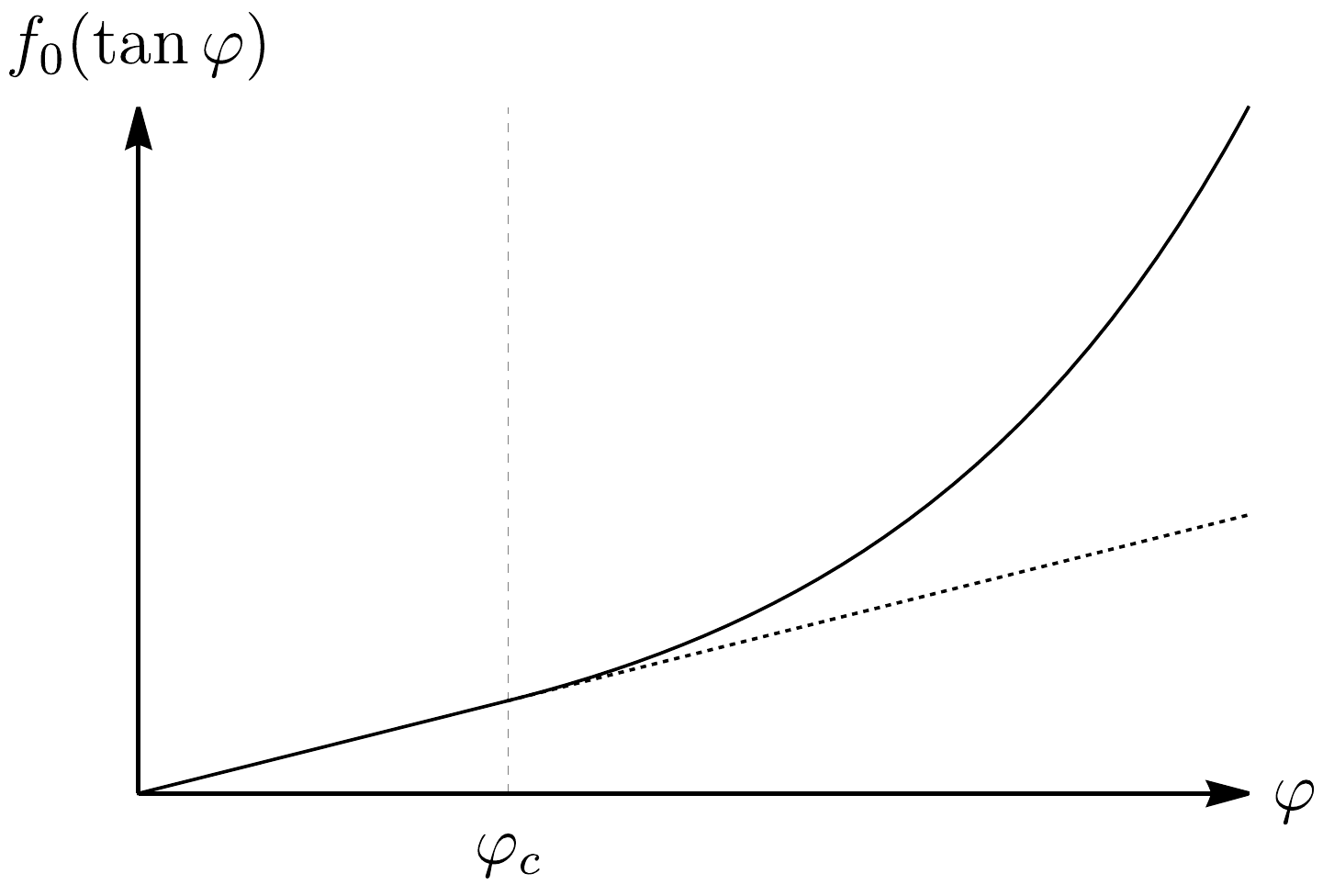} \hspace*{0.2cm}
  \includegraphics[scale=0.55]{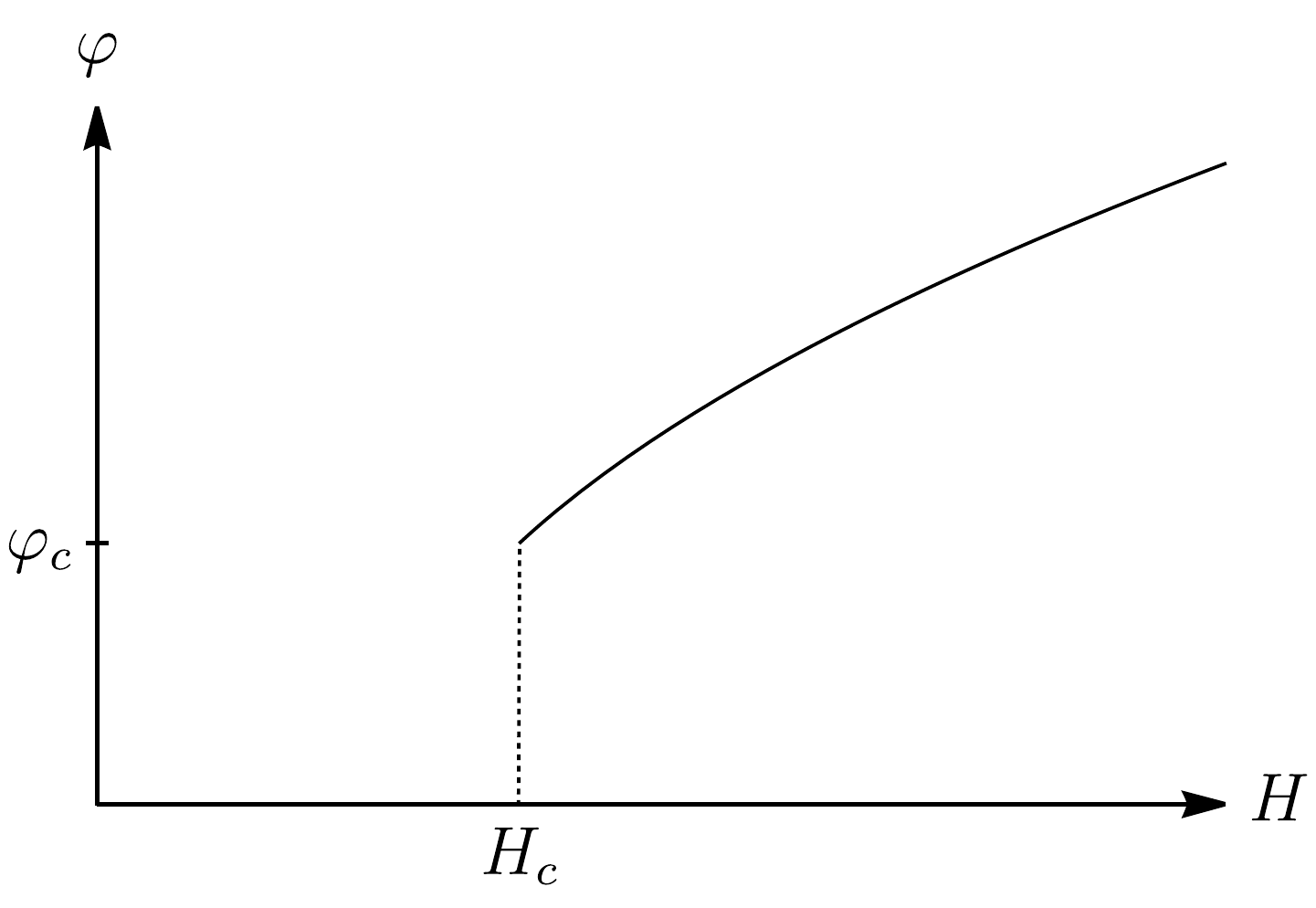} \\
  \includegraphics[scale=0.5]{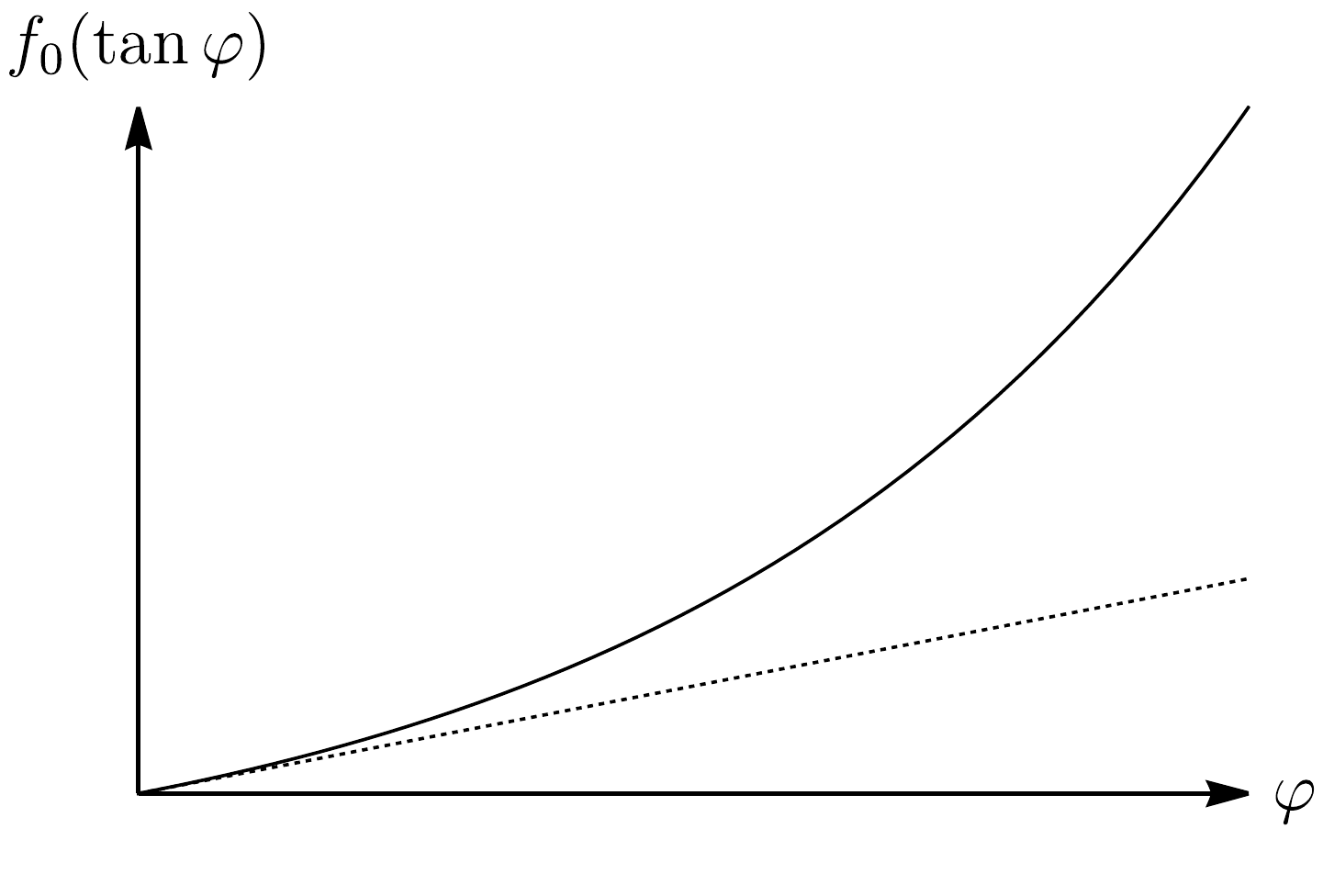} \hspace*{0.5cm}
  \includegraphics[scale=0.5]{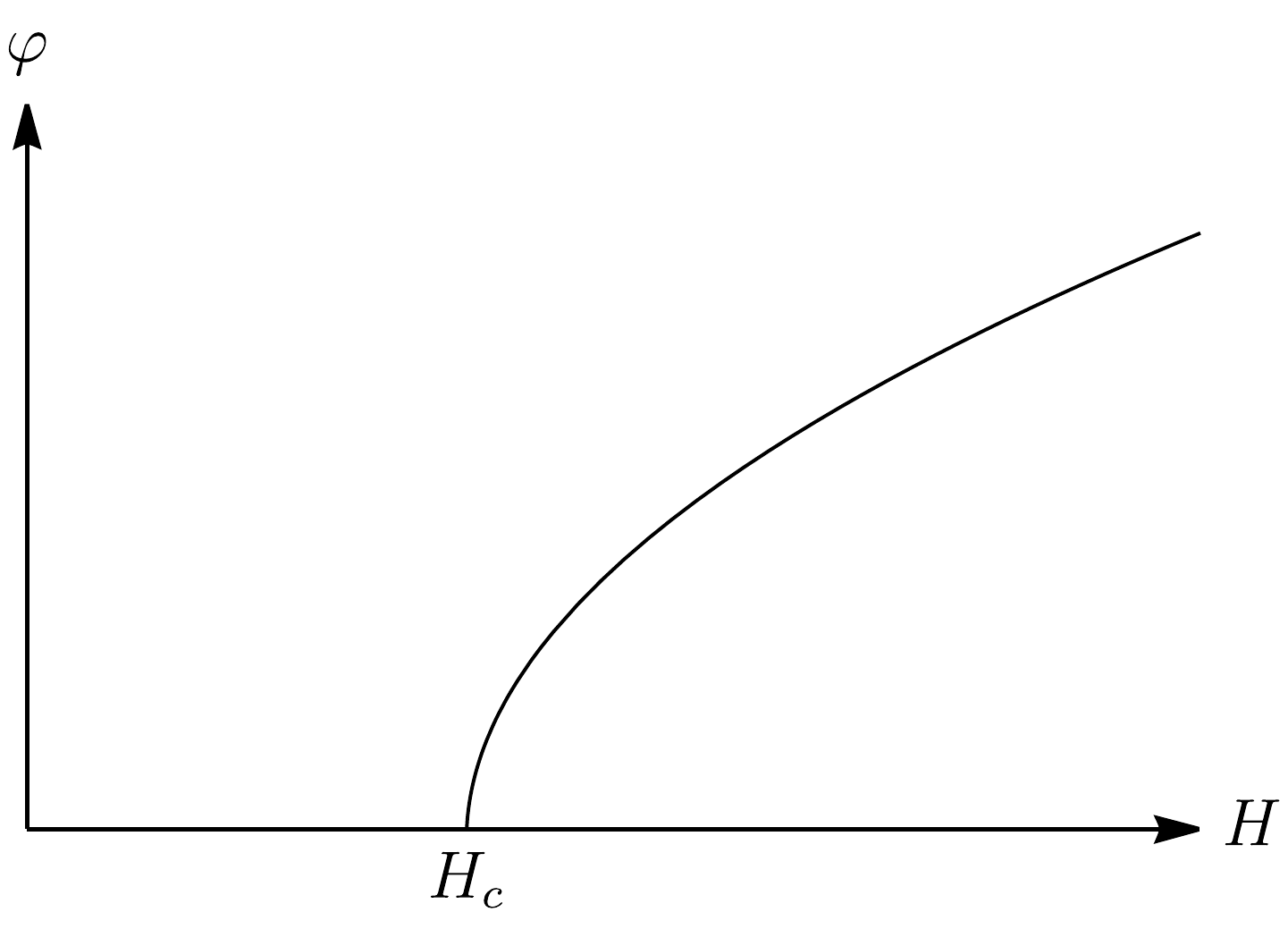} 
 \caption{Left: Free energy $f_0(\tan \phi)$ 
 as a function of the tilt angle $\phi$ with the $z$ axis, in two cases
 (i) \textbf{top:} it is linear along a segment $[0,\phi_c]$, as in 
 \eqref{f00}.
 (ii) \textbf{bottom:} it is slightly curved. The slope at the
 origin is $H_c$ indicated in dashed. Right: corresponding tilt angle versus $H$ curve in the fixed $H$ ensemble.
 (i) \textbf{top:} the tilt exhibits a first order jump at $H_c$, from $\phi=0$ to $\phi=\phi_c$
 (ii) \textbf{bottom:} the tilt raises sharply at $H_c$ but continously, and $\phi_c=0$, see 
 Eq. \eqref{inverselog}.
 }
 \label{fig:appA2}
\end{center}
\end{figure}

\subsection{Periodic columnar potential}

Consider now the model \eqref{en0} in presence of a periodic potential $U(u)$ of period $r_0$ which is nearly zero in between the columns and has a depth of order $U_{\rm min}=-U_0$ at each column location. We first assume the absence of
point disorder $V(u,z)=0$. Let us recall that $\phi$ such that $\tan \phi = \frac{(N-1) r_0}{x} \simeq \frac{N r_0}{x}=1/\theta$ is the angle
of the line with the $z$ axis. In presence of a macroscopic tilt, i.e. $\phi,\theta >0$, we
want to approximate this model by the lattice polymer model of the text with all column energies identical $\epsilon_i= - U_0$, $L=x$, $u(0)=0$ and $u(x)=u_f=(N-1) r_0$.

In the continuum model the jumps from column to column occur as kinks, whose shape is obtained by the minimization
of the energy function \eqref{en0} for a single jump, and
depend on $U(u)$ and $\gamma$, see Fig. \ref{fig:appA1} (Right).
Its precise shape can be calculated (see Section \ref{subsec:exact} below) but is not important here: if $U(u)$ has a single scale $r_0$, dimensional estimates show that the  "kink energy" cost is $E_k \sim r_0 \sqrt{\gamma U_0}$ and the "kink width" (i.e. its length along $x$) is
$\delta x_k \sim r_0 \sqrt{\gamma/U_0}$. The analogy with the OY lattice model is closest when the kink width is "small" so that the jump can be considered as almost instantaneous: it is much smaller than the mean inter-kink distance, whenever $\delta x_k \ll x/N$. i.e. $1/\theta \simeq \tan \phi \ll \sqrt{U_0/\gamma}$. The ground state energy per unit length for a fixed tilt angle is

\be \label{Esx} 
\frac{{\cal E}^0}{x}  \simeq f_0(\tan \phi) - H \tan \phi  \quad , \quad f_0(\tan \phi) \simeq \begin{cases} - U_0  + \frac{E_k}{r_0} \tan \phi 
\quad , \quad \tan \phi \ll \sqrt{\frac{U_0}{\gamma}} \\
\frac{\gamma}{2} (\tan \phi)^2 \quad \hspace*{0.9cm} , \quad \tan \phi \gg \sqrt{\frac{U_0}{\gamma}}  \end{cases}
\ee 
where the convex function $f_0(t)$ can be calculated from the Section \ref{subsec:exact} below, but its
asymptotics are easy to obtain, as indicated. In the first regime the minimum energy configuration is made of $N$ well separated
kinks, i.e. one has $N \delta x_k \ll x$, while in the second regime the line spends little time on the columns and one obtains the usual elastic energy (as in the absence of $U(u)$). The general form
of \eqref{Esx} is quite similar to \eqref{f00}, leading to a similar behavior in the 
fixed $H$ ensemble (when $u(L)$ is free to move). The minimization of \eqref{Esx} 
over the angle leads to $\phi=0$ for $H < H_c^0 = \frac{E_k}{r_0}$ and 

a tilting transition at $H_c^0$ (with transverse Meissner effect for $H<H_c^0$).
For large $H-H_c^0$ one can use the asymptotics of $f_0$ and 
$\tan \phi \simeq \frac{H-H_c^0}{\gamma}$. The precise form of the curve of the optimal 
$\tan \phi$ versus $H-H_c^0$ is obtained by studying in more details the 
minimization equation $f'_0(\tan \phi) = H$. Depending on $U(u)$ it leads
to either a jump in $\phi$, or, for a smooth potential, a
singularity of the form \eqref{log2} (see \eqref{inverselog} below).

 Note that this "delocalization
transition" for a periodic potential $U(u)$ in $d=1$ exists only at $T=0$.
For any finite $T>0$ the eigenstates in the associated quantum mechanics
for $H=0$ are Bloch waves hence they are
delocalized and there is no localized phase. For a localized
phase to exist one needs quenched disorder in the column strength, i.e.
a random potential $U(u)$, see below.

We now use the results of the present paper on the one-way OY model. Let us first continue with the above model of a periodic $U(u)$ and add point disorder.
It is difficult to estimate when the effect of backward jumps can be neglected.
As a try, consider the insertion of a kink-antikink pair (which costs an energy $2 E_k$) 
of size  (i.e. the separation) $x_p$ between two forward kinks. 
The typical energy gain from the point disorder by forming this pair is $2 \sigma \sqrt{x_p}$
(here $\sigma$ measures the strength of the point disorder, i.e. the Brownian motions in the effective OY are now taken as
$B_j(x) \to \sigma B_j(x)$). Comparing the gain and the cost one sees that it becomes favorable to
create such an excitation in the ground state if $x_p > x_p^* = E_k^2/\sigma^2$.
If $x_p^*$ is itself larger than the mean inter-kink distance $x/N$, then 
such defects may not be favored, which leads to the condition (i) 
$\tan \phi \gg \tan \phi_0 = r_0 \sigma^2/E_k^2$. For $\phi \lesssim \phi_0$ the ground
state is unstable to backward jumps \cite{footnotebackward}.

Let us assume that the kink energy $E_k$ is large, equivalently the point disorder is weak, so that $\phi_0 \ll 1$. In presence of point disorder we must now minimize, approximately,
\be \label{Esx2} 
\frac{{\cal E}^0}{x}  \simeq f_0(\tan \phi) - 2 \frac{\sigma}{\sqrt{r_0}} \sqrt{\tan \phi} - H \tan \phi 
\ee
where we used the result \eqref{TW1}. Examining the minimization equation,
$H = f'_0(t)-\frac{\sigma}{\sqrt{r_0 t}}$ with $t=\tan \phi$, we see that now there is no transition.
In fact there is a solution, $\phi=\phi_0>0$ even at $H=0$, where $\phi_0 \ll 1$ was
defined above. This is not too surprising: the point disorder favors
the wandering of the line and in $d=1+1$ we expect that it overcomes at large scale
the effect of the periodic potential
(a bit as $T>0$ does). In the OY model the only way to wander that is by tilting. The spontaneous tilt 
is an artefact of the one-way model, and $\phi \sim \phi_0 \ll 1$ is precisely
the region where we cannot use it because in the two way model kink-antikink pairs will proliferate (from condition (i) above). However, we can use it for $\phi=\mathcal{O}(1)$. From the minimization equation, we find that, in the whole region
$H \lesssim H_c^0=E_k/r_0$, $\tan \phi \approx \frac{\sigma^2}{r_0 (H_c^0- H)^2}$, and crossover
for $H \gtrsim H_c^0$ to a tilt response nearly identical to the one is absence of point disorder.

\subsection{Columnar disorder on top of periodic}

Let us now smoothly deform the above periodic potential so that the local minima remain at
positions $u=j r_0$ but with a random value $U(j r_0) = U_0 - a_j$, introducing columnar
disorder, so as to mimic the OY model with drifts. Let us consider the case of a single
active column $a_1>0$, $a_{j \neq 1}=0$. From the results in Sections \ref{subsec:rmt} 
and \ref{sec:singlesingle} and restoring units, the energy to minimize is either given by \eqref{Esx2} if 
$\tan \phi > a_1^2 r_0/\sigma^2$ or, if $\tan \phi < a_1^2 r_0/\sigma^2$ it is given by (from \eqref{loc})
\be \label{Esx3} 
\frac{{\cal E}^0}{x}  \simeq f_0(\tan \phi) - a_1 - \frac{\sigma^2}{a_1 r_0} \tan \phi - H \tan \phi  \quad , \quad 
\tan \phi < a_1^2 r_0/\sigma^2 
\ee
The minimization equation is now 
$H = f'_0(t)- \min(\frac{\sigma^2}{a_1 r_0}, \frac{\sigma}{\sqrt{r_0 t}})$, with $t=\tan \phi$.
Let us recall that from \eqref{Esx} $f'_0(0)=E_k/r_0= H_c^0$ 
and $f'_0(t) \simeq \gamma t$ for $t \gg \sqrt{U_0/\gamma}$.
We see that now there is a threshold field 
\be
H_c = H_c^0 - \frac{\sigma^2}{a_1 r_0}  = \frac{E_k}{r_0} (1 - \frac{\sigma^2}{a_1 E_k}) 
\ee 
provided $H_c>0$, that is $a_1 > \sigma^2/E_k$. For $H<H_c$ the minimum is at $\phi=0$ and
there is a true localized phase, with zero tilt, i.e. with transverse Meissner effect. For $H>H_c$ the
tilt angle versus field curve can be obtained from the minimization equation. 
Since for small $\phi$ the disorder term in \eqref{Esx3} is linear in $\tan \phi$ we expect
a similar behavior as in the two previous subsection: either a jump from $\phi=0$ to
$\phi=\phi_c$, or, for a smooth potential, a sharp but slightly smoothed jump of amplitude 
$\phi_c$, with $\tan \phi_c = a_1^2 r_0/\sigma^2$. In the fixed $\phi$ ensemble (which can
handle both cases), for $\phi<\phi_c$ the line occupies a finite fraction $\ell^0/x = \mathcal{O}(1)$ of the single active column (this fraction being unity for $\phi=0$), and 
this fraction vanishes at the delocalization transition at $\phi=\phi_c$ (where $\phi_c$ is the
value of $\phi=\phi(H)$ at $H=H_c$ in the fixed $H$ ensemble). These arguments 
are of course quite heuristic and it would be of great interest to analyze these observables
in more details.

Note that in the localized phase, the macroscopic segment of the line along 
the active column $a_1$ is stable to a large size single kink-antikink pair, since its energy cost
is now $2 E_k + a_1 x_p - \omega \sigma \sqrt{x_p}$, where $\omega$ a Gaussian random variable.
These pairs are never favorable for large $x_p$ and if $E_k \gg \sigma^2/a_1$ they are never favorable at any scale $x_p$ \cite{footnotepairs}.

\subsection{Exact solution for an elastic line in a columnar potential at $T=0$} 
\label{subsec:exact} 

Here we study a single elastic line, described by the model \eqref{en0}, in a general potential $U(u)$
at $T=0$ and in the absence of point disorder, following closely the analysis in \cite{BalentsSimon1995} and adding a few remarks and results useful in the present context. 
This method allows to determine the function $f_0(t)$ introduced in the previous subsections.
For fixed endpoints, the minimization equation
\be
E_0[u_f,x,U,\gamma] = \min_{u(z),u(0)=0,u(x)=u_f} \int_0^x \rmd u [ \frac{\gamma}{2} (\frac{\rmd u(z)}{\rmd z})^2 + U(u(z)) ]
\ee 
has the solution 

\be
E_0[u_f,x,U,\gamma]  = \sqrt{\gamma} \int_0^{u_f} \rmd u 
\frac{\epsilon + 2 U(u)}{\sqrt{2 (\epsilon + U(u))}} =
- \epsilon x + \sqrt{\gamma} \int_0^{u_f} \rmd u \sqrt{2 (\epsilon + U(u))} \label{EU} 
\ee 
where $\epsilon$ is determined by the condition
\be
\sqrt{\gamma}  \int_0^{u_f} \rmd u \frac{1}{\sqrt{2 (\epsilon + U(u))}} = x   \label{xU} 
\ee 
and the optimal $u(z)$ is the solution of $\gamma \frac{d^2 u}{dz^2} = U'(u)$, i.e. 
of the classical mechanics problem in the
inverted potential $-U(u)$, at total "energy" $\epsilon$, thus determined by $\sqrt{\gamma}  \int_0^{u(z)} du \frac{1}{\sqrt{2 (\epsilon + U(u))}} = z$. In the fixed $H$ ensemble, where $u_f$ is free (we keep
$u(0)=0$), one minimizes $E_0 - H u_f$ over $u_f$. This is equivalent to optimize over $\epsilon$, which leads to $H = \sqrt{\gamma} \sqrt{2 (\epsilon + U(u_f))}$. There may be multiple
solutions to this equation, and a more useful equation will be given below.

{\bf Periodic potential.} Consider first the periodic potential $U(u)$ described above. 
Let us start with $u_f=r_0$ and $x \to +\infty$, which leads to the standard "kink energy" $E_k$
for an ideal jump between 2 columns. In that case $\epsilon = - U_{\min}=U_0$ and 
$E_0= - U_0 x + E_k$ with $E_k=\sqrt{\gamma} \int_0^{r_0} du 
\sqrt{2 (U_0 + U(u))}$. Let us consider now 
$u_f=(N-1) r_0 + \delta u_f$ where $0< \delta u_f < r_0$.
One can rewrite exactly \eqref{EU} and \eqref{xU} 
by replacing $\int_0^{u_f} du \to (N-1) \int_0^{r_0} du + \int_0^{\delta u_f} du$.
We are interested in the limit where both $x,N$ are large, with a fixed tilt angle 
$t = \tan \phi \simeq N r_0/x$, for which one obtains
\be
\frac{1}{t} \simeq \frac{\sqrt{\gamma} }{r_0} \int_0^{r_0} \frac{du}{\sqrt{2 (\epsilon + U(u))}} 
\quad , \quad 
\frac{E_0 - H u_f}{x}  \simeq - \epsilon  + \frac{t \sqrt{\gamma}}{r_0} \int_0^{r_0} du \sqrt{2 (\epsilon + U(u))} - H  t \label{double} 
\ee
The minimization over $t=\tan \phi$ leads to 
\be
H = \frac{\sqrt{\gamma}}{r_0} \int_0^{r_0} du \sqrt{2 (\epsilon + U(u))}  \label{Hmin} 
\ee 
which is different from the result $H= \sqrt{\gamma} \sqrt{2 (\epsilon + U(\delta u_f))}$ 
obtained if one takes the derivative w.r.t. $\delta u_f$. This is because there are
multiple extrema and one should minimize over both $N$ (choose the column) and $\delta u_f$ 
(fine structure inside one column). Eq. \eqref{Hmin} corresponds to the former, and 
together with \eqref{double} determines $t=\tan \phi$ as a function of $H$ (the $\delta u_f$
then adjusts to be at a local minimum). As $\tan \phi \to 0$ 
one has $\epsilon \to U_0$ and one recovers the 
small $\phi$ estimate for $f_0(t)$ in \eqref{Esx}. 
For large $\tan \phi$ one has $\epsilon \gg |U(u)|, U_0$ and one can
neglect the potential $U(u)$ altogether, and 
one finds $E_0 \simeq \epsilon x$ with $\epsilon \simeq \frac{\gamma}{2} (\tan \phi)^2$
recovering the second estimate for $f_0(t)$ in \eqref{Esx}.

The equation \eqref{Hmin} determines $\epsilon$ as a function of $H$. Since $\epsilon \geq U_0=-U_{\min}$, there are no solution to \eqref{Hmin} for $H<H_c$ where $H_c= \frac{\sqrt{\gamma}}{r_0} \int_0^{r_0} du \sqrt{2 (U_0 + U(u))}= E_k/r_0$, a signature of the tilting transition at $H_c$. The precise behavior of the tilt angle (jump or no
jump) depends on the exact form of the potential. If $\epsilon$ in \eqref{double} was constant equal to $- U_0$, the energy would be exactly linear and a jump occurs. For a smooth potential $U(u)= -U_0 + U_2 u^2$
one obtains $r_0/\sqrt{\tilde \epsilon} = \sinh( \frac{r_0}{t} \sqrt{\frac{2 U_2}{\gamma}})$,
where $\tilde \epsilon=(\epsilon-U_0)/U_2$. 
Hence at small angle, $\tilde \epsilon \simeq 4 r_0^2 \exp(-  \frac{2 r_0}{t} \sqrt{\frac{2 U_2}{\gamma}})$.
We can use the general relation $\frac{dH}{d\epsilon}=\frac{1}{t}$ from \eqref{Hmin}-\eqref{double}. 
Integrating we find for $H > H_c$
\be \label{inverselog}
H - H_c \simeq \frac{4 \tilde U_2}{t} \exp\left(-  \frac{2}{t} \sqrt{\frac{2 \tilde U_2}{\gamma}}\right) \quad , \quad 
t =\tan \phi \simeq \frac{ 2 \sqrt{2 \tilde U_2/\gamma} }{|\log (\frac{H-H_c}{4 \tilde U_2})| }
\ee 
and $\phi=0$ for $H<H_c$, where $\tilde U_2=r_0^2 U_2$ is an energy scale (per unit length) characteristic of the
local curvature of the well near its minimum. The jump in the tilt angle is smoothed on this scale.

\bigskip

{\bf Random potential.}
Consider now $U(u)$ a random potential. If ergodicity applies for
large $x$ and $u_f$ at fixed angle $\theta =x/u_f = 1/t$ with $t=\tan \phi$, it means that we can replace
the translational averages over the potential $U(u)$ by disorder averages noted $\langle \dots \rangle_U$.
To mimic the notations in the OY model, we can introduce the one-point density $\rho(a)$ 
of minus the potential $a=-U$, as
$\rho(a)= \frac{1}{u_f} \int_0^{u_f} du \delta(U(u)+a)=\langle \delta(U(u)+a) \rangle_U$. 
We assume that this density has an upper edge $a_e=-U_{\min}$. The above equations then lead to 
the system of equations
\be \label{solu1} 
\theta = \sqrt{\gamma} \, \int da \frac{\rho(a)}{\sqrt{2 (\epsilon - a)}}  \quad , \quad  \frac{E_0}{x}= - \epsilon + 
 \sqrt{\gamma} \frac{1}{\theta} \, \int da \rho(a)  \sqrt{2 (\epsilon -a)} - \frac{H}{\theta} \quad , \quad 
 H = \sqrt{\gamma} \int da \rho(a) \sqrt{2 (\epsilon-a)} 
\ee 
where $\epsilon > a_e= - U_{\rm min}$ is determined from the first equation as a function of $\theta$.
The first two of these equations are
reminiscent of the system \eqref{eq1}, \eqref{eqtheta} for the OY model at fixed tilt angle, with the identification
$z^*=\epsilon$ and $\mu=\frac{E_0}{N}$. They are of course different since the
models are different (here an elastic line here without point disorder and there
a discrete one way model with point disorder). 
The last equation in \eqref{solu1} leads to a tilting transition in the fixed $H$ ensemble at $H_c=\sqrt{\gamma} \int da \rho(a) \sqrt{2 (a_e-a)}$. For $H<H_c$ the tilt angle is zero, and one cannot use Eqs. \eqref{solu1}
(one cannot use the ergodicity). For $H \geq H_c$ the behavior of the tilt angle depends on
how $\rho(a)$ vanishes near its upper edge $a_e$. For $\rho(a) \sim (a_e-a)^{2 k+1}$ 
there is a finite jump from $\phi=0$ to $\phi_c$ with $1/\tan \phi_c=\theta_c=
\sqrt{\gamma} \, \int da \frac{\rho(a)}{\sqrt{2 (a_e - a)}}$ which is finite for $k > - 1/2$,
and one has the critical behavior \cite{BalentsSimon1995} 
$\theta-\theta_c \propto \phi-\phi_c \propto  (H-H_c)^{\max(1,2k+1)}$
in the delocalized phase. As $k \to 0$ the jump in $\phi$ vanishes and for 
$k=0$ (which corresponds to the smooth potential)
one finds again a behavior similar to \eqref{inverselog} with $\phi_c=0$, $\theta_c=+\infty$. 

Note that for $k>0$, 
as in \eqref{laloc} \eqref{delocdeloc}, one can define an occupation length measure 
which acquires a delta contribution for $\phi<\phi_c$
\be 
\ell(a)  = \begin{cases} N \frac{\rho(a)}{\sqrt{\epsilon-a}} \hspace*{3.9cm} , \quad  \phi>\phi_c \\
N \frac{\rho(a)}{\sqrt{a_e-a}}  + (x- N \theta_c) \delta(a-a_e) \quad , \quad \phi<\phi_c
\end{cases} 
\ee 
so that $\int \rmd a \, \ell(a) = x$.

\section{Dyson Brownian motion and Airy process}

\label{app:dbm} 

\subsection{DBM without drift and the Gaussian $\beta$ ensemble} 

\label{app:dbm0} 

Consider $W(x)$ a Hermitian Brownian motion in $x$, or Brownian motion in the space
of $N \times N$ Hermitian matrices. The stochastic evolution equation for the process
of the eigenvalues $\lambda_i(x)$, $i=1,\dots,N$, of $W(x)$, i.e. the Dyson Brownian motion,
reads
\be \label{dbm1} 
d\lambda_i(x) = a \sum_{j \neq i} \frac{\rmd x}{ \lambda_i(x)-\lambda_j(x)} + b \, db_i(x) 
\ee
where $b_i(x)$ are $N$ independent unit Brownian motions.
This is the non-stationary DBM. 
We have introduced two arbitrary parameters $a,b$ so that \eqref{dbm1} is actually
the $\beta$-DBM, the case $\beta=2$ relevant for Hermitian random matrices
corresponds to $a=b^2$ (the general case is $\beta=2 a/b^2$, see below).
The choice made here in the text is $a=b=1$, which
corresponds to a choice of normalization for $W(x)$. Here we define $W(0)=0$, i.e. $\lambda_i(0)=0$
(see below for non zero initial condition). In the large $N$ limit, the density of eigenvalues normalized to unity corresponding to \eqref{dbm1} is a semi-circle with the edges at $\pm 2 \sqrt{N a x}$. \\

Let us also recall the stationary version of the DBM, or Orstein-Uhlenbeck version, which can be obtained from the
non-stationary one via a Lamperti transformation. Defining $\Lambda_i(X)=\frac{\lambda_i(x)}{\sqrt{x}}$
and performing the "time change" $x = e^{c X}$ one finds that 
\be \label{dbm2}
d\Lambda_i(X) = - \frac{1}{2} c \Lambda_i(X) dX + a c \sum_{j \neq i} \frac{dX}{ \Lambda_i(X)-\Lambda_j(X)} + b \sqrt{c} \, d{\sf b}_i(X) 
\ee 
where ${\sf b}_i(X)$ are are $N$ independent unit Brownian motions in $X$. 
The stationary JPDF of the $\Lambda_i(X)$ is, for any $N$, the equilibrium measure
\be \label{stat}
P_0(\Lambda) = \frac{1}{Z_N} \prod_{1 \leq i < j \leq N} |\Lambda_i-\Lambda_j|^\beta  e^{- \frac{1}{2 b^2} \sum_i \Lambda_i^2}  \quad , \quad \beta=\frac{2 a}{b^2} 
\ee
where $Z_N$ is a normalization. For large $N$, the density has the semi-circle shape with the edges at $\pm 2 \sqrt{a N}$, see Fig.~\ref{fig:semiCircle}. Note that to recover the DBM associated to the Gaussian $\beta$-ensemble with support at equilibrium $[-2,2]$ one would choose instead $a=\frac{1}{N}$ and $b^2=\frac{2}{\beta N}$.\\

\begin{figure}[t!]
\centerline{\includegraphics[scale=0.7]{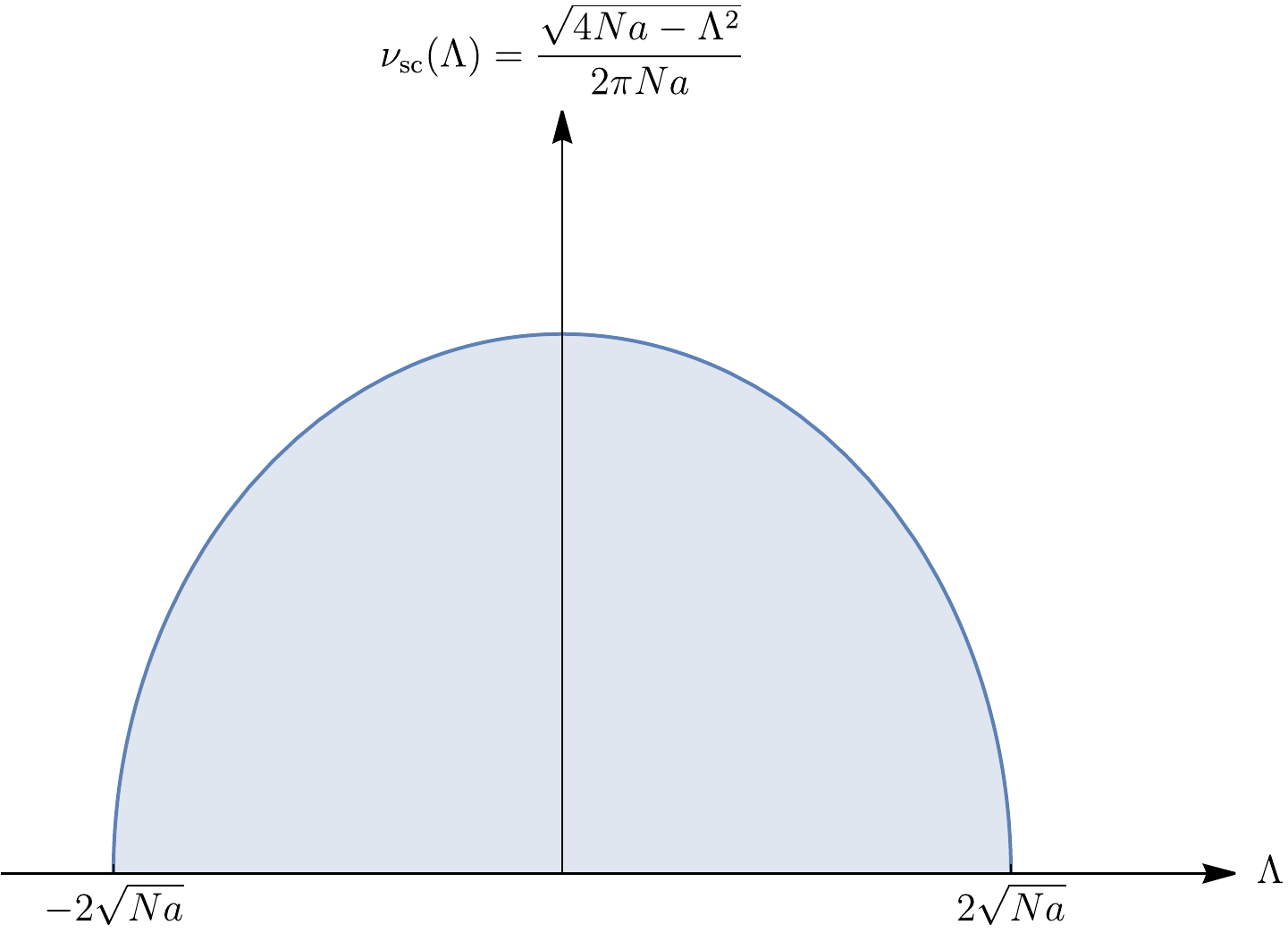}}
\caption{Semi-circular law with support $[-2\sqrt{Na},2\sqrt{Na}]$.}
\label{fig:semiCircle} 
\end{figure}

The equilibrium solution \eqref{stat} for the stationary DBM also provides the $x$-dependent JPDF $P(\lambda,x)$ for the non stationary DBM in \eqref{dbm1} with initial condition $\lambda_i(0)=0$,
via the simple scaling $P(\lambda,x) = x^{- N/2} P_0(\lambda/\sqrt{x})$.\\

Note that the probability that the DBM remains below a barrier $\lambda_1(t) < W \sqrt{t}$ for all $t<x$ decays as $t^{- \beta_c(N,W)}$, where $\beta_c(N,W)$ was calculated in \cite{Gautie}. This thus has a direct translation for the polymer model with $a_i=0$ in terms of events such that the ground state energy remains larger than $-W \sqrt{t}$ for all $t<x$.

\subsection{DBM with initial condition and the deformed Gaussian unitary ensemble}

Consider now ${\cal M}(x)$ defined as
\be
{\cal M}(x) = {\rm diag}(c_1,\dots,c_N) + W(x) 
\ee 
and denote now $\lambda_i(x)$ the eigenvalues of ${\cal M}(x)$. Then the $\lambda_i(x)$
follow the DBM evolution equation \eqref{dbm1} with initial conditions $\lambda_i(0)=c_i$. 
For a fixed $x$, the eigenvalues of ${\cal M}(x)$ have the same law as the
law at time $x$ of a standard DBM with initial condition $c_1,\dots c_N$.
It is given by \eqref{JPDFDeformed} with $a_j=c_j/x$.

Note that the DBM for $\beta=2$ is equivalent to $N$ Brownian walks conditioned not to cross
until infinite time \cite{Grabiner1999,tao2012topics,Gautie}. In the text we have chosen $c_i= x a_i$. That case can equivalently be seen as $N$ Brownian motions all started at zero, conditioned never to intersect, and with drifts $a_i$ in the time interval $[0,x]$. Indeed the formula 
\eqref{JPDFDeformed} has a simple interpretation. The determinant containing the $a_i$'s
is the expression of the Karlin Mc Gregor formula \cite {KarlinMcGregor} for non intersecting paths
on $[0,x]$, which is still valid in presence of particle-dependent drifts (as can
be seen in the simplest way from the path integral formula \cite{Gautie2}). The second determinant, i.e. the vandermonde $\Delta_N(\lambda)$, arises 
from the non-intersection conditioning for all later times.\\

In the text we have also introduced in \eqref{tildeM} the eigenvalues $\tilde \lambda(x)=\lambda(x)/x$ of the matrix $\tilde M(x)={\cal M}(x)/x$. The
equation \eqref{tildeM} can be interpreted as a process in $x$, and its r.h.s as 
a process in the variable $\tau=N/x$ as
\be \label{eq23} 
\tilde M(x) = {\rm diag}(a_1,\dots a_j)  + \tilde W(\tau) 
\ee
where we have replaced $\sqrt{\tau} \, \tilde V \to \tilde W(\tau)$, where $\tilde W(\tau)$ is a Hermitian Brownian motion in $\tau$, with parameters $a=b^2=1/N$. Hence 
the process $\tau \to \tilde \lambda(x=N/\tau)$ is a DBM with
initial conditions at $\tau=0$ given by the $\{ a_i \}$ (which is the final
condition for the process $x \to \tilde \lambda(x)$).

\subsection{Resolvent of the Dyson Brownian motion and the Burgers equation}

The DBM $\lambda_i(x)$ is often studied using the resolvent $g_x(z)=\frac{1}{N} \sum_{i=1}^N \frac{1}{z- \lambda_i(x)}$.
Using Ito's rule, the stochastic equation \eqref{dbm1} leads to (denoting $\lambda_i(x)$
simply as $\lambda_i$)
\be
d g_x(z) = \frac{1}{N} \sum_i d\lambda_i \, \partial_{\lambda_i} \frac{1}{z-\lambda_i} + 
\frac{b^2 dx}{2 N} \sum_i \partial_{\lambda_i}^2 \frac{1}{z-\lambda_i}
\ee
After standard manipulations, i.e. $\partial_{\lambda_i} \equiv - \partial_z$,
$\frac{1}{N} \sum_{i \neq j} \frac{1}{\lambda_i-\lambda_j} \frac{1}{z-\lambda_i} 
= \frac{1}{2} ( N g_x(z)^2 + \partial_z g_x(z))$, one obtains 
\be \label{burg1} 
\partial_x g_x(z)= - \frac{1}{2} a N \partial_z g_x(z)^2 + \frac{1}{2} (b^2 - a) \partial_z^2 g_x(z) 
- \frac{b}{\sqrt{N}} \partial_z \eta_{z,x} 
\ee 
where $\eta_{z,x} dx = \frac{1}{\sqrt{N}} \sum_i \frac{db_i(x)}{z-\lambda_i}$ is Gaussian of correlator ${\rm Cov}(\eta_{z,x} \eta_{z',x'}) = - \delta(x-x') \frac{g_x(z)-g_x(z')}{z-z'}$.
Note that for $\beta=2$ one has $a=b^2$ and the diffusion term is absent. 
Considering for instance the Gaussian $\beta$-ensemble with support $[-2 \sqrt{x},2 \sqrt{x}]$, i.e. 
the choice $a=\frac{1}{N}$ and $b^2=\frac{2}{\beta N}$, it is clear that in the large $N$ limit (for fixed $\beta$) all terms in the r.h.s. of \eqref{burg1} except the first one are subdominant in $N$. In the text, to study the $\beta=2$ DBM of the $\lambda(x)$, we choose instead $a=b=1$, but the conclusion remains (it is a simple change in scale), i.e. at large $N$ one has $\partial_x g_x(z)= - N g_x(z) \partial_z g_x(z)^2$.\\

In the text of the paper we study a regime where $x = N \theta$ with $\theta=\mathcal{O}(1)$.
We can thus define $\lambda_i= N \hat \lambda_i$ where $\hat \lambda_i = \mathcal{O}(1)$. 
Then if one defines $\hat g_\theta(\hat z)= \frac{1}{N} \sum_i \frac{1}{\hat z - \hat \lambda_i}$
we have $g_x(z)=\frac{1}{N} \hat g_\theta(\hat z=z/N)$ which leads to
\be \label{burg2} 
\partial_\theta \hat g_\theta(\hat z)= - \hat g_\theta(\hat z) \partial_{\hat z} \hat g_\theta(\hat z) 
- \frac{1}{N} \partial_{\hat z} \hat \eta_{\hat z,\theta} 
\ee 
with ${\rm Cov}(\hat \eta_{\hat z,\theta} \hat \eta_{\hat z,\theta} ) = - \delta(\theta-\theta') \frac{\hat g_\theta(\hat z)-\hat g_\theta(\hat z')}{\hat z- \hat z'}$. 
The rescaled process $\theta \to \hat \lambda_i$ is thus a $\beta=2$ DBM with $a=b^2=\frac{1}{N}$.

On the other hand, from the discussion around \eqref{eq23} we see that the 
process $\tau \to \tilde \lambda_i(x=N/\tau)$ is also a $\beta=2$ 
DBM with the same parameter 
$a=b^2=\frac{1}{N}$. Its resolvant $G_\tau(z)=\frac{1}{N} \sum_{i=1}^N \frac{1}{z-\tilde \lambda_i(x)}$ thus also satisfies the equation \eqref{burg1} with these parameters,
i.e. it satisfies the same equation as \eqref{burg2} with $g_\theta(\hat z)$
replaced by $G_\tau(z)$. This is true for any $N$, and for large $N$ it yields 
the equation \eqref{burg0} of the text. This is quite remarkable since the process 
$\tau \to \tilde \lambda_i(x=N/\tau)$ is the  "time-inverted" process of 
the DBM process $\theta \to \hat \lambda_i$. The fact that it is also a DBM with the same
parameters originates from the following property, which we quote with slightly different notations:\\

{\it Time inversion of the Dyson Brownian motion}.
Let $t \to \mu_i(t)$, $t \geq 0$, a DBM \eqref{dbm1} (with parameters $a,b$) and initial condition 
$\mu_i(0)=0$. Then $\tau \to \rho_i(\tau) = \tau \mu_i(1/\tau)$, $\tau \geq 0$, is also a DBM with the same 
parameters and the same initial condition $\rho_i(0)=0$. 

For $N=1$ and $\mu_1(t)= b(t)$ a Brownian motion, this property is well known as the time inversion of the Brownian \cite[IV.2]{HandbookBrownian} (the same property
extends to a Bessel process of any index \cite{ShigaWatanabeBessel}).
If $\mu_1(t)=b(t) + a t$ is a Brownian with a drift, the property is that $\rho_1(\tau)=\tau \mu_1(1/\tau)$ is a Brownian
(without drift) started at $a$, i.e. $\rho_1(\tau=0)=a$. One can then use the fact that the $\beta=2$ DBM is equivalent to a collection of $N$ Brownian conditioned to never cross, and one sees that
the property immediately carries through. Note that this property extends to any $\beta$. 
For $N=2$, since the sum is a BM, and the difference is an independent Bessel process, it follows from \cite{ShigaWatanabeBessel}.

Note that under the Lamperti transformation discussed above this property 
of the non-stationary DBM becomes simply the time reversal of the stationary DBM,
i.e. that the process $T \to \Lambda_i(-T)$ is an identical copy of the stationary DBM
$\Lambda_i(T)$.

\subsection{Convergence to the Airy process} 

\label{app:sub:convergence} 

In this section we recall how the process $x \mapsto \lambda_1(x)$ of the
largest eigenvalue of the matrix ${\cal M}(x)$ studied in the text, can be described, as a random function, by the 
Airy$_2$ process. We will make connection and use the properties of the 
Airy$_2$ process with respect to the stationary DBM given in Ref. \cite{Prolhac_2011}.
In Ref. \cite{Prolhac_2011} the stationary Hermitian BM for a hermitian matrix ${\mathsf M}(X)$ was considered 
\be
d {\mathsf M} (X) = - \frac{1}{N} {\mathsf M}(X) dX + \frac{1}{\sqrt{N}} d{\cal B}(X)
\ee 
where $X$ is the Lamperti variable (the same as in Section \ref{app:dbm0}),
with $d{\cal B}_{ij}(X) \rmd {\cal B}_{kl}(X)= \frac{\rmd X}{2} (\delta_{ik}\delta_{jl}+\delta_{il} \delta_{jk})$.
Note that ${\mathsf M}(X)=\frac{{\cal M}(x)}{\sqrt{N}}$ where ${\cal M}(x)$ obeys Eq. (1.6) in \cite{Prolhac_2011}
(with $X\equiv t$).
One easily sees that the eigenvalues $\Lambda_i(X)$ of ${\mathsf M}(X)$ satisfy 
\eqref{dbm2} with $c=\frac{1}{N}$, $a=b=1$. As claimed there, at large $N$ one
has the following convergence of the largest eigenvalue (noted $\Lambda_1(X)$) of the
DBM to the Airy$_2$ process
\be
\Lambda_1(X + \delta X) \simeq 2 \sqrt{N} + N^{-1/6} {\cal A}_2(\frac{\delta X}{N^{2/3}}) 
\ee 

We can now perform the Lamperti mapping in reverse, $\lambda(x)=\sqrt{x} \Lambda( \frac{\log x}{c} )$, and we obtain, upon expanding to leading order in the fluctuations
\be
\begin{split}
 \lambda_1(x+\delta x) &\simeq 2 \sqrt{N(x + \delta x)} + \sqrt{x} N^{-1/6} {\cal A}_2(\frac{\delta x}{c x N^{2/3}}) 
\\
& = 2 \sqrt{N(x + \delta x)} + \sqrt{x} N^{-1/6} {\cal A}_2(N^{1/3} \frac{\delta x}{2 x}) 
\end{split}
\ee
which holds for large $N$ in the scaling region in $x,N,\delta x$ where the argument of ${\cal A}_2$ is of order $\mathcal{O}(1)$. 
Note that if we set $x = N^{1/3}$ and $\delta x=2 s$, this is consistent with 
\cite[Theorem 5.3]{Katori_2007}. The result is more general and the convergence to the extended
Airy$_2$ point process holds for all eigenvalues. 

\section{O'Connell-Yor Polymer models} 

\label{app:defpoly}

\subsection{Stochastic equation}

Let us for completeness give the stochastic evolution equation for the
free energy of the OY model. In the OY model (minus) the free energy $F_N(t)$ satisfies the Ito evolution 
\be
dF_N(t) = e^{F_{N-1}(t) - F_{N}(t)} dt + a_N dt + dB_N(t) 
\ee 
One can define the partition sum $z_N(t)=e^{-\alpha t} e^{F_N(t)}$ which, using Ito rule, satisfies
\bea
dz_N(t) = (z_{N-1}(t)+ (\frac{1}{2}-\alpha) z_{N}(t)) dt + z_N(t) (a_N dt + dB_N(t) )
\eea 
with $z_0(t)=0$ (equivalently $dF_1(t)=dB_1(t)$). The initial condition studied here is $z_N(0)=\delta_{N1}$. Convenient choices for $\alpha$ are $\alpha=1/2$ or $\alpha=3/2$. 

\subsection{Definition of the many line model}
\label{app:defmany} 

The O'Connell-Yor model with $M$ lines, extended to arbitrary drifts, is defined as follows. 
One consider $M$ paths $\pi_1, \dots, \pi_M$ which live only on the columns $j=1,\dots,N$. The paths are non-crossing. Their starting points are on columns 
$1,\dots,M$ and their endpoints are on columns $N-M+1,\dots,N$. The path $\pi_k$ jumps from column $j$ to $j+1$ at $x_j^{(k)}$. There are no backward jumps, i.e. from $j \to j-1$. One has 
\be
x_0^{(k)}=0<x_1^{(k)}<\dots<x_{N-M+1}^{(k)}=x \quad , \quad k=1,\dots,M 
\ee
The non-crossing conditions on the paths furthermore imply that the paths starting upward must jump before, hence
\be
x_j^{(M)} < x_j^{(M-1)} < \dots < x_j^{(2)} < x_j^{(1)} \quad , \quad j=1,\dots,N-M+1 
\ee 
The set of paths is now parametrized by ${\bf x} =\{x^{(k)}\}_{0 \leq j \leq N-M+1, 1 \leq k \leq M }$ and one defines its total energy as
\begin{equation}
    \begin{split}
     E_{N,M}({\bf x}) = \sum^M_{k = 1} \sum^{N-M + 1}_{j = 1} \left[ B_{j  }(x^{(k)}_j, x^{(k)}_{j - 1}) -a_j (x^{(k)}_j - x^{(k)}_{j - 1}) \right],
    \end{split}
  \end{equation}
Finally the optimal energy, i.e the ground state energy for the system of $M$ lines is
\be
{\cal E}^0_{N,M}(x) =\min_{{\bf x}} E_{N,M}({\bf x})
\ee 
where the minimum is over all allowed configurations. The following figure
shows the interlacing condition obeyed by the $x_j^{(k)}$.

\begin{figure}[h]
\centering
\begin{tikzpicture}

\draw (0,0) node {$x_1^{(1)}$};
\draw (2,0) node {$x_2^{(1)}$};
\draw (4,0) node {$\dots$};
\draw (6,0) node {$x_{N-M}^{(1)}$};
\draw (0,-1.5) node {$x_1^{(2)}$};
\draw (2,-1.5) node {$x_2^{(2)}$};
\draw (4,-1.5) node {$\dots$};
\draw (6,-1.5) node {$x_{N-M}^{(2)}$};
\draw (0,-2.75) node {$\vdots$};
\draw (2,-2.75) node {$\vdots$};
\draw (4,-2.75) node {$\vdots$};
\draw (6,-2.75) node {$\vdots$};
\draw (0,-4.25) node {$x_1^{(M)}$};
\draw (2,-4.25) node {$x_2^{(M)}$};
\draw (4,-4.25) node {$\dots$};
\draw (6,-4.25) node {$x_{N-M}^{(M)}$};

\draw (1,0) node [rotate=0] {$\leq$};
\draw (3,0) node [rotate=0] {$\leq$};
\draw (5,0) node [rotate=0] {$\leq$};

\draw (1,-1.5) node [rotate=0] {$\leq$};
\draw (3,-1.5) node [rotate=0] {$\leq$};
\draw (5,-1.5) node [rotate=0] {$\leq$};

\draw (1,-4) node [rotate=0] {$\leq$};
\draw (3,-4) node [rotate=0] {$\leq$};
\draw (5,-4) node [rotate=0] {$\leq$};

\draw (0,-0.75) node [rotate=90] {$\leq$};
\draw (0,-2.25) node [rotate=90] {$\leq$};
\draw (0,-3.5) node [rotate=90] {$\leq$};

\draw (2,-0.75) node [rotate=90] {$\leq$};
\draw (2,-2.25) node [rotate=90] {$\leq$};
\draw (2,-3.5) node [rotate=90] {$\leq$};

\draw (6,-0.75) node [rotate=90] {$\leq$};
\draw (6,-2.25) node [rotate=90] {$\leq$};
\draw (6,-3.5) node [rotate=90] {$\leq$};

\draw (8,0) node {Polymer $1$};
\draw (8,-1.5) node {Polymer $2$};
\draw (8,-4) node {Polymer $M$};

\end{tikzpicture}
\caption{Interlacing property of the jumping positions $x_i^{(k)}$.}
\label{figGT}
\end{figure}

\section{Baik-Ben Arous-P\'ech\'e kernel } \label{app:BBP}.

It is useful to recall the formula for the BBP kernel mentionned in the text, which describes a rank $m$ perturbation, in the critical regime. Let $\mathbf{b}=(b_1\leq \dots \leq b_m)\in \mathbb{R}^m$, 
\begin{equation}
K_{\mathrm{BBP},\mathbf{b}}(v,v')=\frac{1}{(2 \I \pi)^2}\int\limits_{e^{-2 \I \pi/3}\infty}^{e^{2 \I \pi/3}\infty}\mathrm{d}w\int\limits_{e^{-\pi i/3}\infty}^{e^{\pi i/3}\infty}\mathrm{d}z\frac{1}{z-w}\frac{e^{z^3/3-zv}}{e^{w^3/3-wv'}}\prod_{k=1}^m\frac{z-b_k}{w-b_k}
\end{equation}

Note that for $m=1$, the kernel reads 
\be \label{bpp11} 
K_{\mathrm{BBP},b_1}(\eta,\eta')= K_{Ai}(\eta,\eta') + Ai(\eta) \int_{0}^{+\infty} \rmd u Ai(\eta'+u)
e^{b_1 u} 
\ee

\section{Ground state energy: critical behavior at the localization transition}
\label{app:trans} 

Here we give details on the derivation of some of the results displayed in Section \ref{subsec:gse}.
We recall that $\theta=x/N$ and that the transition from the delocalized phase to localized phase occurs at $\theta=\theta_c$ where $\theta_c$ is given by \eqref{thetac}. We assume that the density $\rho(a)$ vanishes  
fast enough, i.e. $k>1/4$ in \eqref{dens0}, so that this transition exists. We are interested in the ground state energy per column $\mu= - \lim_{N \to +\infty} \frac{1}{N} {\cal E}^0_N$ and its dependence as a function of $\theta$, $\mu(\theta)$. Let us denote
\be \label{mue} 
\mu_e(\theta)=(\theta-\theta_c)a_e+\mu(\theta_c)
\ee
which, according to \eqref{mutheta}, is the energy per column in the localized phase for $\theta > \theta_c$, but can be defined from \eqref{mue} for any value of $\theta$. We want to calculate the difference $\mu(\theta)-\mu_e(\theta)$ in the delocalized phase, which by definition vanishes in the localized phase (hence provides an order parameter). To this aim we must first calculate $z^*$, and we focus now
on the region near the transition $\theta \to \theta_c^-$. Let us define 
\be
b= a_e-a \quad , \quad z^*-a_e =\epsilon
\ee 
and substract \eqref{eqtheta}  and \eqref{thetac} and rearrange. We obtain
\be \label{rel1}
\theta_c-\theta= \epsilon \int_0^{+\infty} db \frac{2 b + \epsilon}{b^2 (b+\epsilon)^2}  \rho(a_e-b)
\ee 
and we recall that $\rho(a_e-b)\simeq A b^{2 k+1/2}$. We see that if $k>3/4$ the integral 
$J_3=\int_0^{+\infty} db \frac{1}{b^3}  \rho(a_e-b)$ converges and one finds $\theta_c - \theta \sim 2 J_3 \epsilon$ leading to the linear behavior \eqref{zstrans}. However, for $1/4 < k < 3/4$, $J_3$ diverges and the critical behavior changes. At small $\epsilon$ the leading behavior of the r.h.s. of \eqref{rel1} is obtained by rescaling $b \to \epsilon b$ leading to 
\be \label{rel11}
\theta_c-\theta \simeq A  C_k \epsilon^{2 k - 1/2} \quad , \quad C_k =  \int_0^{+\infty} db \frac{2 b +1}{b^2 (b+1)^2} b^{2k + 1/2} = \frac{- \pi}{\cos(2 \pi k)} \frac{1+4 k}{2} 
\ee 
where the integral is convergent for $\frac{1}{4} < k < \frac{3}{4}$. 
Hence we obtain $z^*-a_e \sim ((\theta_c- \theta)/A C_k)^{\frac{2}{4 k - 1}}$
as given in \eqref{new}, and by integration using \eqref{dmu}
\bea
\mu(\theta) - \mu_e(\theta) \simeq - \frac{4 k-1}{4 k+1} A C_k \left(\frac{\theta_c- \theta}{AC_k}\right)^{\frac{4 k + 1}{4 k - 1}} 
\eea

\section{More on the variational calculation}
\label{app:variational} 

\subsection{Single active column in the bulk to leading order: test of the permutation invariance}
\label{app:permutation} 

Let us consider the case of a single active column is in position $j=n$, $\epsilon_n = - a_n <0$ and all other $a_j=0$. Let us determine the ground state energy ${\cal E}^0_N(x)$ and 
the occupation length $\ell_j^0$ to leading order in $N$ at large $N$. In that case one must 
minimize
\be \label{tomin} 
 {\cal E}^0_N(x) = \min_{x_{n-1} < x_n \in [0,x]} \left[B_n(x_n)- B_n(x_{n-1}) - a_n (x_n-x_{n-1}) 
 +G_{n}(0,x_{n-1})
+G_{N-n}(x_{n},x)\right] 
\ee
where $G_m(y,z)=\min\limits_{y=x_1<x_2<\dots<x_{N-1}<x_N=z}\sum\limits_{i=2}^{N}B_i(x_{i},x_{i-1})$ represents the ground state energy of a segment of polymer with no active column, and is defined in \eqref{GG}.
For the simplest application of the variational calculation we now assume that $n \sim N$ as $N \to 
\infty$. We use that $G_m(y,z) \simeq - 2 \sqrt{m(z-y)} + o(m)$ at large $m$. 
One can neglect the fluctuations and other subleading terms and obtain to leading order in $N$
\be
{\cal E}^0_N(x)  \simeq 
\min_{x_{n-1} < x_n \in [0,x]} \left[- a_n (x_n-x_{n-1}) - 2 \sqrt{n x_n} 
- 2 \sqrt{(N-n) x_n} \right]
\ee
One finds that the minimum is attained for
\be
x_{n-1}^0 = \frac{n}{a_n^2}  \quad , \quad x_{n}^0 = x - \frac{N-n}{a_n^2} 
\ee 
hence the occupation length is $\ell_n^0 = x_{n}^0 - x_{n-1}^0= (x- \frac{N}{a_n^2})_+$,
i.e. it is the same (with $a_n \to a_1$) as for the case $n=1$ studied in the text, 
see Eq. \eqref{29}, and the energy
\be
{\cal E}^0_N(x)  \simeq - (a_n x + \frac{N}{a_n} )
\ee
is also the same as for $n=1$. The above is valid in the localized phase, 
$a_n > \sqrt{\frac{N}{x}}$. The simple argument is thus in agreement with the
general property of invariance by permutation of the columns.

\subsection{Two macroscopic groups of columns}
\label{app:2groups} 

Let us turn to another interesting example where $N_1= N p$ columns have the same $\epsilon_j = - a_1$ and
$N_2=N (1-p)$ have $\epsilon_j = - a_2$. Then denoting $X$ the total length in the region $a_1$ one obtains to leading order in $N$
\bea
&& {\cal E}^0_N(x) = - \max_{0<X<x} [ a_1 X - G_{N_1}(0,X) + a_2(x-X) - G_{N_2}(X,x) ] \\
&& \simeq - \max_{0<X<x} [ a_1 X + 2 \sqrt{p N X} + a_2(x-X) + 2 \sqrt{(1-p) N (x-X) }  ]
\eea
With no loss of generality we can choose $a_2=0$ and $a_1>0$, and study
\bea \label{gse} 
&& {\cal E}^0_N(x) 
\simeq - N \max_{0<\tilde X<\theta} [ a_1 \tilde X + 2 \sqrt{p \tilde X} + 2 \sqrt{(1-p) (\theta -\tilde X) }  ] 
\eea
where $\tilde X=X/N$ and $\theta=x/N$.
The minimization equation is 
\be
 \frac{\sqrt{1-p}}{ \sqrt{\theta - \tilde X} } - \frac{\sqrt{p}}{ \sqrt{\tilde X} }  = a_1
\ee 
Since the derivative of the l.h.s is increasing from $]-\infty,+\infty[$ there is always a
unique root $\tilde X \in [0,\theta]$ which, reported into \eqref{gse} gives the ground state
energy as a function of $a_1$ and $\theta$. There is thus no localization phase transition
in that case. This is because one is looking here at the largest eigenvalue of the matrix
${\cal M}(x)$. However it is known that there can be a phase transition in the middle
of the spectrum (see \cite{claeys2019critical} for references), which is thus relevant
for the $M = \mathcal{O}(N)$ line problem. 

\subsection{Delocalized phase: distribution of the occupation length}

\label{app:sub:delococc} 

{\bf Preliminary remark}. Let us define the positive random variables $v,\omega$ and 
their scaled versions $\tilde v, \tilde \omega$ as
\be
v = \frac{\sigma^2}{\mu} \tilde v  = 
\max_{y \geq 0}[ \sigma B(y) - \mu y ]   \quad , \quad \omega
= \frac{\sigma^2}{\mu^2} \tilde  \omega = \argmax_{y \geq 0} [ \sigma B(y) - \mu y ]  
\ee 
The JPDF of $v$ and $\omega$ is known. One has \cite[Chapter IV, item 32]{HandbookBrownian}  the Laplace transform
\be
\mathbb{E}[ e^{- r \tilde v - s \tilde \omega} ] = \frac{2}{\sqrt{2 s  + 1} + r + 1} 
\ee 
which leads to the the JPDF and the marginals, with $\tilde v>0,\tilde \omega>0$
\be
P(\tilde v, \tilde \omega) = \sqrt{\frac{2}{\pi}} \frac{\tilde v}{\tilde \omega^{3/2}} 
e^{- \frac{(\tilde v + \tilde \omega)^2}{2 \tilde \omega}}  \quad , \quad 
P(\tilde \omega) = \sqrt{\frac{2}{\pi \tilde \omega}} e^{- \frac{\tilde \omega}{2}} - {\rm Erfc}( \sqrt{\frac{\tilde \omega}{2}}) \quad , \quad P(\tilde v)= 2 e^{- 2 \tilde v} 
\ee 
which gives the formula of the text \eqref{pmu}. Note that the PDF of the value of the maximum, $v$, is
simply an exponential distribution of parameter $2 \mu$. 

\bigskip

Let us now ask what is the distribution of the occupation length in the delocalized
phase. Let us start with the case where all $a_j=0$ and focus on the first column.
The calculation has similarities with the one in Section \ref{subsubsec:critical}, but is different. Using the estimate \eqref{TW2} with $\delta x=-\ell_1$, one has (anticipating that $\ell_1=\mathcal{O}(1)$ )
\bea
 \ell_1^0 \; &&=\argmin_{\ell_1\in [0,x]} \left[ B_1(\ell_1) + G_{N-1}(\ell_1,x) \right] \\
&& \simeq \argmin_{\ell_1\in [0,x]} \bigg[B_1(\ell_1) 
- 2 \sqrt{N (x - \ell_1)} - \frac{\sqrt{x}}{N^{1/6}}
{\cal A}_2(- \frac{N^{1/3} \ell_1}{2 x})
\bigg] \\
&& \simeq \argmin_{\ell_1\in [0,x]} \bigg[B_1(\ell_1) + \frac{1}{\sqrt{\theta}} \ell_1 + \tilde B(\ell_1) \bigg] \\
&&= \argmin_{\ell_1\in [0,x]} \bigg[\sqrt{2} B(\ell_1) + \frac{1}{\sqrt{\theta}} \ell_1 \bigg] 
\eea
where in the first equation of the last line the second term comes from the expansion of the second one in the line above, and the
Brownian motion $\tilde B(\ell_1)$ comes from the usual estimate of the Airy process near zero. Hence using the preliminary remark above we find that the occupation length has the same
distribution as was found in the text in \eqref{occ12} in some limit of the critical regime (but here with a different
scale $\mathcal{O}(1)$) 
\be \label{minmin} 
\ell_1^0 \simeq 2  \tilde \omega \, \theta = 2  \tilde \omega \, \frac{x}{N}  \quad , \quad P(\tilde \omega) = \sqrt{\frac{2}{\pi \tilde \omega}} e^{- \frac{\tilde \omega}{2}} - {\rm Erfc}( \sqrt{\frac{\tilde \omega}{2}}) 
\ee 
We note that $\mathbb{E} [\tilde \omega] = \frac{1}{2}$ hence
\be
\overline{\ell_1^0}^B = \frac{x}{N} 
\ee 
which is consistent with all columns having the same mean occupation length in that case.
Although we will not do it in details, it is clear that the same variational formula with two Brownian on each sides
will arise if one looks at any other column, with the same result.\\

It is more difficult to study the same question in presence of many active columns,
e.g. when all $a_j$ are non-zero, in the delocalized phase. However in the case of
a single active column it is easy to obtain the result. 
One can indeed extend the above calculation to the case where $a_1>0$ and all other $a_{j \geq 2} =0$.
It amounts to add the term $- a_1 \ell_1$ into \eqref{minmin} and we see that it simply
changes $\frac{1}{\sqrt{\theta}}  \to \frac{1}{\sqrt{\theta}}  - a_1$ in the last line, this leads
to 
\be \label{minmin2} 
\ell_1^0 \simeq \frac{2 \theta}{(1- a_1 \sqrt{\theta})^2} \tilde \omega 
\ee 
which is valid for $\theta<\theta_c=\frac{1}{a_1^2}$, i.e. in the delocalized phase.
It shows how the occupation length diverges upon approaching the transition from
the delocalized phase side.\\

If we now get closer to the transition and set $\frac{\theta}{\theta_c}=1+ \frac{\delta}{N^{1/3}}$ we find
\be \label{minmin20} 
\ell_1^0 \simeq \frac{8 N^{2/3}}{a_1^2 \delta^2} \tilde \omega 
\ee 
which perfectly agrees with the result obtained in \eqref{occ12}, using that
$\omega$ there equals $4 \tilde \omega$. 
The two regimes, (i) inside the delocalized phase where $\ell_1^0=\mathcal{O}(1)$ and, (ii) inside the critical regime where $\ell_1^0=\mathcal{O}(N^{2/3})$, thus match very smoothly, with the same random
variable $\tilde \omega$. \\

{\bf Remark.}
Consider now the overlap $\vert\psi_1(j)\vert ^2$, whose mean value is related to
the average occupation length via Eq. \eqref{meanconnection}. Consider the
case where all $a_j=0$, where ${\cal M}(x)$ is a GUE matrix. The PDF of the overlap can be obtained remembering that for the GUE, the eigenvectors are independent from the spectrum and are uniformly distributed on the unit sphere of $\mathbb{C}^N$.  As $N\to +\infty$, the real and imaginary parts of their components become independent Gaussians, hence
\be
\vert\psi_1(j)\vert ^2 \, \inlaw \, \frac{u_j^2 + v_j^2}{2 N} \,  \inlaw \, \frac{1}{2 N} \, \chi \quad , \quad \chi = \chi_{\beta=2}^2 \quad , \quad 
P(\chi)=\frac{1}{2} e^{-\chi /2} \Theta(\chi) 
\ee
where $u_j,v_j$ are independent standard Gaussians and the normalizing factor is determined from the mean of the constraint
$\sum_{j=1}^N \vert\psi_1(j)\vert ^2 = 1$. Hence at large $N$ the overlap
$\vert\psi_1(j)\vert ^2$ is $1/(2 N)$ times a chi-square $\chi_\beta^2$ distributed random variable with a parameter $\beta=2$ (see Refs.~\cite{diaconis1987dozen,gorin2018kpz} for more details).
It is useful to recall the Laplace transform $\mathbb{E}[e^{-z\chi_\beta^2}]=\frac{1}{(1+2z)^{\frac{\beta}{2}}}$ for $z>0$). Since $\mathbb{E}[\chi]=2$ one finds that 
$\overline{  \vert\psi_1(j)\vert ^2}^V=\frac{1}{N}$ which is consistent with 
\eqref{meanconnection} and $\overline{ \ell_j^0 }^B=\frac{x}{N}$. However we see explicitly 
that the PDF's of $N \vert\psi_1(j)\vert ^2$ and of $\ell^0_j$ in \eqref{minmin} are different.

\end{widetext}

\newpage{\pagestyle{empty}\cleardoublepage}


\begin{thebibliography}{100}
\bibitem{footnoteGinibre} This is reminiscent, but quite different from the Saturn effect in the real Ginibre ensemble 
of non hermitian random matrices, where there is no localisation, see e.g. \cite[Fig. 1]{BaikBothner}.
\bibitem{footnotenew1} This follows from \cite[Theorem 3.1(4)]{O_Connell_2012}, since the Whittaker functions are invariant under permutations of the index, as seen for example from the formula (9) there, and using the formulas in section 4 there for the transformation $\mathcal{T}$W as logarithmic partition sums (see also Proposition 3.1 of 
\cite{OConnellRSK2013}).
\bibitem{footnotePeche} Note that for $b_1=0$, hence $\pi_1=1$, it agrees with the result in 
\cite{Peche_2005}, i.e. 
$\lambda'_1 \simeq 2+N^{-2/3}\rchi$, where $\rchi$ is distributed 
as $F_{k+2}^{TW}$ with $k=1$, given in Eq. (5) there.
\bibitem{footnoterestriction} This is because we are working for a fixed set of $a_j$. For i.i.d. variables
there is an additional term $N^{1/2} f(z)$ in the saddle point equations below.
At the saddle point of interest this eventually leads to a correction
$N^{1/2} f'''(z^*)$ which is subdominant w.r.t. $N \varphi^{(3)}(z) w^3$.
\bibitem{footnote2} For non-coinciding points, there are formulas such as 
\cite[Theorem 3.4]{NeilMultiLayer}, and its discrete / zero-temperature versions. See also
Refs. \cite{BorodinShiftInvariance,DauvergneShift} for interesting properties upon shifting the endpoints.
\bibitem{footnoteBoro} Note that the constant $c$ is missing in Theorem 1.3 (b) in
\cite{borodin2014free}. It can be restored from the derivation
in Section 2.1 there.
\bibitem{footnotebackward} Note that this does not guarantee the stability to larger scale excitations.
The effect of the distribution of inter-kink distance has also been neglected.
\end{thebibliography}
\end{document}